\documentclass{aa}
\usepackage{graphicx}               
\usepackage{placeins}               
\graphicspath{{figures/}}           
\usepackage{txfonts}                
\usepackage{lscape}                 
\usepackage{booktabs}
\usepackage{tabularx}
\newcolumntype{Y}{>{\centering\arraybackslash}X}

\usepackage[
    colorlinks=true,
    urlcolor=cyan,
    linkcolor=blue,
    citecolor=blue
]{hyperref}
\usepackage{comment}
\usepackage{multirow}
\usepackage{natbib}
\bibpunct{(}{)}{;}{a}{}{,}          
\defcitealias{jtecers_2023}{JTERS23}
\defcitealias{carter_2024}{Carter \& May et al. (2024)}

\usepackage{amsmath}
\usepackage{mathabx}
\usepackage{siunitx}
\usepackage[version=4]{mhchem}
\DeclareSIUnit{\solarradius}{R_\sun}
\DeclareSIUnit{\solarmass}{M_\sun}
\DeclareSIUnit{\jupiterradius}{R_J}
\DeclareSIUnit{\jupitermass}{M_J}
\DeclareSIUnit{\au}{au}
\DeclareSIUnit{\gauss}{G}
\DeclareSIUnit{\erg}{erg}
\DeclareSIUnit{\year}{yr}
\DeclareSIUnit{\day}{d}
\DeclareSIUnit{\dex}{dex}
\DeclareSIUnit{\bar}{bar}

\usepackage[nolist, nohyperlinks]{acronym}
\acrodef{hst}[HST]{the Hubble Space Telescope}
\acrodef{ers}[ERS]{Early Release Science}
\acrodef{cia}[CIA]{collision-induced absorption}
\acrodef{vmr}[VMR]{volume-mixing ratio}
\acrodef{pt}[p-T]{pressure-temperature}

\usepackage[normalem]{ulem}
\usepackage{color}

\begin{document}

\title{Knobs and dials of retrieving JWST transmission spectra}
\subtitle{II. Impacts of pipeline-level differences on retrieval posteriors}
\titlerunning{Knobs and dials of retrieving JWST transmission spectra II}

\author{
S. Schleich \inst{1}
\and
S. Boro Saikia \inst{1}
\and
Q. Changeat \inst{2,3}
\and
M. Güdel \inst{1}
\and
A. Voigt \inst{4}
\and
I. Waldmann \inst{3}
}

\institute{
Department of Astrophysics, 
University of Vienna,
Türkenschanzstrasse 17, 1180 Vienna, Austria\\
\email{simon.schleich@univie.ac.at}
\and		
Kapteyn Astronomical Institute, 
University of Groningen, PO Box 800, 
9700 AV Groningen, The Netherlands
\and
Department of Physics and Astronomy, 
University College London, 
Gower Street, WC1E 6BT London, United Kingdom
\and
Department of Meteorology and Geophysics, 
University of Vienna,
Josef-Holaubek-Platz 2, Vienna, Austria
}

\date{Received --; Accepted --}


\abstract
{
Since the launch of the \emph{James Webb} Space Telescope (JWST), observations of exoplanetary atmospheres have experienced a revolution in data quality.
As atmospheric parameter inferences heavily depend on the underlying data set, a re-evaluation of current methodologies is warranted to assess the reliability of these results.
}
{
We investigate the impact of variations in input spectra on atmospheric retrievals for the hot Jupiter WASP-39~b using JWST transit data.
Specifically, we analyse the reliability of parameter estimation results from random perturbations of the underlying spectrum, and their sensitivity to the use of three transmission spectra derived from the same observational data.
}
{
Using the NIRSpec PRISM observation from a single transit of WASP-39~b, we perform retrievals with the \texttt{TauREx} framework. 
As an input baseline, we use a transmission spectrum derived in our work using the \texttt{Eureka!} data reduction pipeline.
To investigate the reliability of these retrieval results, we analyse the behaviour of parameter posterior distributions under deviations of this spectrum.
To mimic random noise, we perform a set of retrievals on scattered instances of the spectrum produced in this work.
We compare this to differences resulting from retrievals based on existing spectra reduced from the same raw observation.
}
{
Our analysis identifies three types of parameter posterior distributions: (1) Stable, Gaussian distributions for species constrained across the entire spectrum (e.g. \ce{H2O, CO2}); (2) Uniform posteriors with upper bounds for species with weak constraints (e.g. \ce{CO, CH4}); and (3) Unstable, heavy-tailed posteriors for species constrained only by minor spectrum features (e.g. \ce{SO2, C2H2}). We find that other parameters, like the planetary radius and pressure-temperature profile, are stable under spectral perturbations.
}
{
Parameter posterior distributions are different for atmospheric retrievals performed on independently reduced transmission spectra derived from the same raw data.
This makes robust interpretation difficult, particularly for skewed distributions.
Based on this, we advocate for careful assessment and selection of credible interval sizes to reflect this.
}
%

%
%

\keywords{
Methods: statistical
--
Planets and satellites: atmospheres
--
Planets and satellites: composition
--
Techniques: spectroscopic
}

\maketitle
\section{Introduction}\label{sec:introduction}
With the launch of the \emph{James Webb} Space Telescope \citep[JWST,][]{gardner_2023}, the frontier of exoplanetary sciences has been pushed forward significantly.
Among the many advances JWST has brought are observations of exoplanetary atmospheres of a quality far beyond previously available observatories.
Together with the ever increasing inventory of known exoplanets, these advancements are starting to enable the inference of population-level planetary parameters \citep{fu_2025}, as well as detailed studies of the atmospheres of individual exoplanets.

One of the most deeply studied exoplanets with JWST is WASP-39~b \citep{faedi_2011}, a Saturn-mass hot Jupiter selected for the early release science (ERS) programme for transiting exoplanets. 
WASP-39~b has been observed will all four instruments of JWST, which has lead to the detection of several atmospheric trace species, such as \ce{CO2}, \ce{H2O}, \ce{Na}, and \ce{K} \citep[e.g.][]{ahrer_2023, alderson_2023, rustamkulov_2023, feinstein_2023}.
It has also lead to the first detection of \ce{SO2} in the atmosphere of an exoplanet, a product of photochemical processes \citep{alderson_2023, tsai_2023, powell_2024}, marking one of the early milestones in exoplanetary sciences achieved with JWST.
However, putting precise constraints on identified atmospheric characteristics has proven to be difficult, as the results of characterisation techniques are sensitive to model setup assumptions and the steps taken in the data reduction process \citep[e.g.][]{constantinou_2023, lueber_2024}.
The jump in data quality with JWST also poses new challenges, which need to be addressed to appropriately adjust the methodology used to infer the atmospheric properties of exoplanets.

The most prevalent technique used to characterise exoplanetary atmospheres is called \emph{atmospheric retrieval}.
This method has been used to infer atmospheric properties from data of  different observational methods, including transmission, emission, and phase curve spectroscopy, as well as direct imaging.
We refer to, for instance, \citet{madhusudhan_2019} or \citet{barstow_2020} for comprehensive reviews on this topic.
With the increased spectral resolution, precision, and wavelength coverage of JWST, the assumptions and approaches used in atmospheric retrievals need to be adjusted in a manner that reflects the increased information content in these state-of-the-art observations.

Atmospheric retrieval is a data-driven inverse modelling technique, and the parameters inferred from it depend on two main factors.
One of these is the forward model used to represent the atmospheric observation.
In atmospheric retrievals, these models are commonly constructed as one-dimensional vertical slices of the probed atmospheric region, possibly under-representing the inherent three-dimensional nature of a planetary atmosphere \citep{blecic_2017,caldas_2019,espinoza_2021,pluriel_2022}.
Adjusting forward models to properly reflect the information contained in current observational data is a key factor to avoid characterisation biases from oversimplified assumptions \citep[e.g.][]{changeat_2019,al-refaie_2022, schleich_2024}.

The other factor is the underlying observational data, on which the parameter estimates of the atmospheric forward model are optimised.
Assumptions and techniques applied at different stages of the data reduction process can influence the resulting atmospheric spectrum, propagating into the results of atmospheric characterisation efforts.
At the highest level, combining atmospheric spectra from multiple instruments introduces the problem of agreement between mean transit depths.
Treatment of these potential offsets between spectra of the same planet can propagate into different conclusions about its atmospheric nature \citep{madhusudhan_2023,edwards_2024-1}.
When only considering individual instruments, assumptions such as temporal and chromatic binning \citep{morello_2022-1, davey_2025}, as well as the characterisation of stellar limb-darkening \citep{morello_2017,keers_2024} act as another source of bias influencing the reliability of atmospheric characterisation results. At the lowest level, individual data reduction pipelines and techniques could introduce disagreements into derived atmospheric spectra, which propagates into atmospheric characterisation results through discrepancies in estimated parameter values \citep[e.g.][]{mugnai_2024}.
Being aware of, and accounting for, all these sources of biases will be imperative when maximising the potential for atmospheric characterisation that JWST is providing to us.    

Our goal in this work is to investigate how random and systematic differences in instances of a transmission spectrum propagate into the results of atmospheric retrievals.
Firstly, we perform end-to-end data reduction of NIRSpec PRISM observation of the hot Jupiter WASP-39~b using the open-source pipeline \texttt{Eureka!} \citep{bell_2022} to derive a transmission spectrum.
We perform forward model tuning on the basis of this spectrum, and investigate the impact of random data perturbations by applying standardised atmospheric retrievals to scattered instances of this spectrum.
We also analyse the results of the atmospheric retrieval achieved on two more transmission spectra of WASP-39~b.
These spectra were derived from the same underlying data used in our work, a single-transit observation with the NIRSpec PRISM instrument configuration.
Next to the spectrum produced in this work, we consider the \texttt{Eureka!}-derived transmission spectrum presented in \citet{rustamkulov_2023}, the first reported transmission spectrum considering the full wavelength range of NIRSpec PRISM and treating partial saturation.
Additionally, we use the transmission spectrum presented in \citetalias{carter_2024}, which was produced in an effort to homogenise the analysis of all available near-infrared observations of WASP-39~b.
We note that \citetalias{carter_2024} adopted the spectral time series data from \citet{rustamkulov_2023} which was reduced with the \texttt{FIREFLy} pipeline.

\section{Observational data}\label{sec:obs-data}

WASP-39~b is a Saturn-mass hot Jupiter ($M_\mathrm{p} = \SI{0.281}{\jupitermass}$ and $R_\mathrm{p} = \SI{1.279}{\jupiterradius}$) orbiting a late G-type star at a period of approximately \SI{4}{\day} \citep{faedi_2011}. 
It is part of the JWST early release science (ERS) programme for transiting exoplanets (PID: 1366, PI: N. Batalha, Co-PIs: J. Bean and K. Stevenson) as a target for transmission spectroscopy.  
The JWST panchromatic transmission spectroscopy observations of this target include transits observed with all near-infrared (NIR) instruments of JWST.
A follow-up observation stipulated by the identification of \ce{SO2} in the atmosphere of WASP-39~b \citep{alderson_2023,tsai_2023} also added a mid-infrared (MIR) transmission spectrum \citep{powell_2024}.
This makes the panchromatic transmission spectrum of WASP-39~b one of the most extensive ones produced by JWST so far.

The data set we analyse in this work is the singular transit NIRSpec PRISM observation of WASP-39~b, taken on 10 July 2022 (14:05 -- 23:38 UT). 
The raw observational data (non-calibrated Stage 1b, or \texttt{.uncal}-files) were queried from the Mikulski Archive for Space Telescopes (MAST).

\begin{table}
    \centering
    \caption{System parameters for WASP-39.}
    \begin{tabularx}{\hsize}{YcY}
        \toprule\toprule
        Parameter & Value & Assoc. unit \\
        \midrule
        \multicolumn{3}{c}{\textbf{WASP-39}} \\
        \midrule
        $M_*$                 & $0.918 \pm 0.047$ & \si{\solarmass} \\
        $R_*$                 & $1.013 \pm 0.022$ & \si{\solarradius} \\
        $T_\mathrm{eff}$      & $5485 \pm 50$     & \si{\kelvin} \\
        $[\ce{Fe} / \ce{H}]$  & $0.01 \pm 0.09$   & - \\
        $\log_{10}{g}$        & $4.41 \pm 0.15$   & \si{\centi\meter\per\square\second} \\
        \midrule
        \multicolumn{3}{c}{\textbf{WASP-39~b}} \\
        \midrule
        $M_\mathrm{p}$ & $0.281 \pm 0.032$                  & \si{\jupitermass}\\
        $R_\mathrm{p}$ & $1.279 \pm 0.040$                  & \si{\jupiterradius}\\
        $P$            & $4.0552941 \pm 3.4 \times 10^{-6}$ & \si{\day}\\
        \bottomrule
    \end{tabularx}
    \label{tab:wasp39-system-pars}
    \tablefoot{
        Stellar and planetary parameters are taken from \protect\citet{mancini_2018}.
    }
\end{table}

\subsection{Data reduction}

We use the open-source pipeline \texttt{Eureka!} \citep{bell_2022} to perform end-to-end data reduction on the raw JWST data products.
\texttt{Eureka!} acts both as a wrapper for the official \texttt{jwst} pipeline \citep{bushouse_2024} in its first stages, and as a framework to perform light-curve fitting.
\texttt{Eureka!} is highly modular, supporting the fine-tuning of data reduction steps to ensure optimal precision in the produced data products.
We refer to Appendix~\ref{app:data-reduction} for a detailed description of the data reduction steps taken in this work, and summarise the individual stages below.

The first three stages of \texttt{Eureka!} are concerned with detector-level data processing, as well as calibration and reduction. These stages transform the raw observational data into reduced dynamic light-curves.
We perform stages 1 and 2 with mainly default assumptions.
For the \texttt{jump\_rejection} step in stage 1, we choose a threshold of $10\,\sigma$ to counteract excessive pixel flagging connected to the low number of groups in each integration \citep[][]{rustamkulov_2023}.
To mitigate the effects of $1/f$-noise, we perform group-level background subtraction (GLBS) in this stage.
The \texttt{refpix} step is omitted in this stage, as there are no reference pixels on the subarray used for this observation \citep{birkmann_2022}.
We also omit the \texttt{flat\_field} step in stage 2, which did not work as intended at the time of data reduction \citep[e.g.][]{alderson_2023,sarkar_2024}.
In stage 3, we restrict the extracted detector region to $x > 160$ based on a conservative saturation threshold of 60\%.
We perform the spectral extraction with a combination of $(6, 9)$ for the pixel-width of the aperture and background, respectively.

The final three stages of \texttt{Eureka!} process the dynamic light-curve data, and perform light-curve fitting to extract a transmission spectrum.
In stage 4, we extract the spectroscopic light-curves at a detector-pixel level.
We flag individual channels with a noise level higher than a factor of 1.75 compared to noise-budget simulations as deviations, excluding them from further analysis.
The light-curve fitting in stage 5 is done using the \texttt{batman} Python package \citep{kreidberg_2015}.
We fit a combined astrophysical and systematics model to each spectroscopic, as well as the integrated white light-curve.
We use pre-calculated limb-darkening coefficients from the \texttt{ExoTiC-LD} Python package \citep{grant_2022}.
Lastly, we bin the transmission spectrum into fixed groups of 3 pixels. This accounts for the typical instrument resolution element size of 2.2 pixels for NIRSpec \citep{jakobsen_2022}.
We show the final transmission spectrum produced in this work in Fig.~\ref{fig:existing-spectra-comparison}.
For clarity, we refer to the transmission spectrum produced in this work as SP-TW (`\emph{Sp}ectrum - \emph{T}his \emph{w}ork') from here onwards.

\subsection{Panchromatic perturbation of the spectrum}
The results of atmospheric retrievals are anchored to the underlying data set used in the inference process.
As these data guide the parameter estimation, in a Bayesian inference framework they are assumed to be the `true' state.
However, the parameter estimates derived from a forward model have a non-uniform dependence on the data points of a transmission spectrum.

One method used to judge the importance of individual spectral data points in the parameter estimation process is a leave-one-out cross-validation (LOO-CV) technique computing the expected log pointwise predictive density (elpd$_\mathrm{LOO}$).
LOO-CV works by fitting a given model to a data set with one data point removed \citep{gelman_2014}.
In the analysis of pre-JWST data, this method requires on the order of several tens of retrievals, each time performing an atmospheric retrieval while excluding an individual spectral data point.
However, for current state-of-the-art data, this requirement would be increased by several orders of magnitude, accounting for the increased resolution and wavelength coverage of JWST.
Current applications of LOO-CV make use of the PSIS approximation \citep{vehtari_2017} to avoid this computational boundary \citep[e.g.][]{welbanks_2023, murphy_2025}.

Additionally, atmospheric absorbers produce correlated signals within an atmospheric spectrum.
A comprehensive investigation of the posterior dependence on the underlying spectrum would also require a validation analysis under all possible combinations of excluded data points.
This would be computationally extremely expensive when considering data sets as provided by JWST, with hundreds or thousands of individual data points.

As an initial test of the stability of our retrieval results to perturbations of the underlying data, we therefore opt to produce fully scattered instances of SP-TW (shown in Fig.~\ref{fig:existing-spectra-comparison}, which we consider to be `true').
We successively create randomised instances of SP-TW by drawing new transit depth values using a normal distribution $\mathcal{N}(\mu_{\mathrm{td}, \lambda}, \sigma^2_{\mathrm{td}, \lambda})$, where $\mu_{\mathrm{td}, \lambda}$ and $\sigma_{\mathrm{td}, \lambda}$ represent the baseline transit depth mean and standard deviation, respectively.
We attach the existing transit depth uncertainties, $\sigma_{\mathrm{td}, \lambda}$, to this new transit depth values to create scattered instances of SP-TW.
While this method is not sensitive to possible correlated noise, it provides insight into the stability of the posterior distributions under deviations following a normal distribution.
Considering perturbations from data reduction assumptions, we interpret these scattered instances as deviations following random noise.
These spectra are used as input data for the standardised atmospheric retrieval analysis described in Sect.~\ref{ssec:stat-scatter}.

\subsection{Existing transmission spectra}\label{ssec:existing-spectra}

\begin{figure*}
    \centering
    \includegraphics[width=\hsize]{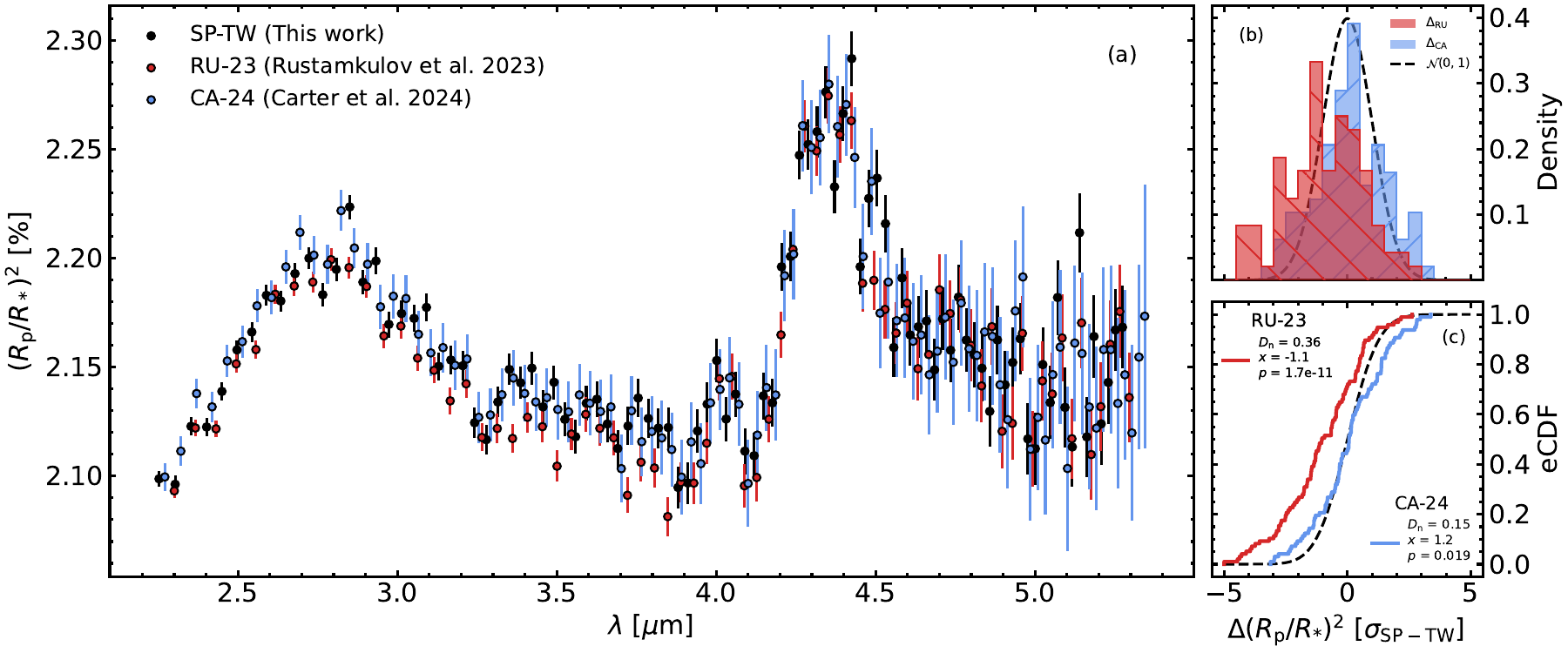}
    \caption{
        Comparison of transmission spectra used in this work. Data associated with the spectrum produced in this work (SP-TW) are shown in black, while data associated with \protect\citet{rustamkulov_2023} (RU-23) and \protect\citetalias{carter_2024} (CA-24) are shown in red and blue, respectively.
        (a) Transmission spectra, showing wavelength (in $\si{\mu\meter}$) on the x-axis and transit depth (in \%) on the y-axis.
        (b) Residual distribution of RU-23 and CA-24, normalised to the transit depth uncertainty of SP-TW. For display purposes, the residuals are binned in steps of $0.5\,\sigma_\mathrm{SP-TW}$.
        (c) Empirical cumulative distribution functions (eCDF) for the residuals of RU-23 and CA-24. The values of a one-sample K-S test for a standard normal distribution are given in the legend of the figure.
        In (b) and (c), the black dashed line represents the PDF and CDF of $\mathcal{N}(0,1)$, respectively.
    }
    \label{fig:existing-spectra-comparison}
\end{figure*}

While we present our own data reduction results for the transmission spectrum of WASP-39~b from a NIRSpec PRISM observation, previous analyses of the same data exist as well.
Using Bayesian inference to estimate model parameters is inherently dependent on the underlying data set.
As recently shown by \citet{mugnai_2024} and \citet{edwards_2024-1}, data-reduction based variations in transmission spectra from Hubble Space Telescope (HST) observations can affect the derived exoplanet atmospheric properties.
This leads to diverging conclusions about individual-, as well as population-level characterisation results.
We are conscious of this potential biasing effect when only considering the transmission spectrum produced in this work. 
We therefore investigate the results of applying the atmospheric retrieval setup described above to existing transmission spectra derived from the same underlying observational data.

\citet{rustamkulov_2023} presented a reduction of the NIRSpec PRISM observation of WASP-39~b with a specific focus on recovering the signal in the saturated region of the detector.
Out of the four different pipeline results presented in their work, we use the \texttt{Eureka!}-based data reduction (referred to as RU-23 from here onwards)\footnote{\url{https://zenodo.org/records/7388032}}. 
Compared to the data reduction performed in this work (as described in Appendix~\ref{app:data-reduction}), RU-23 was derived with several different assumptions.
A detailed description of these data reduction steps can be found in the `Methods' section of \citet{rustamkulov_2023}, but the main differences are listed in Table~\ref{tab-app:dr-diff}.

Additionally, a recent study by \citetalias{carter_2024} reanalysed the full range of observations taken with the near-infrared instruments of JWST as part of the ERS programme.
From the results presented in their work, we use the NIRSpec PRISM transmission spectrum at native instrument resolution, derived with fixed limb-darkening coefficients (referred to as CA-24 from here onwards)\footnote{\url{https://zenodo.org/records/10161743}}.
We note that they base their re-reduction on the spectroscopic time series from the `baseline' reduction result presented in \citet{rustamkulov_2023}, which was derived with the \texttt{FIREFLy} data reduction pipeline.

We show a comparison between the three spectra used in this work in Fig.~\ref{fig:existing-spectra-comparison}.
In its original form, both RU-23 and CA-24 cover the full wavelength range available to NIRSpec PRISM by treating the saturated part of the spectrum.
We constrain RU-23 and CA-24 to the non-saturated wavelength range recovered in our work (approximately \SIrange{2.2}{5.3}{\micro\meter} for SP-TW).
All three spectra show a closely comparable wavelength-dependent behaviour of the transit depth (left panel of Fig.~\ref{fig:existing-spectra-comparison}).
Two absorption peaks, a broad feature centred at approximately \SI{2.7}{\micro\meter}, and a narrower feature centred at approximately \SI{4.4}{\micro\meter} are identifiable in all three cases.
We note that the wavelength maps for all three spectra do not fully coincide.
In all three cases, we are comparing transmission spectra not at the native resolution of the detector (pixel-level), but at a resolution binned to account for the resolution element size connected to the dispersion element.
Differences in the binning will therefore result in slightly offset wavelength bin centres.
To evaluate the differences between the individual spectra, we calculate the point-wise residuals by linearly interpolating RU-23 and CA-24 onto the wavelength map of SP-TW.
The resulting residual distributions are shown in the right panels of Fig.~\ref{fig:existing-spectra-comparison}.
Applying a one-sample Kolmogorov-Smirnov (K-S) test to the normalised residual distributions shows that neither of them are consistent with a standard normal distribution.
The $p$-value of the K-S test applied to the residuals of RU-23 can reject the hypothesis of a standard normal distribution at more than $3\,\sigma$.
This can also clearly be seen in the distribution itself, which has a visible offset from 0, and in the associated empirical cumulative distribution function (eCDF), which is clearly shifted from the CDF of a standard normal distribution (panel c of Fig.~\ref{fig:existing-spectra-comparison}).
In the transmission spectrum, this is visible from \SIrange{2.8}{3.9}{\micro\meter}, where RU-23 shows smaller transit depth values than both CA-24 and SP-TW.
The same test shows that the residuals of CA-24 are non-Gaussian distributed up to a level of 2\,$\sigma$.
In the eCDF of the CA-24 residuals, this is visible through tails of larger residuals.

This implies that both spectra show systematic differences compared to the spectrum derived in our work.
We note that the differences in transmission spectra are most pronounced at shorter wavelengths.
This is a result of the smaller transit depth errors in these regions.
In general, the differences in transit depth vanish toward longer wavelengths, as the error-bar size for all spectra increases significantly with wavelength.

\section{Methods}\label{sec:methods}

To generate atmospheric forward models, and perform parameter estimations, we use the fully-Bayesian inference framework \texttt{TauREx} \citep{waldmann_2015-1,al-refaie_2021}, specifically \texttt{TauREx3.1} \citep{al-refaie_2022}.
\texttt{TauREx} has been used to perform atmospheric retrievals on a variety of exo-atmospheric spectra, ranging from hot Jupiters to terrestrial planets, and encompassing transmission, emission, and phase curve spectroscopy \citep[e.g.][]{tsiaras_2018,changeat_2021,edwards_2021,saba_2022,edwards_2024,voyer_2025}.

To perform parameter estimation with \texttt{TauREx}, we use nested sampling implemented through \texttt{MultiNest} \citep{feroz_2009,buchner_2014}.
In all retrieval cases, we use homogenised values of 700 live points and an error tolerance of 0.5 for the natural logarithm of the evidence.

\subsection{Atmospheric retrieval}\label{ssec:method-retrieval}

We define the extent of the atmospheric pressure domain in our retrievals through 110 layers uniformly distributed within $\log_{10} (p \, \left[\si{\bar}\right]) \in \left[1; -9\right]$, using \ce{H2 and He} as background gases (in a ratio $\mathrm{He} / \mathrm{H}_2 = 0.13$).
We represent the vertical chemical profiles of molecular species through homogeneous \acp{vmr}.
As shown in \citet{schleich_2024}, in atmospheric retrievals of transmission spectra with a data quality of this observation, using \ac{pt} profiles with too few points can introduce a bias in the associated molecular abundances.
We therefore choose the \ac{pt} profile in our retrievals as a heuristic multipoint profile with four fixed pressure nodes.
These pressure nodes are placed at $\log_{10}(p\,[\si{\bar}]) \in \{1, -3, -7, -9\}$.

In the radiative transfer calculations of our forward model, we consider absorption cross-sections from the Exomol project \citep{tennyson_2020, chubb_2021}, as well as the HITRAN \citep{gordon_2022} and HITEMP \citep{rothman_2010} archives. 
We include collision-induced absorption (CIA) from \ce{H2-H2} and \ce{H2-He} pairs, as well as Rayleigh scattering as included in \texttt{TauREx} \citep{cox_2015}. We refer to Table~\ref{tab:opacity-sources} for individual references of the opacity sources.
We consider clouds in our atmospheric forward model through a flat-opacity layer at a specific pressure, $\log_{10} \left(p_\mathrm{cloud}\right)$.

During atmospheric retrievals, we perform parameter estimations for the planetary reference radius, $R_\mathrm{p}$, individual molecular \acp{vmr}, $X_\mathrm{VMR}$, temperature values at individual pressure nodes, $T_\mathrm{i}$, and the cloud-top pressure, $p_\mathrm{cloud}$. These parameters, together with their associated priors used in the inference process, are listed in Table~\ref{tab:retrieval-pars}.

\begin{table}
    \centering
    \caption{Priors for atmospheric retrievals performed in this work.}
    \begin{tabularx}{\hsize}{XYYY}
        \toprule\toprule
        Parameter          & Prior type & Prior range   & Assoc. unit \\
        \midrule
        $R_\mathrm{p}$     & Uniform    & $[1.0; 1.5]$   & R$_\mathrm{J}$ \\
        $X_\mathrm{VMR}$   & LogUniform & $[-12; -0.1]$  & - \\
        $T_\mathrm{i}$     & Uniform    & $[300; 3000]$  & \si{\kelvin} \\
        $p_\mathrm{cloud}$ & LogUniform & $[1; -9]$      & \si{\bar} \\
        \bottomrule
    \end{tabularx}
    \label{tab:retrieval-pars}
    \tablefoot{
        $T_\mathrm{i}$ refers to the temperature nodes within the 4-point p-T profile used in the retrievals, where $T_0$ is equivalent to the bottom of the atmospheric domain. The VMRs for atmospheric trace gases are associated with constant vertical profiles.
    }
\end{table}

\subsection{Model tuning}

We tune our atmospheric forward model by looking for additional molecules as opacity contributions.
To do this, we evaluate the performance of a baseline model (containing \ce{CO2, CO, H2O, and CH4}) against an atmospheric forward model extended by an additional molecular opacity source.
We judge model performance and preference on several metrics.
These are described in more detail in Appendix~\ref{app:model-tuning}, but their application is summarised below.

We judge model preference on the Bayes factor, $B_{\mathrm{m}0}$, between the extended model (indexed with $m$) and the baseline (indexed with `0').
Based on the formalism suggested by \citet{kass_1995}, we consider threshold values of the natural logarithm of the Bayes factor as given in Table~\ref{tab:bayes-factor-thresholds}.
Specifically, we consider a molecular contribution as significantly preferred if $\ln B_{\mathrm{m}0} > 3$ (corresponding to a posterior odds ratio of 20:1 in favour of the extended model to the baseline).
We then create an atmospheric forward model containing all molecules that fulfill this Bayes factor criterion.
We also run retrievals of intermediately constructed models, covering all unique combinations of molecules indicated as preferred in the initial step.

While the Bayes factor evaluates the marginalised likelihood of each model, we also assess model performance based on point-estimates to provide comparative metrics.
Firstly, we use the corrected Akaike information criterion (cAIC, henceforth referred to as $\Psi$) for all models run in the tuning process, which is calculated from the maximised likelihood of each model.
Within a set of competing models, $\Psi$ is used to determine a relative model preference metric, $\Delta_m = \Psi_\mathrm{min} - \Psi$.
Following the prescription of \citet{burnham_2004}, we assume that models with $\Delta_m < 2$ show considerable support compared to the model defining $\Psi_\mathrm{min}$.
Secondly, we calculate the reduced $\chi$-square metric, $\bar{\chi}_\nu^2$, connected to each model. Similarly to $\Psi$, $\bar{\chi}_\nu^2$ is calculated from a point-estimate of the posterior distribution.
In this case, we use the median solution.

We point out that, in contrast to the description in \citet{benneke_2013}, we build the molecular parameter space of our model from the bottom up.
After identifying initially favoured additional contributions, we then analyse the full parameter space against reduced combinations.
For a discussion of this method, and drawbacks associated with it, we refer to Appendix~\ref{app:model-tuning}.

\begin{table}
    \centering
    \caption{Threshold values for evaluating the Bayes factor.}
    \begin{tabularx}{\hsize}{XXc}
        \toprule\toprule
        $\ln B_{\mathrm{m}0}$    & Posterior odds    & Evidence for model $m$ \\
        \midrule
        1 -- 3          & 3 -- 20           & Positive \\
        3 -- 5          & 20 -- 150         & Strong \\
        > 5             & > 150             & Very strong \\
        \bottomrule
    \end{tabularx}
    \label{tab:bayes-factor-thresholds}
    \tablefoot{
        Threshold values of $\ln B_{m0}$ correspond to the formalism suggested by \protect\citet{kass_1995}. We note that $\vert \ln B_{\mathrm{m}0} \vert < 1$ presents an inconclusive statement about model preference, and that $\ln B_{\mathrm{m}0} < 0$ correspond to evidence in favour of the baseline model.
    }
\end{table}

\subsection{Parameter estimation results}
To evaluate the parameter estimation results from \texttt{TauREx}, we use the weighted trace values inferred by \texttt{MultiNest}.
Commonly, parameter estimation results from Bayesian inference networks are reported as credible intervals of sizes equivalent to frequentist confidence intervals.
More specifically, this means that, for example, a `1\,$\sigma$' credible interval will contain approximately 68\% of the posterior samples.
This correlation between the frequentist and Bayesian result statistics holds if the posterior distributions are close to normal distributions.
However, posterior distributions produced in Bayesian inference processes can often be asymmetric, or far from normal distributions in other ways, making the relation between `$\sigma$'-equivalent intervals more difficult.
Credible intervals from asymmetric posterior distributions also do not allow a simple scaling when derived using `1$\,\sigma$' edges.
In these cases, considering intervals of the equivalent width of 1\,$\sigma$ can induce false confidence in the parameter estimation results.
This compounds with limitations on parameter estimation accuracy stemming from current stellar and planetary models used in the characterisation of exoplanet atmospheres \citep[e.g.][]{rackham_2018,macdonald_2020} and from current opacity models \citep[e.g.][]{niraula_2022,niraula_2023}, as well as from inherent model uncertainties \citep{barstow_2020-1,nixon_2024}.

For the retrievals performed here, we report credible intervals encompassing 95\% of the weighted marginalised posterior samples, centred on the posterior distribution median.
We refer to this as `CCI$_{95}$' (centred credible interval) henceforth.

\section{Results and discussion}\label{sec:results}

\begin{figure*}
    \centering
    \includegraphics[width=\hsize]{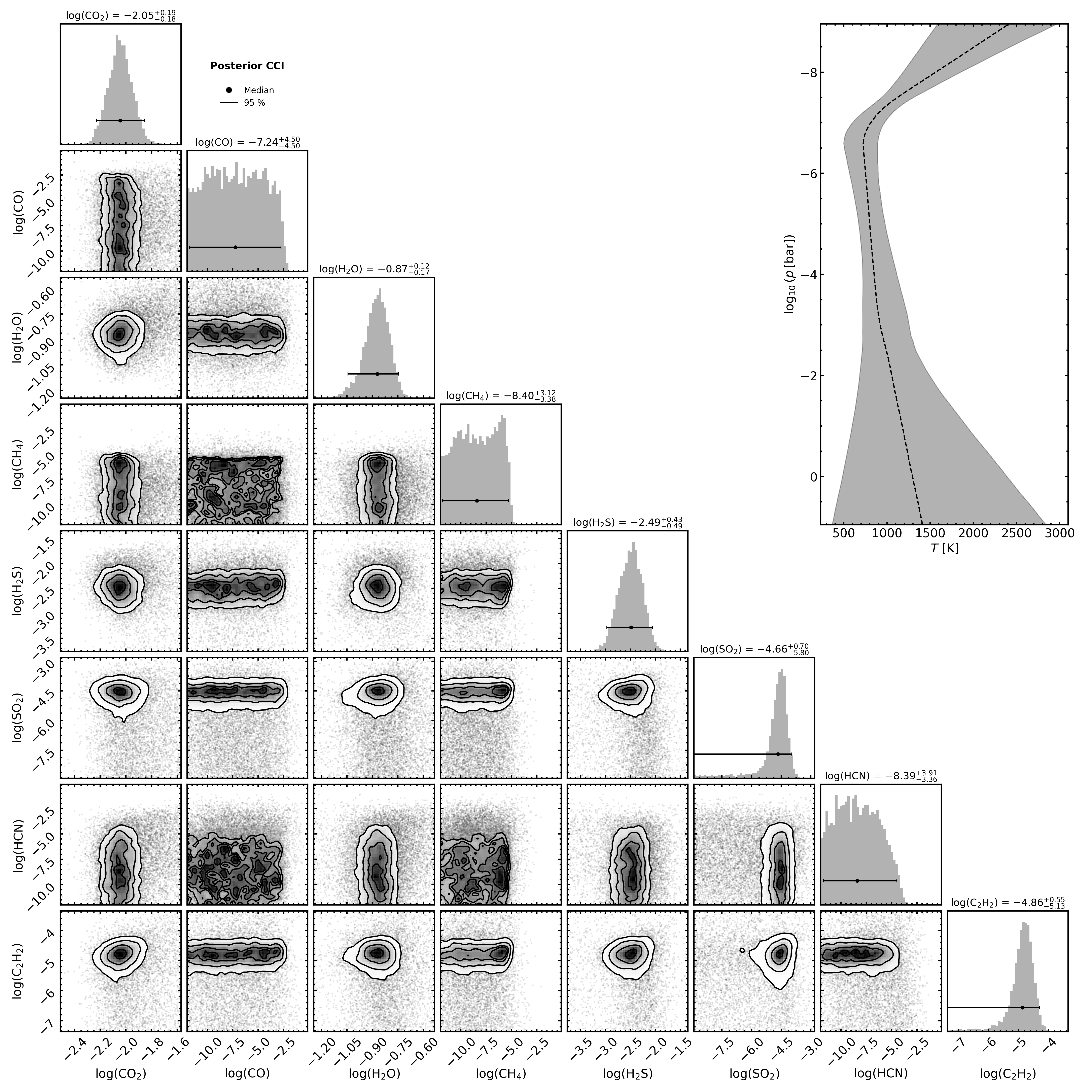}
    \caption{
        Retrieval results of the fiducial model applied to SP-TW (the spectrum produced in our work), showing the posterior distributions of the molecular mixing ratios.
        Marginalised posterior distributions (main diagonal) show the parameter estimate median (points) and CCI$_{95}$ (error bar).
        The inset plot on the top right shows the median retrieved $p$-$T$ profile (dashed line) and CCI$_{95}$ (shaded region).
    }
    \label{fig:corner-thiswork}
\end{figure*}

We perform iterative atmospheric forward model tuning on the transmission spectrum produced in this work (SP-TW).
Our baseline model contains molecular absorption cross-sections for \ce{H2O, CO2, CO, and CH4}.
These species are chosen as major spectrally active \ce{C}- and \ce{O}-bearing species in \ce{H2-He} dominated atmospheres \citep[e.g.][]{molliere_2022}.
We then search for additional molecular absorption by individually adding the species listed in Table~\ref{tab:opacity-sources} to the forward model, and calculating the Bayes' factor relative to the baseline model.
As additional metrics to analyse model preference, we also consider the corrected Akaike Information criterion (cAIC), and the reduced $\chi^2$ value.
We provide a detailed description of the model-tuning process in Appendix~\ref{app:model-tuning}, but summarise the final result below.

We find strong preference (i.e. $\ln B_\mathrm{m0} > 5.0$, or a posterior probability of more than \num{150}:\num{1}) for models that individually include \ce{H2S}, \ce{SO2}, \ce{HCN}, and \ce{C2H2}.
For an extended model containing all four of these molecules as additional sources of absorption, we find a Bayes' factor of $\ln B_\mathrm{m0} = 23.58$.
We also analyse intermediate model iterations based on all combinations of these four molecules.
A full list of all associated model metrics is given in Table~\ref{tab-app:model-tuning-metric}.
Models containing \ce{H2S} produce the biggest increase in posterior probability compared to the baseline model.
Including \ce{H2S} also produces the biggest improvement in the value of $\bar{\chi}^2$.
When considering all three model preference metrics used in this work, we see that the model containing $\{\ce{H2S, SO2, HCN, C2H2}\}$, and the model containing $\{\ce{H2S, SO2, C2H2}\}$ perform on an equivalent level.
They show, respectively, a Bayes factor of \num{23.58} and \num{24.03} and a $\bar{\chi}^2$ value of \num{2.66} and \num{2.63}.
The value of $\Delta_\mathrm{m}$ between them is \num{2.70} (in favour of the model containing $\{\ce{H2S, SO2, C2H2}\}$).
For this work, we adopt the fully extended model containing molecular absorption contributions from $\{\ce{CO2, CO, H2O, CH4, H2S, SO2, HCN, C2H2}\}$ as the fiducial model.

We emphasise that we do not make any claims on the detection or detection significance of atmospheric constituents from our model selection process beyond the posterior probability associated with the Bayes factor.
Using $\sigma$-values derived from model comparison Bayes factors to report the detection of atmospheric constituents runs the risk of misrepresenting the relative nature of $\ln B_\mathrm{m0}$ \citep[][]{schmidt_2025,welbanks_2025}.
In this work, we analyse the impact of transmission spectrum perturbations on parameter posterior distributions derived from atmospheric retrievals.
The inclusion of \ce{HCN} provides an additional point of comparison when performing atmospheric retrievals on the existing transmission spectra of WASP-39~b, which is the reason we choose the fully extended model over the one not containing \ce{HCN}.

\subsection{Atmospheric retrieval of SP-TW}\label{ssec:retireval-sptw}
\begin{figure*}
    \centering
    \includegraphics[width=\hsize]{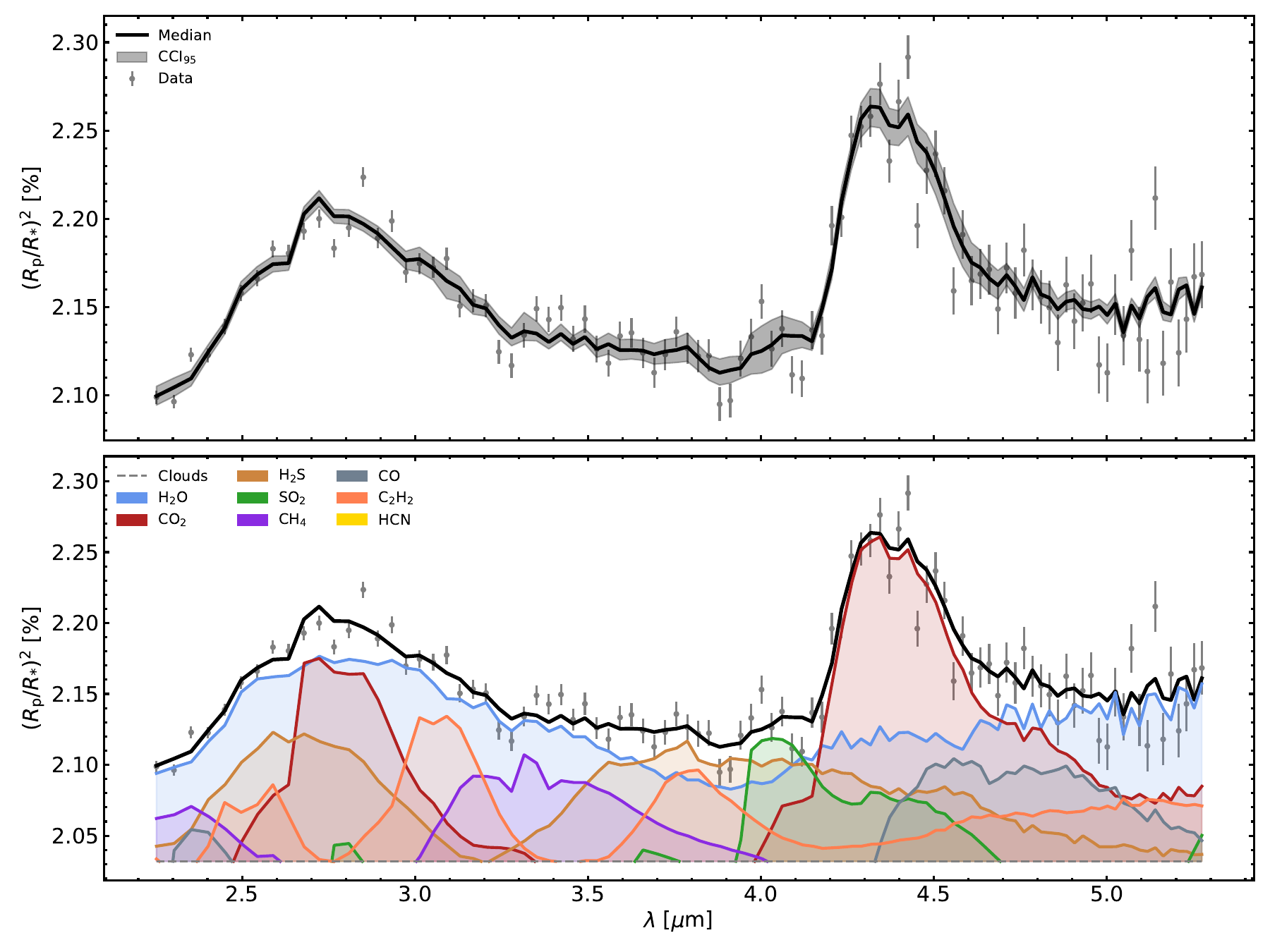}
    \caption{
        Transmission spectrum with model fit solution from model tuning process. Both panels show wavelength (in \si{\micro\meter}) on the x-axis against transit depth (in \%) on the y-axis, as well as the data points and error bars (grey) from the spectrum produced in this work (SP-TW).
        (Top) Median model solution (solid black line) and corresponding 95\% CCI (shaded area).
        (Bottom) Contributions of individual molecular opacity sources (colour-coded by molecule) and the flat-opacity cloud deck (dashed grey line).
    }
    \label{fig:thiswork-bestfit-contrib}
\end{figure*}

In Fig.~\ref{fig:corner-thiswork}, we show the parameter posterior distributions of the molecular mixing ratios of our fiducial model, as well as the retrieved $p$-$T$ profile.
Figure~\ref{fig:thiswork-bestfit-contrib} illustrates the resulting model and uncertainty, as well as the individual opacity contributions based on the finalised model setup.
A detailed list of the parameter estimates for all model parameters is given in Table~\ref{tab:estimated-pars-different-spectra}.

The main molecules contributing to the transmission spectrum are \ce{H2O, CO2, and H2S}.
\ce{H2O} shows a broad absorption feature from \SIrange{2.3}{3.5}{\micro\meter}, with an inferred mixing ratio of $\log_{10}(X_{\ce{H2O}}) \in [-1.04, -0.75]$.
Similarly, \ce{CO2} shows a prominent absorption feature centred at \SI{4.4}{\micro\meter}, and a secondary one centred at \SI{2.8}{\micro\meter}.
We find $\log_{10}(X_{\ce{CO2}}) \in [-2.23, -1.86]$.
In addition to these two constituents, we find a contribution from \ce{H2S}, with $\log_{10}(X_{\ce{H2S}}) \in [-2.97, -2.05]$.
In the spectrum, \ce{H2S} produces a broad feature from \SIrange{3.5}{4.5}{\micro\meter}, and a more narrow feature centred at \SI{2.6}{\micro\meter}.
All three of these species are constrained within \SI{0.5}{\dex} (for \ce{H2O and CO2}) and \SI{1}{\dex} (for \ce{H2S}) in the CCI$_{95}$, indicating that the underlying spectrum provides a significant amount of information to confidently constrain the molecular mixing ratios.

We only find upper limits for the mixing ratios of \ce{CO} and \ce{CH4}.
This is represented in Fig.~\ref{fig:corner-thiswork} by their flat posterior distributions with sharp edges at the estimated upper limits.
\ce{CO} has no significant absorption signature in the wavelength range of SP-TW.
Consequently, the corresponding upper limit of the CCI$_{95}$ $(\log_{10}(X_{\ce{CO}}) < -2.74)$ is a very broad constraint, encompassing most of the prior range.
In contrast to this, \ce{CH4} has a well-known absorption feature centred at \SI{3.4}{\micro\meter}, within the range of SP-TW.
Therefore, the posterior distribution shows a more constraining upper limit of $\log_{10}(X_{\ce{CH4}}) < -5.28$ in its mixing ratio, given the lack of this feature.

Our fiducial model also includes \ce{SO2, C2H2, and HCN}.
The corresponding parameter estimates are $\log_{10}(X_{\ce{SO2}}) \in [-10.46, -3.96]$, $\log_{10}(X_{\ce{C2H2}}) \in [-9.99, -4.30]$, and $\log_{10}(X_{\ce{HCN}}) \in [-11.75, -4.48]$, respectively.
All three of these posterior CCIs are very wide (approximately \SIrange{5.5}{7.0}{\dex}), implying that none of these mixing ratios are well constrained beyond upper limits.
However, we point out the difference in the shapes of their respective posterior distributions (shown in Fig.~\ref{fig:corner-thiswork}).
Equivalent to \ce{CO} and \ce{CH4}, the posterior distribution of \ce{HCN} represents an upper limit with $\log_{10}(X_{\ce{HCN}}) < -4.48$, indicated by a distribution that is close to uniform in shape up to this boundary.
In contrast to that, the posteriors of \ce{SO2} and \ce{C2H2} appear to be close to normal distributions.
This is reflected by the fact that the median values of $\log_{10}(X_{\ce{SO2}}) = -4.66$ and $\log_{10}(X_{\ce{C2H2}}) = -4.86$ are not centred in the CCI$_{95}$, but rather skewed toward their upper edge.
In these cases, calculating a CCI of width `1\,$\sigma$' would provide a wrong sense of confidence in the parameter estimation.
In the example of \ce{SO2}, the equivalent `1\,$\sigma$' CCI (expressed in commonly used point estimate and uncertainty values) is $\log_{10}(X_{\ce{SO2}}) = -4.66^{+0.37}_{-0.85}$.
This would scale the actual lower boundary of the CCI$_{95}$ to almost `7\,$\sigma$', rather than `2\,$\sigma$', highlighting the importance of properly calculating parameter estimation ranges, rather than scaling apparent `$\sigma$' values.

To contextualise these parameter estimates, we compare the results reported in our work to previously published analyses.
We note that the previous works we compare our results to here have all  reported retrievals performed with a `free' chemistry approach, enabling a clear comparison of the parameter estimates.
Before the launch of JWST, one of the main atmospheric species accessible to transmission spectroscopy performed with HST (the previous state-of-the-art) was \ce{H2O}.
\citet{tsiaras_2018} used two transits of WASP-39~b from HST WFC3 to infer an \ce{H2O} mixing ratio of $\log_{10}(X_{\ce{H2O}}) = -5.94 \pm 0.61$.
In contrast to that, \citet{wakeford_2018} combined HST STIS and WFC3 measurements with \textit{Spitzer} IRAC and VLT FORS2 data to derive $\log_{10}(X_{\ce{H2O}}) = -1.37^{+0.05}_{-0.13}$.
Analysing the same data set, \citet{welbanks_2019} report a prior-dependent value of $\log_{10}(X_{\ce{H2O}}) = -0.65^{+0.14}_{-1.83}$.
The \ce{H2O} mixing ratio reported from our retrievals is in agreement with the latter two, but the range of retrieved abundances from these analyses cover 5 orders of magnitude, indicating a poor overall agreement on the constraints of the \ce{H2O} mixing ratio from HST-era observations.

With the advent of JWST, a significantly larger portion of possible atmospheric constituents have become accessible through absorption signatures in the near- and mid-infrared.
\citet{constantinou_2023} performed atmospheric retrievals on the cut-off NIRSpec PRISM spectrum of WASP-39~b \citep{jtecers_2023}.
From a spectrum derived with \texttt{Eureka!}, they infer a \ce{H2O} mixing ratio of $-3.29^{+0.59}_{-0.56}$.
Additionally, they report precise posterior constraints on \ce{CO2}, \ce{SO2}, \ce{CO}, and \ce{H2S}.
Compared to these, the results reported from our retrieval are generally higher by two to three orders of magnitude in the cases of \ce{H2O, CO2, and H2S}.
However, we note that the results reported by \citet{constantinou_2023} 
also show a variation of approximately \SI{1}{\dex} when considering spectra from different pipelines.
The values of \ce{SO2 and CO} fall within the broad constraints reported from our retrievals.
\citet{lueber_2024} studied the information content in the panchromatic transmission spectrum of WASP-39~b.
While using the full-range NIRSpec PRISM transmission spectrum presented in \citet{rustamkulov_2023}, they find a (cloud-model dependent) value of $\log_{10}(X_{\ce{H2O}}) = -3.10^{+0.20}_{-0.19}$.
Similar to \citet{constantinou_2023}, they report a much more precise posterior constraint on \ce{CO and SO2} than our results suggest, with values of $-2.85^{+0.17}_{-0.28}$ and $-5.68^{+0.31}_{-0.62}$, respectively, which fall within the broad constraints reported from our retrievals.
Values for the mixing ratios of \ce{CO2} and \ce{H2S} are smaller by two orders of magnitude compared to our results.

We note that all constraints shown here to contextualise our results were reported as point-estimates with `uncertainties' equivalent to a 1\,$\sigma$ CCI.
While we cannot feasibly reproduce the corresponding posterior distributions to derive CCI$_{95}$ values for a direct comparison, we point out that, especially in cases where the reported uncertainties are asymmetric (such as the \ce{H2O} mixing ratio from \citealt{welbanks_2019}), calculating CCI boundaries would be necessary rather than scaling the reported uncertainties. 
This would potentially result in closer agreement between the here listed results and the parameter estimates from our retrievals.

Lastly, Fig.~\ref{fig:corner-thiswork} also illustrates the retrieved $p$-$T$ profile, represented by the median profile (dashed line) and corresponding CCI$_{95}$ envelope (shaded area).
The temperature values in the middle of the atmospheric pressure domain (given as $T_{p_1}$ and $T_{p_2}$ in Table~\ref{tab:estimated-pars-different-spectra}) are constrained within approximately \SI{500}{\kelvin} and indicate a close-to isothermal behaviour in this region of the atmosphere.
In contrast to that, the temperature nodes at the bottom and top of the atmosphere ($T_{p_0}$ and $T_{p_3}$ in Table~\ref{tab:estimated-pars-different-spectra}) are less well constrained.
The temperature at the top-most pressure node ($p = 10^{-9}$ \si{\bar}) has a CCI width of \SI{1500}{\kelvin} (half the prior space).
The atmosphere is fully transparent at this pressure level, indicating that there is little information to constrain the temperature of this region.
While the thermospheres of hot Jupiters are expected to be heated by the absorption of XUV radiation \citep{fortney_2021}, the retrieval model setup indicates that this temperature increase could be a model degeneracy with the simplified vertical chemical structure \citep{schleich_2024}.
The temperature at the bottom-most pressure node ($p_0 = \SI{10}{\bar}$) is even less constrained with a CCI width if \SI{2400}{\kelvin} encompassing almost the entire prior range.
The top of the flat-opacity cloud deck is constrained to $\log_{10} (p_\mathrm{cloud} [\si{\bar}]) < -3.29$.
Consequently, the lower region of the atmosphere is not accessible in the transmission spectrum of WASP-39~b, resulting in an unconstrained posterior for this temperature node.

\subsection{Sensitivity to random scatter}\label{ssec:stat-scatter}

\begin{figure}
    \centering
    \includegraphics[width=\hsize]{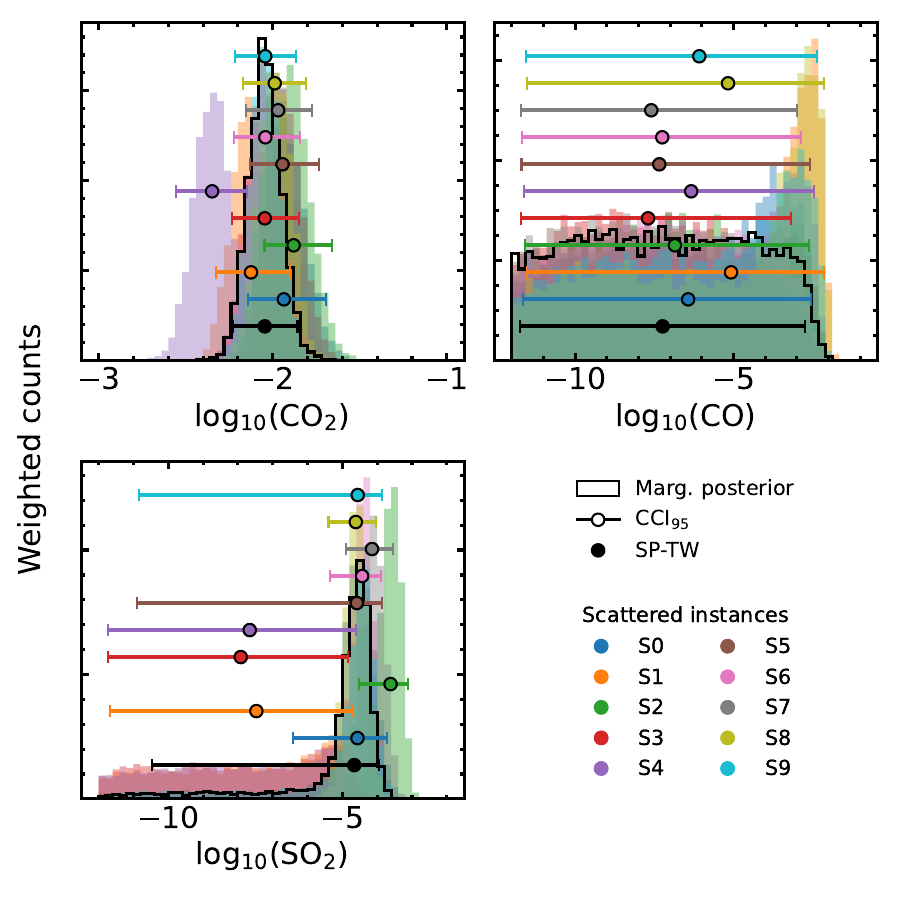}
    \caption{
        Posterior distributions of select forward model parameters for atmospheric retrievals on scattered instances of SP-TW, showing inferred parameter values (x-axis) against weighted counts (y-axis) in all panels.
        The parameter posterior distributions are the \acp{vmr} of \ce{CO2} (top left), \ce{CO} (top right), and \ce{SO2} (bottom right).
        Marginalised posteriors and CCIs from the initial instance of SP-TW are shown in black, while the results from the scattered instances of SP-TW are colour-coded.
    }
    \label{fig:posterior_scatter_comparison}
\end{figure}

As shown in Fig.~\ref{fig:corner-thiswork}, the posterior distributions of the \ce{SO2} and \ce{C2H2} mixing ratios have `tails' towards low abundances.
As the parameter inference process in a Bayesian framework is guided by the underlying data set, this could indicate that the abundances of these molecules are estimated over only few data points.
To test the robustness of the reported parameter estimates to perturbations of the underlying spectrum, we conduct the same homogenised atmospheric retrievals on \num{10} self-scattered instances of SP-TW.
We find that the resulting posterior distributions and derived parameter estimates can be categorised into three types -- (1) stable and well constrained, (2) stable and unconstrained, and (3) unstable and skewed.
A selection of of these are illustrated in Fig.~\ref{fig:posterior_scatter_comparison}.
We show a full overview of all marginalised posterior distributions in Fig.~\ref{fig-app:scatter_posterior_full}.
We also list the results from a two-sample K-S test in Table~\ref{tab-app:scatter-posterior-pval}, which compares the marginalised posterior distributions of the scattered instances with the ones from the original instance of SP-TW.

For posterior distributions previously identified as `well constrained', we find that the parameter estimates are stable against the perturbations of the spectrum.
This is shown in the top left panel of Fig.~\ref{fig:posterior_scatter_comparison} by the posterior distribution of $\log_{10}(X_{\ce{CO2}})$.
In all cases, the shape of the posterior distribution remains close to being Gaussian, and the CCIs agree with the parameter estimation results retrieved from the baseline instance of SP-TW.
This can be explained by the fact that for all stable, well-constrained cases (\ce{H2O, CO2, and H2S}), the spectral features are mapped onto a broad wavelength range (as can be seen in Fig.~\ref{fig:thiswork-bestfit-contrib}).

Parameter estimation results identified as upper or lower limits are similarly stable against these perturbations.
This is illustrated in the top right panel of Fig.~\ref{fig:posterior_scatter_comparison} for \ce{CO}.
As it has no significant absorption features in the wavelength range of SP-TW, perturbing the spectrum will have no significant influence on the parameter inference of $\log_{10}(X_{\ce{CO}})$.

Finally, we find posterior distributions and parameter estimates that are unstable under perturbations of the spectrum.
This is shown for \ce{SO2} in the bottom left panel of Fig.~\ref{fig:posterior_scatter_comparison}.
The mixing ratio parameter estimates of \ce{SO2 and C2H2} depend on small regions of the transmission spectrum.
Subsequently, scattered instances that influence these regions specifically will produce strong variations in the extent of the associated parameter estimation result.
In addition to the tailed posterior distributions described before, in the case of \ce{SO2 and C2H2}, we find several instances with narrow parameter estimates, mirroring the `well constrained' mixing ratios of \ce{H2O, CO2, and H2S}.
We also find several instances representing upper limits, with broad ranges for the CCI$_{95}$ and a centred median.
We point out that this behaviour can also, to a lesser extend, be seen in the posterior distributions of \ce{CH4}.
While in its initial instance, the inferred mixing ratio of \ce{CH4} is interpreted as an upper limit, we find several instances where the CCI$_{95}$ resembles the tailed cases we have found for \ce{SO2} and \ce{C2H2}.
As \ce{CH4} has an absorption feature at \SI{3.4}{\micro\meter}, these instances represent cases where the scattering induces a clearer identification of this feature.

Results from a two-sample K-S test on these distributions indicates that none of the marginalised posteriors share an underlying distribution with the marginalised posteriors from the retrieval performed on the unperturbed SP-TW (Table~\ref{tab-app:scatter-posterior-pval}).
However, performing a statistical test on the marginalised posterior distributions neglects information contained in the covariances of these parameters.
In reporting atmospheric retrieval results, parameter estimates derived from posterior distributions are the more pertinent conclusions.
We find that, even in the unstable cases of \ce{SO2, C2H2, and CH4}, all CCI$_{95}$ values agree with the initial results from SP-TW (Fig.~\ref{fig-app:scatter_posterior_full}).
While random perturbations of the input data do not produce disagreement in parameter estimation results, the associated parameter constraints could be overconfidently small.
We therefore argue that the assignment of specific \ac{vmr} values for atmospheric trace species should be interpreted with caution, especially when the credible interval of the parameter posterior is heavily skewed.

\subsection{Homogenised atmospheric retrievals on existing transmission spectra}\label{ssec:retrieval-comp}

\begin{figure*}
    \centering
    \includegraphics[width=\hsize]{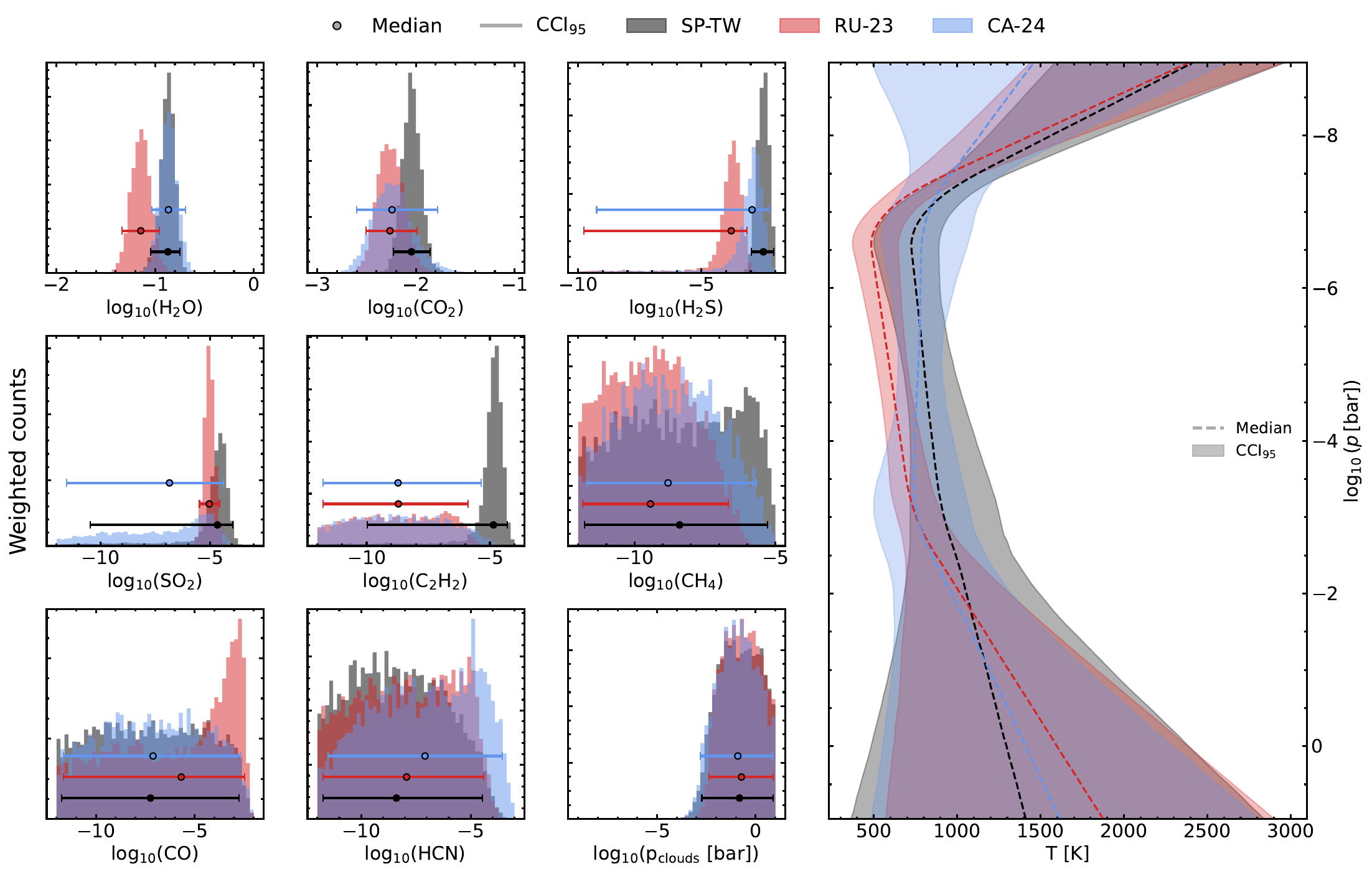}
    \caption{
        Results of atmospheric retrieval performed on three transmission spectra derived from the same observation.
        Results achieved with SP-TW (the spectrum produced in our work), as well as with RU-23 and CA-24 are shown in black, red, and blue, respectively.
        (Left) The grid of smaller panels shows the marginalised posterior distributions of the molecular mixing ratios and cloud-top pressure 
        (Right) Retrieved 4-point pressure temperature profiles, where dashed lines indicate the posterior median, and shaded area the corresponding CCI$_{95}$.
    }
    \label{fig:posterior-comparison}
\end{figure*}

As shown in Fig.~\ref{fig:existing-spectra-comparison}, the three transmission spectra of WASP-39~b considered in this work show systematic differences.
While a random perturbation of SP-TW did not produce disagreement in the resulting atmospheric retrieval results, we now investigate the impact of systematic differences stemming from variations in data reduction assumptions.

As these three spectra were derived from the same underlying raw observation, we assume that they should contain the same information on the nature of the atmosphere of WASP-39~b.
Under this assumption, we do not perform additional model tuning on RU-23 and CA-24 (we show a comparison of the values of $\ln B_{\mathrm{m}0}$ derived from retrievals based on each of the three spectra considered in our work in Fig.~\ref{fig-app:bf-comparison}).
Instead, we perform atmospheric retrievals on RU-23 and CA-24 with a homogenised setup from the model tuning on SP-TW.
Figure~\ref{fig:posterior-comparison} shows a comparison between the parameter posteriors of the molecular mixing ratios and cloud-top pressure for all three retrieval cases, as well as the associated $p$-$T$ profiles.
A comprehensive list of all parameter estimates is given Table~\ref{tab:estimated-pars-different-spectra}.

Generally, all three retrieval cases produce results that agree within the CCI$_{95}$ values.
The mixing ratios of \ce{H2O and CO2} are well constrained, with posterior distributions that are close to Gaussian and a parameter estimation range that spans \SIrange{0.5}{1}{\dex}.
Similarly, the mixing ratios of \ce{CO, CH4, and HCN} are constrained to upper limits based on the retrieval performed on all three spectra.
We note that in the case of RU-23, the upper limit on \ce{CH4} is approximately one order of magnitude smaller than in the other two cases.
When comparing all three spectra in Fig.~\ref{fig:existing-spectra-comparison}, RU-23 shows lower transit depths in the region of the methane absorption feature at \SI{3.4}{\micro\meter}, which could explain this reduced upper limit.
Additionally, the posterior distribution of \ce{CO} derived from RU-23 shows a stronger peak toward the upper edge of the parameter estimation range, indicated by the positively shifted median of the posterior distribution.

We find the biggest differences in the molecular mixing ratios previously categorised as unstable, skewed cases (\ce{SO2 and C2H2}).
In the case of \ce{C2H2}, the tailed posterior from SP-TW is contrasted by two uniform posterior distributions denoting upper limits of $\log_{10}(X_{\ce{C2H2}}) < -5.89$ and $\log_{10}(X_{\ce{C2H2}}) < -5.36$ in the case of RU-23 and CA-24, respectively.
In the case of \ce{SO2}, we see the largest variety in posterior behaviour.
The skewed posterior distribution from SP-TW is contrasted by an unconstrained mixing ratio in the case of CA-24 (with $\log_{10}(X_{\ce{SO2}}) < -4.38$), and a much narrower constraint from RU-23 (with $\log_{10}(X_{\ce{SO2}}) \in [-5.48, -4.57]$).
For both \ce{C2H2} and \ce{SO2}, all three retrieved parameter estimates are still in agreement in this case, but the behaviour of the associated posterior distributions mirrors the unstable and skewed case identified in the retrievals of the scattered instances of SP-TW (bottom left panel of Fig.~\ref{fig:posterior_scatter_comparison}).

We also find one case of a disagreement in the parameter estimation results, which is the mixing ratio of \ce{H2S}.
For both RU-23 and CA-24, a narrow posterior on its mixing ratio is replaced by a skewed distribution with with $\log_{10}(X_{\ce{H2S}}) \in [-9.77, -3.15]$ and with $\log_{10}(X_{\ce{H2S}}) \in [-9.25, -2.28]$, respectively.
The parameter estimate of \ce{H2S} from RU-23 does not overlap with the posterior constraints from SP-TW, although the disagreement is smaller than \SI{0.2}{\dex}.

Finally, the retrieved $p$-$T$ profiles agree between all three cases.
The close-to isothermal structure in the middle of the atmospheric domain is preserved, and all cases struggle to constrain the temperature values at the top and bottom of the atmosphere.
We point out that the retrievals of SP-TW and RU-23 indicate a thermal inversion in the upper layers of the atmosphere.
The retrieval of CA-24 has a broader posterior of the temperature at the top of the atmosphere, being consistent with both an isothermal behaviour, as well as an increasing temperature profile.
The retrieved cloud-top pressure is in close agreement for all three retrievals, which as a flat-opacity layer masks regions below the value of $p_\mathrm{cloud}$ as inaccessible for all three spectra.

\subsection{Limitations}

The range of reported mixing ratios for the atmospheric constituents of WASP-39~b is very large.
An immediate comparison between our results and previous analyses of the NIRSpec PRISM data set shows higher mixing ratios of the dominant atmospheric trace species from our retrievals ($\sim14\%$ \ce{H2O} and $\sim1\%$ \ce{CO2}).
Accounting for previous analyses of HST observations, as well as of observations by the other instruments of JWST shows instrument- and model-dependent discrepancies as large as 5 orders of magnitude for these main trace species \citep[e.g.][]{tsiaras_2018,wakeford_2018,lueber_2024}.
Exoplanet atmospheres are inherently more complex than retrieval models usually account for, reducing a multi-dimensional problem into a one-dimensional atmospheric slice.
In our work, we address the differences in characterisation results stemming from the underlying data set through a relative result comparison from homogenised atmospheric retrievals.
As such, we circumvent the problem of disagreeing results by applying the same model setup to data derived from the same underlying observation. 
We leave the solution to addressing this tension in results to future work.

We do note that using homogenised atmospheric retrievals on all three spectra overlooks the model-tuning possibility with respect to RU-23 and CA-24.
All transmission spectra considered in this work were derived from the same raw data.
We therefore make the assumption that the model tuning process is independent of the underlying transmission spectrum.
This will not necessarily be the case, as the model-tuning process is guided by the data.
As shown in Fig.~\ref{fig:existing-spectra-comparison}, the spectra show systematic differences, which might propagate into the molecule selection process.
While a full comparison to a flexible model-tuning approach is beyond the scope of this work, we note that individualised model setups for SP-TW, RU-23, and CA-24 could produce a smaller range of overlapping molecular constituents.
An example of this is the unconstrained nature of the \ce{C2H2} posterior distribution in both the case of the RU-23 and CA-24 retrieval.

We also note that our work does not address the importance of individual data reduction steps on the results of atmospheric retrievals.
Previous work has reported on the impact of using different data reduction pipelines \citep[e.g.][]{mugnai_2024, powell_2024, davenport_2025}.
Our work highlights the differences in parameter estimates from spectra independently derived using the same pipeline.
Follow-up investigations into the importance of individual steps during data reduction could provide even more insights into the stability of atmospheric retrieval posteriors.

Finally, as the parameter estimation `uncertainty' represents a subjective choice of the credible interval size, we caution against over-interpreting disagreements on the level of `1\,$\sigma$'.
As illustrated in Sect.~\ref{ssec:retireval-sptw}, skewed posterior distributions will result in much broader directly calculated `$\sigma$-equivalent' CCIs, compared to values scaled from `1\,$\sigma$'.

\begin{table*}
    \centering
    \caption{Estimated parameters for atmospheric retrievals performed on the spectra used in this work.}
    \renewcommand{\arraystretch}{1.1}
    \begin{tabularx}{\hsize}{llYYYYYY}
        \toprule\toprule
        \multirow{3}{*}{Parameter} & \multirow{3}{*}{Unit} & \multicolumn{6}{c}{Input spectrum} \\ 
        \cmidrule(lr){3-8}
        &  & \multicolumn{2}{c}{\textbf{SP-TW}} & \multicolumn{2}{c}{\textbf{RU-23}} & \multicolumn{2}{c}{\textbf{CA-24}} \\ 
        \cmidrule(lr){3-4}\cmidrule(lr){5-6}\cmidrule(lr){7-8}
        &  & 95\% CCI & Median & 95\% CCI & Median & 95\% CCI & Median \\
        \midrule
        $R_\mathrm{p}$             & R$_\mathrm{J}$    & $[1.21, 1.29]$ & $1.26$  & $[1.19, 1.28]$ & $1.24$  & $[1.22, 1.30]$ & $1.26$ \\
        $\log_{10}(\ce{CO2})$         & -              & $[-2.23, -1.86]$ & $-2.05$ & $[-2.51, -1.99]$ & $-2.26$ & $[-2.60, -1.78]$ & $-2.24$ \\
        $\log_{10}(\ce{CO})$          & -              & $[-11.74, -2.74]$ & $-7.24$ & $[-11.63, -2.46]$ & $-5.68$ & $[-11.63, -2.68]$ & $-7.11$ \\
        $\log_{10}(\ce{H2O})$         & -              & $[-1.04, -0.75]$ & $-0.87$ & $[-1.34, -0.95]$ & $-1.15$ & $[-1.04, -0.69]$ & $-0.87$ \\
        $\log_{10}(\ce{CH4})$         & -              & $[-11.78, -5.28]$ & $-8.40$ & $[-11.84, -6.66]$ & $-9.43$ & $[-11.69, -5.69]$ & $-8.81$ \\
        $\log_{10}(\ce{H2S})$         & -              & $[-2.97, -2.05]$ & $-2.49$ & $[-9.77, -3.15]$ & $-3.79$ & $[-9.25, -2.28]$ & $-2.95$ \\
        $\log_{10}(\ce{SO2})$         & -              & $[-10.46, -3.96]$ & $-4.66$ & $[-5.48, -4.57]$ & $-5.03$ & $[-11.55, -4.38]$ & $-6.85$ \\
        $\log_{10}(\ce{HCN})$         & -              & $[-11.75, -4.48]$ & $-8.39$ & $[-11.74, -4.42]$ & $-7.93$ & $[-11.51, -3.55]$ & $-7.09$ \\
        $\log_{10}(\ce{C2H2})$        & -              & $[-9.99, -4.30]$ & $-4.86$ & $[-11.76, -5.89]$ & $-8.71$ & $[-11.76, -5.36]$ & $-8.73$ \\
        $\log_{10}(p_\mathrm{cloud})$ & $\si{\bar}$ & $[-3.27, 0.88]$ & $-1.18$  & $[-3.65, 0.88]$ & $-1.28$  & $[-3.20, 0.86]$ & $-1.09$  \\
        $T_{p_0}$                     & $\si{\kelvin}$ & $[366.33, 2841.23]$ & $1407.14$     & $[571.13, 2901.82]$ & $1874.77$    & $[481.25, 2805.83]$ & $1615.97$ \\
        $T_{p_1}$                     & $\si{\kelvin}$ & $[711.31, 1296.84]$ & $912.26$       & $[607.68, 895.15]$ & $729.77$         & $[446.74, 1088.07]$ & $732.90$ \\
        $T_{p_2}$                     & $\si{\kelvin}$ & $[440.38, 909.23]$ & $700.23$       & $[321.21, 650.24]$ & $450.71$         & $[497.63, 1128.41]$ & $789.30$ \\
        $T_{p_3}$                     & $\si{\kelvin}$ & $[1554.85, 2965.68]$ & $2402.63$     & $[1427.31, 2962.16]$ & $2382.54$     & $[486.12, 2617.69]$ & $1462.25$ \\
        \bottomrule
    \end{tabularx}
    \label{tab:estimated-pars-different-spectra}
    \tablefoot{
        SP-TW, RU-23, and CA-24 refer to the transmission spectra presented in this work, \protect\citet{rustamkulov_2023}, and \protect\citetalias{carter_2024}, respectively. 
        The reported values for individual parameters are given as a 95\% centred credible interval, followed by the median value.
        The pressure nodes associated with the four temperature values are located at $\log_{10}(p\,[\mathrm{bar}]) = \{ 1, -3, -7, -9 \}$.
    }
\end{table*}

\section{Conclusion}\label{sec:conclusion}

Parameter estimation processes in Bayesian inference networks are guided by observational data.
In this work, we investigated the impact of data perturbations on the retrieval posteriors of atmospheric parameters from the transmission spectrum of the hot Jupiter WASP-39~b.

We produced a transmission spectrum from a NIRSpec PRISM observation of WASP-39~b, and selected an atmospheric forward model based on this data set.
From a baseline model containing absorption contributions of \ce{H2O, CO2, CO, and CH4}, we construct a fiducial model with additional contributions from \ce{H2S, SO2, C2H2, and HCN}.
To investigate the reliability of the reported parameter posteriors, we performed homogenised atmospheric retrievals on several additional transmission spectra of WASP-39~b.
We performed these retrievals on self-scattered instances of the transmission spectrum produced in our work, which mimics potentially random variations caused by assumptions made during data reduction.
We also used two previously published transmission spectra of WASP-39~b, which were derived from the same underlying observation.
We find that several forward model parameters (the planetary reference radius, cloud-top pressure, and $p$-$T$ profile) show no significant variations under the perturbations of the transmission spectrum.
The $p$-$T$ profile is well constrained in the probed region of the atmosphere.
The retrieved temperature values at the top and bottom of the atmospheric domain are unconstrained.

In the parameter posteriors of the molecular mixing ratios, we identify three types of behaviour:
\begin{enumerate}
    \item Well constrained posteriors that are close to Gaussian distributions (\ce{H2O, CO2, and H2S}), 
    resulting in parameter estimates which are stable under the perturbations characterised by the selection of spectra we use.
    \item Posteriors constrained by upper limits (\ce{CO and HCN}), which result in parameter estimates that are also stable under these perturbations.
    \item Skewed posterior distributions with heavy tails (\ce{SO2, C2H2, and CH4}), which produce unstable parameter estimates under the cases considered in our work.
\end{enumerate}

When compared to our reference of performing retrievals on the spectrum produced in our work (SP-TW), we find general agreement between the inferred parameter values of atmospheric retrievals performed on different instances of the transmission spectrum of WASP-39~b.
However, we emphasise the impact of unstable posterior distributions on the interpretation of these parameter constraints.
Heavily skewed parameter posteriors, from which small credible intervals (CIs), such as a `1\,$\sigma$'-equivalent, are derived can provide a misleading sense of accuracy in the inferred values.
Directly calculating CIs can reveal these tails clearly, and help identify unstable forward model parameters.

\section*{Data and software statement}
An online repository with all data products produced and used in this work, as well as supplementary plots, can be found at \url{https://doi.org/10.5281/zenodo.15697940}).
We also gratefully acknowledge the use of open-source packages for the Python programming language:
\textbf{corner} \citep{foreman-mackey_2016},
\textbf{matplotlib} \citep{hunter_2007},
\textbf{numpy} \citep{harris_2020},
\textbf{scipy} \citep{virtanen_2020}.

\begin{acknowledgements}
We would like to thank the anonymous referee for their insightful comments and feedback.
S. Schleich thanks J. Davey for an insightful discussion on error bar asymmetries in atmospheric retrievals, and K. H. Yip for insights on distributions distance metrics.
This project was funded by the FGGA Emerging Field Grant 2021. 
We acknowledge financial support by the University of Vienna and Österreichische Forschungsgemeinschaft (ÖFG).
This work is based in part on observations made with the NASA/ESA/CSA James Webb Space Telescope. The data were obtained from the Mikulski Archive for Space Telescopes at the Space Telescope Science Institute, which is operated by the Association of Universities for Research in Astronomy, Inc., under NASA contract NAS 5-03127 for JWST. These observations are associated with program \#1366.
The authors acknowledge the Transiting Exoplanet Community Early Release Science Program team for developing their observing program with a zero-exclusive-access period.
The computational results have been achieved in part using the Austrian Scientific Computing (ASC) infrastructure.
\end{acknowledgements}

\bibliography{references.bib}

@article{abel_2011,
  title = {Collision-{{Induced Absorption}} by {{H2 Pairs}}: {{From Hundreds}} to {{Thousands}} of {{Kelvin}}},
  author = {Abel, Martin and Frommhold, Lothar and Li, Xiaoping and Hunt, Katharine L. C.},
  year = {2011},
  month = jun,
  journal = {J. Phys. Chem. A},
  volume = {115},
  number = {25},
  pages = {6805--6812},
  publisher = {American Chemical Society},
  issn = {1089-5639},
  doi = {10.1021/jp109441f}
}

@article{davenport_2025,
  title = {{{TOI-421}} b: {{A Hot Sub-Neptune}} with a {{Haze-free}}, {{Low Mean Molecular Weight Atmosphere}}},
  author = {Davenport, Brian and Kempton, Eliza M.-R. and Nixon, Matthew C. and Ih, Jegug and Deming, Drake and Fu, Guangwei and May, E. M. and Bean, Jacob L. and Gao, Peter and Rogers, Leslie and Malik, Matej},
  year = 2025,
  month = may,
  journal = {The Astrophysical Journal Letters},
  volume = {984},
  number = {2},
  pages = {L44},
  publisher = {The American Astronomical Society},
  issn = {2041-8205},
  doi = {10.3847/2041-8213/adcd76},
  langid = {english}
}

@article{benneke_2013,
  title = {{{HOW TO DISTINGUISH BETWEEN CLOUDY MINI-NEPTUNES AND WATER}}/{{VOLATILE-DOMINATED SUPER-EARTHS}}},
  author = {Benneke, Bj{\"o}rn and Seager, Sara},
  year = 2013,
  month = nov,
  journal = {The Astrophysical Journal},
  volume = {778},
  number = {2},
  pages = {153},
  publisher = {The American Astronomical Society},
  issn = {0004-637X},
  doi = {10.1088/0004-637X/778/2/153},
  langid = {english}
}

@article{abel_2012,
  title = {Infrared Absorption by Collisional {{H2}}--{{He}} Complexes at Temperatures up to 9000 {{K}} and Frequencies from 0 to 20~000 Cm-1},
  author = {Abel, Martin and Frommhold, Lothar and Li, Xiaoping and Hunt, Katharine L. C.},
  year = {2012},
  month = jan,
  journal = {The Journal of Chemical Physics},
  volume = {136},
  number = {4},
  pages = {044319},
  issn = {0021-9606},
  doi = {10.1063/1.3676405}
}

@article{adam_2019,
  title = {Variationally {{Computed IR Line List}} for the {{Methyl Radical CH3}}},
  author = {Adam, Ahmad Y. and Yachmenev, Andrey and Yurchenko, Sergei N. and Jensen, Per},
  year = {2019},
  month = jun,
  journal = {J. Phys. Chem. A},
  volume = {123},
  number = {22},
  pages = {4755--4763},
  publisher = {American Chemical Society},
  issn = {1089-5639},
  doi = {10.1021/acs.jpca.9b02919}
}

@article{ahrer_2023,
  title = {Early {{Release Science}} of the Exoplanet {{WASP-39b}} with {{JWST NIRCam}}},
  author = {Ahrer, Eva-Maria and Stevenson, Kevin B. and Mansfield, Megan and Moran, Sarah E. and Brande, Jonathan and Morello, Giuseppe and Murray, Catriona A. and Nikolov, Nikolay K. and {Petit dit de la Roche}, Dominique J. M. and Schlawin, Everett and Wheatley, Peter J. and Zieba, Sebastian and Batalha, Natasha E. and Damiano, Mario and Goyal, Jayesh M. and Lendl, Monika and Lothringer, Joshua D. and Mukherjee, Sagnick and Ohno, Kazumasa and Batalha, Natalie M. and Battley, Matthew P. and Bean, Jacob L. and Beatty, Thomas G. and Benneke, Bj{\"o}rn and {Berta-Thompson}, Zachory K. and Carter, Aarynn L. and Cubillos, Patricio E. and Daylan, Tansu and Espinoza, N{\'e}stor and Gao, Peter and Gibson, Neale P. and Gill, Samuel and Harrington, Joseph and Hu, Renyu and Kreidberg, Laura and Lewis, Nikole K. and Line, Michael R. and {L{\'o}pez-Morales}, Mercedes and Parmentier, Vivien and Powell, Diana K. and Sing, David K. and Tsai, Shang-Min and Wakeford, Hannah R. and Welbanks, Luis and Alam, Munazza K. and Alderson, Lili and Allen, Natalie H. and Anderson, David R. and Barstow, Joanna K. and Bayliss, Daniel and Bell, Taylor J. and Blecic, Jasmina and Bryant, Edward M. and Burleigh, Matthew R. and Carone, Ludmila and Casewell, S. L. and Changeat, Quentin and Chubb, Katy L. and Crossfield, Ian J. M. and Crouzet, Nicolas and Decin, Leen and D{\'e}sert, Jean-Michel and Feinstein, Adina D. and Flagg, Laura and Fortney, Jonathan J. and Gizis, John E. and Heng, Kevin and Iro, Nicolas and Kempton, Eliza M.-R. and Kendrew, Sarah and Kirk, James and Knutson, Heather A. and Komacek, Thaddeus D. and Lagage, Pierre-Olivier and Leconte, J{\'e}r{\'e}my and {Lustig-Yaeger}, Jacob and MacDonald, Ryan J. and Mancini, Luigi and May, E. M. and Mayne, N. J. and Miguel, Yamila and {Mikal-Evans}, Thomas and Molaverdikhani, Karan and Palle, Enric and Piaulet, Caroline and Rackham, Benjamin V. and Redfield, Seth and Rogers, Laura K. and Roy, Pierre-Alexis and Rustamkulov, Zafar and Shkolnik, Evgenya L. and Sotzen, Kristin S. and Taylor, Jake and Tremblin, P. and Tucker, Gregory S. and Turner, Jake D. and {de Val-Borro}, Miguel and Venot, Olivia and Zhang, Xi},
  year = {2023},
  month = feb,
  journal = {Nature},
  volume = {614},
  number = {7949},
  pages = {653--658},
  publisher = {Nature Publishing Group},
  issn = {1476-4687},
  doi = {10.1038/s41586-022-05590-4},
  copyright = {2023 The Author(s)},
  langid = {english}
}

@article{al-refaie_2021,
  title = {{{TauREx}} 3: {{A Fast}}, {{Dynamic}}, and {{Extendable Framework}} for {{Retrievals}}},
  author = {{Al-Refaie}, A. F. and Changeat, Q. and Waldmann, I. P. and Tinetti, G.},
  year = {2021},
  month = aug,
  journal = {ApJ},
  volume = {917},
  number = {1},
  pages = {37},
  publisher = {American Astronomical Society},
  issn = {0004-637X},
  doi = {10.3847/1538-4357/ac0252},
  langid = {english}
}

@article{al-refaie_2022,
  title = {A {{Comparison}} of {{Chemical Models}} of {{Exoplanet Atmospheres Enabled}} by {{TauREx}} 3.1},
  author = {{Al-Refaie}, A. F. and Changeat, Q. and Venot, O. and Waldmann, I. P. and Tinetti, G.},
  year = {2022},
  month = jun,
  journal = {ApJ},
  volume = {932},
  number = {2},
  pages = {123},
  publisher = {American Astronomical Society},
  issn = {0004-637X},
  doi = {10.3847/1538-4357/ac6dcd},
  langid = {english}
}

@article{alderson_2023,
  title = {Early {{Release Science}} of the Exoplanet {{WASP-39b}} with {{JWST NIRSpec G395H}}},
  author = {Alderson, Lili and Wakeford, Hannah R. and Alam, Munazza K. and Batalha, Natasha E. and Lothringer, Joshua D. and Adams Redai, Jea and Barat, Saugata and Brande, Jonathan and Damiano, Mario and Daylan, Tansu and Espinoza, N{\'e}stor and Flagg, Laura and Goyal, Jayesh M. and Grant, David and Hu, Renyu and Inglis, Julie and Lee, Elspeth K. H. and {Mikal-Evans}, Thomas and {Ramos-Rosado}, Lakeisha and Roy, Pierre-Alexis and Wallack, Nicole L. and Batalha, Natalie M. and Bean, Jacob L. and Benneke, Bj{\"o}rn and {Berta-Thompson}, Zachory K. and Carter, Aarynn L. and Changeat, Quentin and Col{\'o}n, Knicole D. and Crossfield, Ian J. M. and D{\'e}sert, Jean-Michel and {Foreman-Mackey}, Daniel and Gibson, Neale P. and Kreidberg, Laura and Line, Michael R. and {L{\'o}pez-Morales}, Mercedes and Molaverdikhani, Karan and Moran, Sarah E. and Morello, Giuseppe and Moses, Julianne I. and Mukherjee, Sagnick and Schlawin, Everett and Sing, David K. and Stevenson, Kevin B. and Taylor, Jake and Aggarwal, Keshav and Ahrer, Eva-Maria and Allen, Natalie H. and Barstow, Joanna K. and Bell, Taylor J. and Blecic, Jasmina and Casewell, Sarah L. and Chubb, Katy L. and Crouzet, Nicolas and Cubillos, Patricio E. and Decin, Leen and Feinstein, Adina D. and Fortney, Joanthan J. and Harrington, Joseph and Heng, Kevin and Iro, Nicolas and Kempton, Eliza M.-R. and Kirk, James and Knutson, Heather A. and Krick, Jessica and Leconte, J{\'e}r{\'e}my and Lendl, Monika and MacDonald, Ryan J. and Mancini, Luigi and Mansfield, Megan and May, Erin M. and Mayne, Nathan J. and Miguel, Yamila and Nikolov, Nikolay K. and Ohno, Kazumasa and Palle, Enric and Parmentier, Vivien and {Petit dit de la Roche}, Dominique J. M. and Piaulet, Caroline and Powell, Diana and Rackham, Benjamin V. and Redfield, Seth and Rogers, Laura K. and Rustamkulov, Zafar and Tan, Xianyu and Tremblin, P. and Tsai, Shang-Min and Turner, Jake D. and {de Val-Borro}, Miguel and Venot, Olivia and Welbanks, Luis and Wheatley, Peter J. and Zhang, Xi},
  year = {2023},
  month = feb,
  journal = {Nature},
  volume = {614},
  number = {7949},
  pages = {664--669},
  publisher = {Nature Publishing Group},
  issn = {1476-4687},
  doi = {10.1038/s41586-022-05591-3},
  copyright = {2023 The Author(s)},
  langid = {english}
}

@article{august_2023,
  title = {Confirmation of {{Subsolar Metallicity}} for {{WASP-77Ab}} from {{JWST Thermal Emission Spectroscopy}}},
  author = {August, Prune C. and Bean, Jacob L. and Zhang, Michael and Lunine, Jonathan and Xue, Qiao and Line, Michael and Smith, Peter C. B.},
  year = {2023},
  month = aug,
  journal = {ApJL},
  volume = {953},
  number = {2},
  pages = {L24},
  publisher = {The American Astronomical Society},
  issn = {2041-8205},
  doi = {10.3847/2041-8213/ace828},
  langid = {english}
}

@article{azzam_2016,
  title = {{{ExoMol}} Molecular Line Lists -- {{XVI}}. {{The}} Rotation--Vibration Spectrum of Hot {{H2S}}},
  author = {Azzam, Ala'a A. A. and Tennyson, Jonathan and Yurchenko, Sergei N. and Naumenko, Olga V.},
  year = {2016},
  month = aug,
  journal = {MNRAS},
  volume = {460},
  number = {4},
  pages = {4063--4074},
  issn = {0035-8711},
  doi = {10.1093/mnras/stw1133}
}

@article{barber_2014,
  title = {{{ExoMol}} Line Lists -- {{III}}. {{An}} Improved Hot Rotation-Vibration Line List for {{HCN}} and {{HNC}}},
  author = {Barber, R. J. and Strange, J. K. and Hill, C. and Polyansky, O. L. and Mellau, G. {\relax Ch}. and Yurchenko, S. N. and Tennyson, Jonathan},
  year = {2014},
  month = jan,
  journal = {MNRAS},
  volume = {437},
  number = {2},
  pages = {1828--1835},
  issn = {0035-8711},
  doi = {10.1093/mnras/stt2011}
}

@article{barstow_2020,
  title = {Outstanding {{Challenges}} of {{Exoplanet Atmospheric Retrievals}}},
  author = {Barstow, Joanna K. and Heng, Kevin},
  year = {2020},
  month = jun,
  journal = {Space Sci. Rev.},
  volume = {216},
  number = {5},
  pages = {82},
  issn = {1572-9672},
  doi = {10.1007/s11214-020-00666-x},
  langid = {english}
}

@article{bell_2022,
  title = {Eureka!: {{An End-to-End Pipeline}} for {{JWST Time-Series Observations}}},
  author = {Bell, Taylor J. and Ahrer, Eva-Maria and Brande, Jonathan and Carter, Aarynn L. and Feinstein, Adina D. and Caloca\vphantom\{\}, Giannina \{Guzman\vphantom\} and Mansfield, Megan and Zieba, Sebastian and Piaulet, Caroline and Benneke, Bj{\"o}rn and Filippazzo, Joseph and May, Erin M. and Roy, Pierre-Alexis and Kreidberg, Laura and Stevenson, Kevin B.},
  year = {2022},
  month = nov,
  journal = {JOSS},
  volume = {7},
  number = {79},
  pages = {4503},
  issn = {2475-9066},
  doi = {10.21105/joss.04503},
  langid = {english}
}

@article{bell_2023,
  title = {Methane throughout the Atmosphere of the Warm Exoplanet {{WASP-80b}}},
  author = {Bell, Taylor J. and Welbanks, Luis and Schlawin, Everett and Line, Michael R. and Fortney, Jonathan J. and Greene, Thomas P. and Ohno, Kazumasa and Parmentier, Vivien and Rauscher, Emily and Beatty, Thomas G. and Mukherjee, Sagnick and Wiser, Lindsey S. and Boyer, Martha L. and Rieke, Marcia J. and Stansberry, John A.},
  year = {2023},
  month = nov,
  journal = {Nature},
  volume = {623},
  number = {7988},
  pages = {709--712},
  publisher = {Nature Publishing Group},
  issn = {1476-4687},
  doi = {10.1038/s41586-023-06687-0},
  copyright = {2023 The Author(s), under exclusive licence to Springer Nature Limited},
  langid = {english}
}

@article{birkmann_2022,
  title = {The {{Near-Infrared Spectrograph}} ({{NIRSpec}}) on the {{James Webb Space Telescope}} - {{IV}}. {{Capabilities}} and Predicted Performance for Exoplanet Characterization},
  author = {Birkmann, S. M. and Ferruit, P. and Giardino, G. and Nielsen, L. D. and Mu{\~n}oz, A. Garc{\'i}a and Kendrew, S. and Rauscher, B. J. and Beck, T. L. and Keyes, C. and Valenti, J. A. and Jakobsen, P. and Dorner, B. and de Oliveira, C. Alves and Arribas, S. and B{\"o}ker, T. and Bunker, A. J. and Charlot, S. and de Marchi, G. and Kumari, N. and {L{\'o}pez-Caniego}, M. and L{\"u}tzgendorf, N. and Maiolino, R. and Manjavacas, E. and Marston, A. and Moseley, S. H. and Prizkal, N. and Proffitt, C. and Rawle, T. and Rix, H.-W. and te Plate, M. and Sabbi, E. and Sirianni, M. and Willott, C. J. and Zeidler, P.},
  year = {2022},
  month = may,
  journal = {A\&A},
  volume = {661},
  pages = {A83},
  publisher = {EDP Sciences},
  issn = {0004-6361, 1432-0746},
  doi = {10.1051/0004-6361/202142592},
  copyright = {{\copyright} ESO 2022},
  langid = {english}
}

@article{buchner_2014,
  title = {X-Ray Spectral Modelling of the {{AGN}} Obscuring Region in the {{CDFS}}: {{Bayesian}} Model Selection and Catalogue},
  author = {Buchner, J. and Georgakakis, A. and Nandra, K. and Hsu, L. and Rangel, C. and Brightman, M. and Merloni, A. and Salvato, M. and Donley, J. and Kocevski, D.},
  year = {2014},
  month = apr,
  journal = {A\&A},
  volume = {564},
  pages = {A125},
  publisher = {EDP Sciences},
  issn = {0004-6361, 1432-0746},
  doi = {10.1051/0004-6361/201322971},
  copyright = {{\copyright} ESO, 2014},
  langid = {english}
}

@article{burnham_2004,
  title = {Multimodel {{Inference}}: {{Understanding AIC}} and {{BIC}} in {{Model Selection}}},
  author = {Burnham, Kenneth P. and Anderson, David R.},
  year = {2004},
  month = nov,
  journal = {Sociological Methods \& Research},
  volume = {33},
  number = {2},
  pages = {261--304},
  publisher = {SAGE Publications Inc},
  issn = {0049-1241},
  doi = {10.1177/0049124104268644},
  langid = {english}
}

@misc{bushouse_2024,
  title = {{{JWST Calibration Pipeline}}},
  author = {Bushouse, Howard and Eisenhamer, Jonathan and Dencheva, Nadia and Davies, James and Greenfield, Perry and Morrison, Jane and Hodge, Phil and Simon, Bernie and Grumm, David and Droettboom, Michael and Slavich, Edward and Sosey, Megan and Pauly, Tyler and Miller, Todd and Jedrzejewski, Robert and Hack, Warren and Davis, David and Crawford, Steven and Law, David and Gordon, Karl and Regan, Michael and Cara, Mihai and MacDonald, Ken and Bradley, Larry and Shanahan, Clare and Jamieson, William and Teodoro, Mairan and Williams, Thomas and {Pena-Guerrero}, Maria},
  year = {2024},
  month = jan,
  doi = {10.5281/zenodo.10569856},
  howpublished = {Zenodo}
}

@article{carter_2024,
  title = {A Benchmark {{JWST}} Near-Infrared Spectrum for the Exoplanet {{WASP-39}} b},
  author = {Carter, A. L. and May, E. M. and Espinoza, N. and Welbanks, L. and Ahrer, E. and Alderson, L. and Brahm, R. and Feinstein, A. D. and Grant, D. and Line, M. and Morello, G. and O'Steen, R. and Radica, M. and Rustamkulov, Z. and Stevenson, K. B. and Turner, J. D. and Alam, M. K. and Anderson, D. R. and Batalha, N. M. and Battley, M. P. and Bayliss, D. and Bean, J. L. and Benneke, B. and {Berta-Thompson}, Z. K. and Brande, J. and Bryant, E. M. and Burleigh, M. R. and Coulombe, L. and Crossfield, I. J. M. and Damiano, M. and D{\'e}sert, J.-M. and Flagg, L. and Gill, S. and Inglis, J. and Kirk, J. and Knutson, H. and Kreidberg, L. and L{\'o}pez Morales, M. and Mansfield, M. and Moran, S. E. and Murray, C. A. and Nixon, M. C. and {Petit dit de la Roche}, D. J. M. and Rackham, B. V. and Schlawin, E. and Sing, D. K. and Wakeford, H. R. and Wallack, N. L. and Wheatley, P. J. and Zieba, S. and Aggarwal, K. and Barstow, J. K. and Bell, T. J. and Blecic, J. and Caceres, C. and Crouzet, N. and Cubillos, P. E. and Daylan, T. and {de Val-Borro}, M. and Decin, L. and Fortney, J. J. and Gibson, N. P. and Heng, K. and Hu, R. and Kempton, E. M.-R. and Lagage, P. and Lothringer, J. D. and {Lustig-Yaeger}, J. and Mancini, L. and Mayne, N. J. and Mayorga, L. C. and Molaverdikhani, K. and Nasedkin, E. and Ohno, K. and Parmentier, V. and Powell, D. and Redfield, S. and Roy, P. and Taylor, J. and Zhang, X.},
  year = {2024},
  month = aug,
  journal = {Nat Astron},
  volume = {8},
  number = {8},
  pages = {1008--1019},
  publisher = {Nature Publishing Group},
  issn = {2397-3366},
  doi = {10.1038/s41550-024-02292-x},
  copyright = {2024 The Author(s)},
  langid = {english}
}

@article{changeat_2019,
  title = {Toward a {{More Complex Description}} of {{Chemical Profiles}} in {{Exoplanet Retrievals}}: {{A Two-layer Parameterization}}},
  author = {Changeat, Q. and Edwards, B. and Waldmann, I. P. and Tinetti, G.},
  year = {2019},
  month = nov,
  journal = {ApJ},
  volume = {886},
  number = {1},
  pages = {39},
  publisher = {The American Astronomical Society},
  issn = {0004-637X},
  doi = {10.3847/1538-4357/ab4a14},
  langid = {english}
}

@article{changeat_2021,
  title = {An {{Exploration}} of {{Model Degeneracies}} with a {{Unified Phase Curve Retrieval Analysis}}: {{The Light}} and {{Dark Sides}} of {{WASP-43}} b},
  author = {Changeat, Q. and {Al-Refaie}, A. F. and Edwards, B. and Waldmann, I. P. and Tinetti, G.},
  year = {2021},
  month = may,
  journal = {ApJ},
  volume = {913},
  number = {1},
  pages = {73},
  publisher = {The American Astronomical Society},
  issn = {0004-637X},
  doi = {10.3847/1538-4357/abf2bb},
  langid = {english}
}

@article{chubb_2020,
  title = {{{ExoMol}} Molecular Line Lists -- {{XXXVII}}. {{Spectra}} of Acetylene},
  author = {Chubb, Katy L and Tennyson, Jonathan and Yurchenko, Sergei N},
  year = {2020},
  month = apr,
  journal = {MNRAS},
  volume = {493},
  number = {2},
  pages = {1531--1545},
  issn = {0035-8711},
  doi = {10.1093/mnras/staa229}
}

@article{chubb_2021,
  title = {The {{ExoMolOP}} Database: {{Cross}} Sections and k-Tables for Molecules of Interest in High-Temperature Exoplanet Atmospheres},
  author = {Chubb, Katy L. and Rocchetto, Marco and Yurchenko, Sergei N. and Min, Michiel and Waldmann, Ingo and Barstow, Joanna K. and Molli{\`e}re, Paul and {Al-Refaie}, Ahmed F. and Phillips, Mark W. and Tennyson, Jonathan},
  year = {2021},
  month = feb,
  journal = {A\&A},
  volume = {646},
  pages = {A21},
  publisher = {EDP Sciences},
  issn = {0004-6361, 1432-0746},
  doi = {10.1051/0004-6361/202038350},
  copyright = {{\copyright} ESO 2021},
  langid = {english}
}

@article{claret_2000,
  title = {A New Non-Linear Limb-Darkening Law for {{LTE}} Stellar Atmosphere Models. {{Calculations}} for -5.0 {$<$}= Log[{{M}}/{{H}}] {$<$}= +1, 2000 {{K}} {$<$}= {{Teff}} {$<$}= 50000 {{K}} at Several Surface Gravities},
  author = {Claret, A.},
  year = {2000},
  month = nov,
  journal = {A\&A},
  volume = {363},
  pages = {1081--1190},
  issn = {0004-6361}
}

@article{coles_2019,
  title = {{{ExoMol}} Molecular Line Lists -- {{XXXV}}. {{A}} Rotation-Vibration Line List for Hot Ammonia},
  author = {Coles, Phillip A and Yurchenko, Sergei N and Tennyson, Jonathan},
  year = {2019},
  month = dec,
  journal = {MNRAS},
  volume = {490},
  number = {4},
  pages = {4638--4647},
  issn = {0035-8711},
  doi = {10.1093/mnras/stz2778}
}

@article{constantinou_2023,
  title = {Early {{Insights}} for {{Atmospheric Retrievals}} of {{Exoplanets Using JWST Transit Spectroscopy}}},
  author = {Constantinou, Savvas and Madhusudhan, Nikku and Gandhi, Siddharth},
  year = {2023},
  month = feb,
  journal = {ApJ},
  volume = {943},
  pages = {L10},
  publisher = {IOP},
  issn = {0004-637X},
  doi = {10.3847/2041-8213/acaead}
}

@book{cox_2015,
  title = {Allen's Astrophyiscal Quantities},
  author = {Cox, Arthur N.},
  year = {2015},
  publisher = {Springer}
}

@article{davey_2025,
  title = {The Effect of Spectroscopic Binning on Atmospheric Retrievals},
  author = {Davey, Jack J and Yip, Kai Hou and {Al-Refaie}, Ahmed F and Waldmann, Ingo P},
  year = {2025},
  month = jan,
  journal = {MNRAS},
  volume = {536},
  number = {3},
  pages = {2618--2644},
  issn = {0035-8711},
  doi = {10.1093/mnras/stae2731}
}

@article{dyrek_2024,
  title = {{{SO2}}, Silicate Clouds, but No {{CH4}} Detected in a Warm {{Neptune}}},
  author = {Dyrek, Achr{\`e}ne and Min, Michiel and Decin, Leen and Bouwman, Jeroen and Crouzet, Nicolas and Molli{\`e}re, Paul and Lagage, Pierre-Olivier and Konings, Thomas and Tremblin, Pascal and G{\"u}del, Manuel and Pye, John and Waters, Rens and Henning, Thomas and Vandenbussche, Bart and Ardevol Martinez, Francisco and Argyriou, Ioannis and Ducrot, Elsa and Heinke, Linus and {van Looveren}, Gwenael and Absil, Olivier and Barrado, David and Baudoz, Pierre and Boccaletti, Anthony and Cossou, Christophe and Coulais, Alain and Edwards, Billy and Gastaud, Ren{\'e} and Glasse, Alistair and Glauser, Adrian and Greene, Thomas P. and Kendrew, Sarah and Krause, Oliver and Lahuis, Fred and Mueller, Michael and Olofsson, Goran and Patapis, Polychronis and Rouan, Daniel and Royer, Pierre and Scheithauer, Silvia and Waldmann, Ingo and Whiteford, Niall and Colina, Luis and {van Dishoeck}, Ewine F. and {\"O}stlin, G{\"o}ran and Ray, Tom P. and Wright, Gillian},
  year = {2024},
  month = jan,
  journal = {Nature},
  volume = {625},
  number = {7993},
  pages = {51--54},
  publisher = {Nature Publishing Group},
  issn = {1476-4687},
  doi = {10.1038/s41586-023-06849-0},
  copyright = {2023 The Author(s), under exclusive licence to Springer Nature Limited},
  langid = {english}
}

@article{edwards_2021,
  title = {Hubble {{WFC3 Spectroscopy}} of the {{Habitable-zone Super-Earth LHS}} 1140 b},
  author = {Edwards, Billy and Changeat, Quentin and Mori, Mayuko and Anisman, Lara O. and Morvan, Mario and Yip, Kai Hou and Tsiaras, Angelos and {Al-Refaie}, Ahmed and Waldmann, Ingo and Tinetti, Giovanna},
  year = {2021},
  month = jan,
  journal = {AJ},
  volume = {161},
  number = {1},
  pages = {44},
  publisher = {The American Astronomical Society},
  issn = {1538-3881},
  doi = {10.3847/1538-3881/abc6a5},
  langid = {english}
}

@article{edwards_2024,
  title = {Measuring {{Tracers}} of {{Planet Formation}} in the {{Atmosphere}} of {{WASP-77A}} b: {{Substellar O}}/{{H}} and {{C}}/{{H Ratios}}, with a {{Stellar C}}/{{O Ratio}} and a {{Potentially Superstellar Ti}}/{{H Ratio}}},
  author = {Edwards, Billy and Changeat, Quentin},
  year = {2024},
  month = feb,
  journal = {ApJL},
  volume = {962},
  number = {2},
  pages = {L30},
  publisher = {The American Astronomical Society},
  issn = {2041-8205},
  doi = {10.3847/2041-8213/ad2000},
  langid = {english}
}

@article{edwards_2024-1,
  title = {On the Difficulties of Obtaining Absolute Transit Depths with {{HST WFC3}}: {{KELT-11}}\,b, an Example},
  author = {Edwards, Billy and Tsiaras, Angelos and Changeat, Quentin and Yip, Kai Hou},
  year = {2024},
  month = jan,
  journal = {RAS Techniques and Instruments},
  volume = {3},
  number = {1},
  pages = {415--436},
  issn = {2752-8200},
  doi = {10.1093/rasti/rzae023}
}

@article{faedi_2011,
  title = {{{WASP-39b}}: A Highly Inflated {{Saturn-mass}} Planet Orbiting a Late {{G-type}} Star},
  author = {Faedi, F. and Barros, S. C. C. and Anderson, D. R. and Brown, D. J. A. and Cameron, A. Collier and Pollacco, D. and Boisse, I. and H{\'e}brard, G. and Lendl, M. and Lister, T. A. and Smalley, B. and Street, R. A. and Triaud, A. H. M. J. and Bento, J. and Bouchy, F. and Butters, O. W. and Enoch, B. and Haswell, C. A. and Hellier, C. and Keenan, F. P. and Miller, G. R. M. and Moulds, V. and Moutou, C. and Norton, A. J. and Queloz, D. and Santerne, A. and Simpson, E. K. and Skillen, I. and Smith, A. M. S. and Udry, S. and Watson, C. A. and West, R. G. and Wheatley, P. J.},
  year = {2011},
  month = jul,
  journal = {A\&A},
  volume = {531},
  pages = {A40},
  publisher = {EDP Sciences},
  issn = {0004-6361, 1432-0746},
  doi = {10.1051/0004-6361/201116671},
  copyright = {{\copyright} ESO, 2011},
  langid = {english}
}

@article{feinstein_2023,
  title = {Early {{Release Science}} of the Exoplanet {{WASP-39b}} with {{JWST NIRISS}}},
  author = {Feinstein, Adina D. and Radica, Michael and Welbanks, Luis and Murray, Catriona Anne and Ohno, Kazumasa and Coulombe, Louis-Philippe and Espinoza, N{\'e}stor and Bean, Jacob L. and Teske, Johanna K. and Benneke, Bj{\"o}rn and Line, Michael R. and Rustamkulov, Zafar and Saba, Arianna and Tsiaras, Angelos and Barstow, Joanna K. and Fortney, Jonathan J. and Gao, Peter and Knutson, Heather A. and MacDonald, Ryan J. and {Mikal-Evans}, Thomas and Rackham, Benjamin V. and Taylor, Jake and Parmentier, Vivien and Batalha, Natalie M. and {Berta-Thompson}, Zachory K. and Carter, Aarynn L. and Changeat, Quentin and {dos Santos}, Leonardo A. and Gibson, Neale P. and Goyal, Jayesh M. and Kreidberg, Laura and {L{\'o}pez-Morales}, Mercedes and Lothringer, Joshua D. and Miguel, Yamila and Molaverdikhani, Karan and Moran, Sarah E. and Morello, Giuseppe and Mukherjee, Sagnick and Sing, David K. and Stevenson, Kevin B. and Wakeford, Hannah R. and Ahrer, Eva-Maria and Alam, Munazza K. and Alderson, Lili and Allen, Natalie H. and Batalha, Natasha E. and Bell, Taylor J. and Blecic, Jasmina and Brande, Jonathan and Caceres, Claudio and Casewell, S. L. and Chubb, Katy L. and Crossfield, Ian J. M. and Crouzet, Nicolas and Cubillos, Patricio E. and Decin, Leen and D{\'e}sert, Jean-Michel and Harrington, Joseph and Heng, Kevin and Henning, Thomas and Iro, Nicolas and Kempton, Eliza M.-R. and Kendrew, Sarah and Kirk, James and Krick, Jessica and Lagage, Pierre-Olivier and Lendl, Monika and Mancini, Luigi and Mansfield, Megan and May, E. M. and Mayne, N. J. and Nikolov, Nikolay K. and Palle, Enric and {Petit dit de la Roche}, Dominique J. M. and Piaulet, Caroline and Powell, Diana and Redfield, Seth and Rogers, Laura K. and Roman, Michael T. and Roy, Pierre-Alexis and Nixon, Matthew C. and Schlawin, Everett and Tan, Xianyu and Tremblin, P. and Turner, Jake D. and Venot, Olivia and Waalkes, William C. and Wheatley, Peter J. and Zhang, Xi},
  year = {2023},
  month = feb,
  journal = {Nature},
  volume = {614},
  number = {7949},
  pages = {670--675},
  publisher = {Nature Publishing Group},
  issn = {1476-4687},
  doi = {10.1038/s41586-022-05674-1},
  copyright = {2023 The Author(s)},
  langid = {english}
}

@article{feroz_2009,
  title = {{{MultiNest}}: An Efficient and Robust {{Bayesian}} Inference Tool for Cosmology and Particle Physics},
  author = {Feroz, F. and Hobson, M. P. and Bridges, M.},
  year = {2009},
  month = oct,
  journal = {MNRAS},
  volume = {398},
  number = {4},
  pages = {1601--1614},
  issn = {0035-8711},
  doi = {10.1111/j.1365-2966.2009.14548.x}
}

@article{fletcher_2018,
  title = {Hydrogen {{Dimers}} in {{Giant-planet Infrared Spectra}}},
  author = {Fletcher, Leigh N. and Gustafsson, Magnus and Orton, Glenn S.},
  year = {2018},
  month = mar,
  journal = {ApJS},
  volume = {235},
  number = {1},
  pages = {24},
  publisher = {The American Astronomical Society},
  issn = {0067-0049},
  doi = {10.3847/1538-4365/aaa07a},
  langid = {english}
}

@article{foreman-mackey_2013,
  title = {Emcee: {{The MCMC Hammer}}},
  author = {{Foreman-Mackey}, Daniel and Hogg, David W. and Lang, Dustin and Goodman, Jonathan},
  year = {2013},
  month = feb,
  journal = {PASP},
  volume = {125},
  number = {925},
  pages = {306},
  publisher = {IOP Publishing},
  issn = {1538-3873},
  doi = {10.1086/670067},
  langid = {english}
}

@article{foreman-mackey_2016,
  title = {Corner.Py: {{Scatterplot}} Matrices in {{Python}}},
  author = {{Foreman-Mackey}, Daniel},
  year = {2016},
  month = jun,
  journal = {JOSS},
  volume = {1},
  number = {2},
  pages = {24},
  issn = {2475-9066},
  doi = {10.21105/joss.00024},
  langid = {english}
}

@article{fortney_2021,
  title = {Hot {{Jupiters}}: {{Origins}}, {{Structure}}, {{Atmospheres}}},
  author = {Fortney, Jonathan J. and Dawson, Rebekah I. and Komacek, Thaddeus D.},
  year = {2021},
  journal = {JGR Planets},
  volume = {126},
  number = {3},
  pages = {e2020JE006629},
  issn = {2169-9100},
  doi = {10.1029/2020JE006629},
  copyright = {{\copyright} 2021. The Authors.},
  langid = {english}
}

@article{fu_2025,
  title = {Statistical {{Trends}} in {{JWST Transiting Exoplanet Atmospheres}}},
  author = {Fu, Guangwei and Stevenson, Kevin B. and Sing, David K. and Mukherjee, Sagnick and Welbanks, Luis and Thorngren, Daniel and Tsai, Shang-Min and Gao, Peter and Lothringer, Joshua and Beatty, Thomas G. and Gapp, Cyril and {Evans-Soma}, Thomas M. and Allart, Romain and Pelletier, Stefan and Thao, Pa Chia and Mann, Andrew W.},
  year = {2025},
  month = jun,
  journal = {ApJ},
  volume = {986},
  number = {1},
  pages = {1},
  publisher = {The American Astronomical Society},
  issn = {0004-637X},
  doi = {10.3847/1538-4357/ad7bb8},
  langid = {english}
}

@article{gardner_2023,
  title = {The {{James Webb Space Telescope Mission}}},
  author = {Gardner, Jonathan P. and Mather, John C. and Abbott, Randy and Abell, James S. and Abernathy, Mark and Abney, Faith E. and Abraham, John G. and Abraham, Roberto and {Abul-Huda}, Yasin M. and Acton, Scott and Adams, Cynthia K. and Adams, Evan and Adler, David S. and Adriaensen, Maarten and Aguilar, Jonathan Albert and Ahmed, Mansoor and Ahmed, Nasif S. and Ahmed, Tanjira and Albat, R{\"u}deger and Albert, Lo{\"i}c and Alberts, Stacey and Aldridge, David and Allen, Mary Marsha and Allen, Shaune S. and Altenburg, Martin and Altunc, Serhat and Alvarez, Jose Lorenzo and {\'A}lvarez-M{\'a}rquez, Javier and de Oliveira, Catarina Alves and Ambrose, Leslie L. and Anandakrishnan, Satya M. and Andersen, Gregory C. and Anderson, Harry James and Anderson, Jay and Anderson, Kristen and Anderson, Sara M. and Aprea, Julio and Archer, Benita J. and Arenberg, Jonathan W. and Argyriou, Ioannis and Arribas, Santiago and Artigau, {\'E}tienne and Arvai, Amanda Rose and Atcheson, Paul and Atkinson, Charles B. and Averbukh, Jesse and Aymergen, Cagatay and Bacinski, John J. and Baggett, Wayne E. and Bagnasco, Giorgio and Baker, Lynn L. and Balzano, Vicki Ann and Banks, Kimberly A. and Baran, David A. and Barker, Elizabeth A. and Barrett, Larry K. and Barringer, Bruce O. and Barto, Allison and Bast, William and Baudoz, Pierre and Baum, Stefi and Beatty, Thomas G. and Beaulieu, Mathilde and Bechtold, Kathryn and Beck, Tracy and Beddard, Megan M. and Beichman, Charles and Bellagama, Larry and Bely, Pierre and Berger, Timothy W. and Bergeron, Louis E. and Bernier, Antoine-Darveau and Bertch, Maria D. and Beskow, Charlotte and Betz, Laura E. and Biagetti, Carl P. and Birkmann, Stephan and Bjorklund, Kurt F. and Blackwood, James D. and Blazek, Ronald Paul and Blossfeld, Stephen and Bluth, Marcel and Boccaletti, Anthony and Jr, Martin E. Boegner and Bohlin, Ralph C. and Boia, John Joseph and B{\"o}ker, Torsten and Bonaventura, N. and Bond, Nicholas A. and Bosley, Kari Ann and Boucarut, Rene A. and Bouchet, Patrice and Bouwman, Jeroen and Bower, Gary and Bowers, Ariel S. and Bowers, Charles W. and Boyce, Leslye A. and Boyer, Christine T. and Boyer, Martha L. and Boyer, Michael and Boyer, Robert and Bradley, Larry D. and Brady, Gregory R. and Brandl, Bernhard R. and Brannen, Judith L. and Breda, David and Bremmer, Harold G. and Brennan, David and Bresnahan, Pamela A. and Bright, Stacey N. and Broiles, Brian J. and Bromenschenkel, Asa and Brooks, Brian H. and Brooks, Keira J. and Brown, Bob and Brown, Bruce and Brown, Thomas M. and Bruce, Barry W. and Bryson, Jonathan G. and Bujanda, Edwin D. and Bullock, Blake M. and Bunker, A. J. and Bureo, Rafael and Burt, Irving J. and Bush, James Aaron and Bushouse, Howard A. and Bussman, Marie C. and Cabaud, Olivier and Cale, Steven and Calhoon, Charles D. and Calvani, Humberto and Canipe, Alicia M. and Caputo, Francis M. and Cara, Mihai and Carey, Larkin and Case, Michael Eli and Cesari, Thaddeus and Cetorelli, Lee D. and Chance, Don R. and Chandler, Lynn and Chaney, Dave and Chapman, George N. and Charlot, S. and Chayer, Pierre and Cheezum, Jeffrey I. and Chen, Bin and Chen, Christine H. and Cherinka, Brian and Chichester, Sarah C. and Chilton, Zachary S. and Chittiraibalan, Dharini and Clampin, Mark and Clark, Charles R. and Clark, Kerry W. and Clark, Stephanie M. and Claybrooks, Edward E. and Cleveland, Keith A. and Cohen, Andrew L. and Cohen, Lester M. and Col{\'o}n, Knicole D. and Coleman, Benee L. and Colina, Luis and Comber, Brian J. and Comeau, Thomas M. and Comer, Thomas and Reis, Alain Conde and Connolly, Dennis C. and Conroy, Kyle E. and Contos, Adam R. and Contreras, James and Cook, Neil J. and Cooper, James L. and Cooper, Rachel Aviva and Correia, Michael F. and Correnti, Matteo and Cossou, Christophe and Costanza, Brian F. and Coulais, Alain and Cox, Colin R. and Coyle, Ray T. and Cracraft, Misty M. and Crew, Keith A. and Curtis, Gary J. and Cusveller, Bianca and Maciel, Cleyciane Da Costa and Dailey, Christopher T. and Daugeron, Fr{\'e}d{\'e}ric and Davidson, Greg S. and Davies, James E. and Davis, Katherine Anne and Davis, Michael S. and Day, Ratna and de Chambure, Daniel and de Jong, Pauline and Marchi, Guido De and Dean, Bruce H. and Decker, John E. and Delisa, Amy S. and Dell, Lawrence C. and Dellagatta, Gail and Dembinska, Franciszka and Demosthenes, Sandor and Dencheva, Nadezhda M. and Deneu, Philippe and DePriest, William W. and Deschenes, Jeremy and Dethienne, Nathalie and Detre, {\"O}rs Hunor and Diaz, Rosa Izela and Dicken, Daniel and DiFelice, Audrey S. and Dillman, Matthew and Disharoon, Maureen O. and Dixon, William V. and Doggett, Jesse B. and Dominguez, Keisha L. and Donaldson, Thomas S. and {Doria-Warner}, Cristina M. and Santos, Tony Dos and Doty, Heather and Robert E. Douglas, Jr and Doyon, Ren{\'e} and Dressler, Alan and Driggers, Jennifer and Driggers, Phillip A. and Dunn, Jamie L. and DuPrie, Kimberly C. and Dupuis, Jean and Durning, John and Dutta, Sanghamitra B. and Earl, Nicholas M. and Eccleston, Paul and Ecobichon, Pascal and Egami, Eiichi and Ehrenwinkler, Ralf and Eisenhamer, Jonathan D. and Eisenhower, Michael and Eisenstein, Daniel J. and Hamel, Zaky El and Elie, Michelle L. and Elliott, James and Elliott, Kyle Wesley and Engesser, Michael and Espinoza, N{\'e}stor and Etienne, Odessa and Etxaluze, Mireya and Evans, Leah and Fabreguettes, Luce and Falcolini, Massimo and Falini, Patrick R. and Fatig, Curtis and Feeney, Matthew and Feinberg, Lee D. and Fels, Raymond and Ferdous, Nazma and Ferguson, Henry C. and Ferrarese, Laura and Ferreira, Marie-H{\'e}l{\'e}ne and Ferruit, Pierre and Ferry, Malcolm and Filippazzo, Joseph Charles and Firre, Daniel and Fix, Mees and Flagey, Nicolas and Flanagan, Kathryn A. and Fleming, Scott W. and Florian, Michael and Flynn, James R. and Foiadelli, Luca and Fontaine, Mark R. and Fontanella, Erin Marie and Forshay, Peter Randolph and Fortner, Elizabeth A. and Fox, Ori D. and Framarini, Alexandro P. and Francisco, John I. and Franck, Randy and Franx, Marijn and Franz, David E. and Friedman, Scott D. and Friend, Katheryn E. and Frost, James R. and Fu, Henry and Fullerton, Alexander W. and Gaillard, Lionel and Galkin, Sergey and Gallagher, Ben and Galyer, Anthony D. and Mar{\'i}n, Macarena Garc{\'i}a and Gardner, Lisa E. and Garland, Dennis and Garrett, Bruce Albert and Gasman, Danny and G{\'a}sp{\'a}r, Andr{\'a}s and Gastaud, Ren{\'e} and Gaudreau, Daniel and Gauthier, Peter Timothy and Geers, Vincent and Geithner, Paul H. and Gennaro, Mario and Gerber, John and Gereau, John C. and Giampaoli, Robert and Giardino, Giovanna and Gibbons, Paul C. and Gilbert, Karoline and Gilman, Larry and Girard, Julien H. and Giuliano, Mark E. and Gkountis, Konstantinos and Glasse, Alistair and Glassmire, Kirk Zachary and Glauser, Adrian Michael and Glazer, Stuart D. and Goldberg, Joshua and Golimowski, David A. and Gonzaga, Shireen P. and Gordon, Karl D. and Gordon, Shawn J. and Goudfrooij, Paul and Gough, Michael J. and Graham, Adrian J. and Grau, Christopher M. and Green, Joel David and Greene, Gretchen R. and Greene, Thomas P. and Greenfield, Perry E. and Greenhouse, Matthew A. and Greve, Thomas R. and Greville, Edgar M. and Grimaldi, Stefano and Groe, Frank E. and Groebner, Andrew and Grumm, David M. and Grundy, Timothy and G{\"u}del, Manuel and Guillard, Pierre and Guldalian, John and Gunn, Christopher A. and Gurule, Anthony and Gutman, Irvin Meyer and Guy, Paul D. and Guyot, Benjamin and Hack, Warren J. and Haderlein, Peter and Hagan, James B. and Hagedorn, Andria and Hainline, Kevin and Haley, Craig and Hami, Maryam and Hamilton, Forrest Clifford and Hammann, Jeffrey and Hammel, Heidi B. and Hanley, Christopher J. and Hansen, Carl August and Hardy, Bruce and Harnisch, Bernd and Harr, Michael Hunter and Harris, Pamela and Hart, Jessica Ann and Hartig, George F. and Hasan, Hashima and Hashim, Kathleen Marie and Hashimoto, Ryan and Haskins, Sujee J. and Hawkins, Robert Edward and Hayden, Brian and Hayden, William L. and Healy, Mike and Hecht, Karen and Heeg, Vince J. and Hejal, Reem and Helm, Kristopher A. and Hengemihle, Nicholas J. and Henning, Thomas and Henry, Alaina and Henry, Ronald L. and Henshaw, Katherine and Hernandez, Scarlin and Herrington, Donald C. and Heske, Astrid and Hesman, Brigette Emily and Hickey, David L. and Hilbert, Bryan N. and Hines, Dean C. and Hinz, Michael R. and Hirsch, Michael and Hitcho, Robert S. and Hodapp, Klaus and Hodge, Philip E. and Hoffman, Melissa and Holfeltz, Sherie T. and Holler, Bryan Jason and Hoppa, Jennifer Rose and Horner, Scott and Howard, Joseph M. and Howard, Richard J. and Huber, Jean M. and Hunkeler, Joseph S. and Hunter, Alexander and Hunter, David Gavin and Hurd, Spencer W. and Hurst, Brendan J. and Hutchings, John B. and Hylan, Jason E. and Ignat, Luminita Ilinca and Illingworth, Garth and Irish, Sandra M. and III, John C. Isaacs and Jr, Wallace C. Jackson and Jaffe, Daniel T. and Jahic, Jasmin and Jahromi, Amir and Jakobsen, Peter and James, Bryan and James, John C. and James, LeAndrea Rae and Jamieson, William Brian and Jandra, Raymond D. and Jayawardhana, Ray and Jedrzejewski, Robert and Jeffers, Basil S. and Jensen, Peter and Joanne, Egges and Johns, Alan T. and Johnson, Carl A. and Johnson, Eric L. and Johnson, Patricia and Johnson, Phillip Stephen and Johnson, Thomas K. and Johnson, Timothy W. and Johnstone, Doug and Jollet, Delphine and Jones, Danny P. and Jones, Gregory S. and Jones, Olivia C. and Jones, Ronald A. and Jones, Vicki and Jordan, Ian J. and Jordan, Margaret E. and Jue, Reginald and Jurkowski, Mark H. and Justis, Grant and Justtanont, Kay and Kaleida, Catherine C. and Kalirai, Jason S. and Kalmanson, Phillip Cabrales and Kaltenegger, Lisa and Kammerer, Jens and Kan, Samuel K. and Kanarek, Graham Childs and Kao, Shaw-Hong and Karakla, Diane M. and Karl, Hermann and Kassin, Susan A. and Kauffman, David D. and Kavanagh, Patrick and Kelley, Leigh L. and Kelly, Douglas M. and Kendrew, Sarah and Kennedy, Herbert V. and Kenny, Deborah A. and {Keski-Kuha}, Ritva A. and Keyes, Charles D. and Khan, Ali and Kidwell, Richard C. and Kimble, Randy A. and King, James S. and King, Richard C. and Kinzel, Wayne M. and Kirk, Jeffrey R. and Kirkpatrick, Marc E. and Klaassen, Pamela and Klingemann, Lana and Klintworth, Paul U. and Knapp, Bryan Adam and Knight, Scott and Knollenberg, Perry J. and Knutsen, Daniel Mark and Koehler, Robert and Koekemoer, Anton M. and Kofler, Earl T. and Kontson, Vicki L. and Kovacs, Aiden Rose and {Kozhurina-Platais}, Vera and Krause, Oliver and Kriss, Gerard A. and Krist, John and Kristoffersen, Monica R. and Krogel, Claudia and Krueger, Anthony P. and Kulp, Bernard A. and Kumari, Nimisha and Kwan, Sandy W. and Kyprianou, Mark and Labador, Aurora Gadiano and Labiano, {\'A}lvaro and Lafreni{\`e}re, David and Lagage, Pierre-Olivier and Laidler, Victoria G. and Laine, Benoit and Laird, Simon and Lajoie, Charles-Philippe and Lallo, Matthew D. and Lam, May Yen and LaMassa, Stephanie Marie and Lambros, Scott D. and Lampenfield, Richard Joseph and Lander, Matthew Ed and Langston, James Hutton and Larson, Kirsten and Larson, Melora and LaVerghetta, Robert Joseph and Law, David R. and Lawrence, Jon F. and Lee, David W. and Lee, Janice and Lee, Yat-Ning Paul and Leisenring, Jarron and Leveille, Michael Dunlap and Levenson, Nancy A. and Levi, Joshua S. and Levine, Marie B. and Lewis, Dan and Lewis, Jake and Lewis, Nikole and Libralato, Mattia and Lidon, Norbert and Liebrecht, Paula Louisa and Lightsey, Paul and Lilly, Simon and Lim, Frederick C. and Lim, Pey Lian and Ling, Sai-Kwong and Link, Lisa J. and Link, Miranda Nicole and Lipinski, Jamie L. and Liu, XiaoLi and Lo, Amy S. and Lobmeyer, Lynette and Logue, Ryan M. and Long, Chris A. and Long, Douglas R. and Long, Ilana D. and Long, Knox S. and {L{\'o}pez-Caniego}, Marcos and Lotz, Jennifer M. and {Love-Pruitt}, Jennifer M. and Lubskiy, Michael and Luers, Edward B. and Luetgens, Robert A. and Luevano, Annetta J. and Lui, Sarah Marie G. Flores and III, James M. Lund and Lundquist, Ray A. and Lunine, Jonathan and L{\"u}tzgendorf, Nora and Lynch, Richard J. and MacDonald, Alex J. and MacDonald, Kenneth and Macias, Matthew J. and Macklis, Keith I. and Maghami, Peiman and Maharaja, Rishabh Y. and Maiolino, Roberto and Makrygiannis, Konstantinos G. and Malla, Sunita Giri and Malumuth, Eliot M. and Manjavacas, Elena and Marini, Andrea and Marrione, Amanda and Marston, Anthony and Martel, Andr{\'e} R. and Martin, Didier and Martin, Peter G. and Martinez, Kristin L. and Maschmann, Marc and Masci, Gregory L. and Masetti, Margaret E. and Maszkiewicz, Michael and Matthews, Gary and Matuskey, Jacob E. and McBrayer, Glen A. and McCarthy, Donald W. and McCaughrean, Mark J. and McClare, Leslie A. and McClare, Michael D. and McCloskey, John C. and McClurg, Taylore D. and McCoy, Martin and McElwain, Michael W. and McGregor, Roy D. and McGuffey, Douglas B. and McKay, Andrew G. and McKenzie, William K. and McLean, Brian and McMaster, Matthew and McNeil, Warren and Meester, Wim De and Mehalick, Kimberly L. and Meixner, Margaret and Mel{\'e}ndez, Marcio and Menzel, Michael P. and Menzel, Michael T. and Merz, Matthew and Mesterharm, David D. and Meyer, Michael R. and Meyett, Michele L. and Meza, Luis E. and Midwinter, Calvin and Milam, Stefanie N. and Miller, Jay Todd and Miller, William C. and Miskey, Cherie L. and Misselt, Karl and Mitchell, Eileen P. and Mohan, Martin and Montoya, Emily E. and Moran, Michael J. and Morishita, Takahiro and {Moro-Mart{\'i}n}, Amaya and Morrison, Debra L. and Morrison, Jane and Morse, Ernie C. and Moschos, Michael and Moseley, S. H. and Mosier, Gary E. and Mosner, Peter and Mountain, Matt and Muckenthaler, Jason S. and Mueller, Donald G. and Mueller, Migo and Muhiem, Daniella and M{\"u}hlmann, Prisca and Mullally, Susan Elizabeth and Mullen, Stephanie M. and Munger, Alan J. and Murphy, Jess and Murray, Katherine T. and Muzerolle, James C. and Mycroft, Matthew and Myers, Andrew and Myers, Carey R. and Myers, Fred Richard R. and Myers, Richard and Myrick, Kaila and Adrian F. Nagle, {\relax IV} and Nayak, Omnarayani and Naylor, Bret and Neff, Susan G. and Nelan, Edmund P. and Nella, John and Nguyen, Duy Tuong and Nguyen, Michael N. and Nickson, Bryony and Nidhiry, John Joseph and Niedner, Malcolm B. and {Nieto-Santisteban}, Maria and Nikolov, Nikolay K. and Nishisaka, Mary Ann and {Noriega-Crespo}, Alberto and Nota, Antonella and O'Mara, Robyn C. and Oboryshko, Michael and O'Brien, Marcus B. and Ochs, William R. and Offenberg, Joel D. and Ogle, Patrick Michael and Ohl, Raymond G. and Olmsted, Joseph Hamden and Osborne, Shannon Barbara and O'Shaughnessy, Brian Patrick and {\"O}stlin, G{\"o}ran and O'Sullivan, Brian and Otor, O. Justin and Ottens, Richard and Ouellette, Nathalie N.-Q. and Outlaw, Daria J. and Owens, Beverly A. and Pacifici, Camilla and Page, James Christophe and Paranilam, James G. and Park, Sang and Parrish, Keith A. and Paschal, Laura and Patapis, Polychronis and Patel, Jignasha and Patrick, Keith and Jr, Robert A. Pattishall and Paul, Douglas William and Paul, Shirley J. and Pauly, Tyler Andrew and Pavlovsky, Cheryl M. and {Pe{\~n}a-Guerrero}, Maria and Pedder, Andrew H. and Peek, Matthew Weldon and Pelham, Patricia A. and Penanen, Konstantin and Perriello, Beth A. and Perrin, Marshall D. and Perrine, Richard F. and Perrygo, Chuck and Peslier, Muriel and Petach, Michael and Peterson, Karla A. and Pfarr, Tom and Pierson, James M. and Pietraszkiewicz, Martin and Pilchen, Guy and Pipher, Judy L. and Pirzkal, Norbert and Pitman, Joseph T. and Player, Danielle M. and Plesha, Rachel and Plitzke, Anja and Pohner, John A. and Poletis, Karyn Konstantin and Pollizzi, Joseph A. and Polster, Ethan and Pontius, James T. and Pontoppidan, Klaus and Porges, Susana C. and Potter, Gregg D. and Prescott, Stephen and Proffitt, Charles R. and Pueyo, Laurent and Neira, Irma Aracely Quispe and Radich, Armando and Rager, Reiko T. and Rameau, Julien and Ramey, Deborah D. and Alarcon, Rafael Ramos and Rampini, Riccardo and Rapp, Robert and Rashford, Robert A. and Rauscher, Bernard J. and Ravindranath, Swara and Rawle, Timothy and Rawlings, Tynika N. and Ray, Tom and Regan, Michael W. and Rehm, Brian and Rehm, Kenneth D. and Reid, Neill and Reis, Carl A. and Renk, Florian and Reoch, Tom B. and Ressler, Michael and Rest, Armin W. and Reynolds, Paul J. and Richon, Joel G. and Richon, Karen V. and Ridgaway, Michael and Riedel, Adric Richard and Rieke, George H. and Rieke, Marcia J. and Rifelli, Richard E. and Rigby, Jane R. and Riggs, Catherine S. and Ringel, Nancy J. and Ritchie, Christine E. and Rix, Hans-Walter and Robberto, Massimo and Robinson, Gregory L. and Robinson, Michael S. and Robinson, Orion and Rock, Frank W. and Rodriguez, David R. and del Pino, Bruno Rodr{\'i}guez and Roellig, Thomas and Rohrbach, Scott O. and Roman, Anthony J. and Romelfanger, Frederick J. and Jr, Felipe P. Romo and Rosales, Jose J. and Rose, Perry and Roteliuk, Anthony F. and Roth, Marc N. and Rothwell, Braden Quinn and Rouzaud, Sylvain and Rowe, Jason and Rowlands, Neil and Roy, Arpita and Royer, Pierre and Rui, Chunlei and Rumler, Peter and Rumpl, William and Russ, Melissa L. and Ryan, Michael B. and Ryan, Richard M. and Saad, Karl and Sabata, Modhumita and Sabatino, Rick and Sabbi, Elena and Sabelhaus, Phillip A. and Sabia, Stephen and Sahu, Kailash C. and Saif, Babak N. and Salvignol, Jean-Christophe and {Samara-Ratna}, Piyal and Samuelson, Bridget S. and Sanders, Felicia A. and Sappington, Bradley and Sargent, B. A. and Sauer, Arne and Savadkin, Bruce J. and Sawicki, Marcin and Schappell, Tina M. and Scheffer, Caroline and Scheithauer, Silvia and Scherer, Ron and Schiff, Conrad and Schlawin, Everett and Schmeitzky, Olivier and Schmitz, Tyler S. and Schmude, Donald J. and Schneider, Analyn and Schreiber, J{\"u}rgen and {Schroeven-Deceuninck}, Hilde and Schultz, John J. and Schwab, Ryan and Schwartz, Curtis H. and Scoccimarro, Dario and Scott, John F. and Scott, Michelle B. and Seaton, Bonita L. and Seely, Bruce S. and Seery, Bernard and Seidleck, Mark and Sembach, Kenneth and Shanahan, Clare Elizabeth and Shaughnessy, Bryan and Shaw, Richard A. and Shay, Christopher Michael and Sheehan, Even and Sheth, Kartik and Shih, Hsin-Yi and Shivaei, Irene and Siegel, Noah and Sienkiewicz, Matthew G. and Simmons, Debra D. and Simon, Bernard P. and Sirianni, Marco and Sivaramakrishnan, Anand and Slade, Jeffrey E. and Sloan, G. C. and Slocum, Christine E. and Slowinski, Steven E. and Smith, Corbett T. and Smith, Eric P. and Smith, Erin C. and Smith, Koby and Smith, Robert and Smith, Stephanie J. and Smolik, John L. and Soderblom, David R. and Sohn, Sangmo Tony and Sokol, Jeff and Sonneborn, George and Sontag, Christopher D. and Sooy, Peter R. and Soummer, Remi and Southwood, Dana M. and Spain, Kay and Sparmo, Joseph and Speer, David T. and Spencer, Richard and Sprofera, Joseph D. and Stallcup, Scott S. and Stanley, Marcia K. and Stansberry, John A. and Stark, Christopher C. and Starr, Carl W. and Stassi, Diane Y. and Steck, Jane A. and Steeley, Christine D. and Stephens, Matthew A. and Stephenson, Ralph J. and Stewart, Alphonso C. and Stiavelli, Massimo and Jr, Hervey Stockman and Strada, Paolo and Straughn, Amber N. and Streetman, Scott and Strickland, David Kendal and Strobele, Jingping F. and Stuhlinger, Martin and Stys, Jeffrey Edward and Such, Miguel and Sukhatme, Kalyani and Sullivan, Joseph F. and Sullivan, Pamela C. and Sumner, Sandra M. and Sun, Fengwu and Sunnquist, Benjamin Dale and Swade, Daryl Allen and Swam, Michael S. and Swenton, Diane F. and Swoish, Robby A. and Litten, Oi In Tam and Tamas, Laszlo and Tao, Andrew and Taylor, David K. and Taylor, Joanna M. and te Plate, Maurice and Tea, Mason Van and Teague, Kelly K. and Telfer, Randal C. and Temim, Tea and Texter, Scott C. and Thatte, Deepashri G. and Thompson, Christopher Lee and Thompson, Linda M. and Thomson, Shaun R. and Thronson, Harley and Tierney, C. M. and Tikkanen, Tuomo and Tinnin, Lee and Tippet, William Thomas and Todd, Connor William and Tran, Hien D. and Trauger, John and Trejo, Edwin Gregorio and Truong, Justin Hoang Vinh and Tsukamoto, Christine L. and Tufail, Yasir and Tumlinson, Jason and Tustain, Samuel and Tyra, Harrison and Ubeda, Leonardo and Underwood, Kelli and Uzzo, Michael A. and Vaclavik, Steven and Valenduc, Frida and Valenti, Jeff A. and Campen, Julie Van and van de Wetering, Inge and Marel, Roeland P. Van Der and van Haarlem, Remy and Vandenbussche, Bart and van Dishoeck, Ewine F. and Vanterpool, Dona D. and Vernoy, Michael R. and Costas, Maria Bego{\~n}a Vila and Volk, Kevin and Voorzaat, Piet and Voyton, Mark F. and Vydra, Ekaterina and Waddy, Darryl J. and Waelkens, Christoffel and Wahlgren, Glenn Michael and Jr, Frederick E. Walker and Wander, Michel and Warfield, Christine K. and Warner, Gerald and Wasiak, Francis C. and Wasiak, Matthew F. and Wehner, James and Weiler, Kevin R. and Weilert, Mark and Weiss, Stanley B. and Wells, Martyn and Welty, Alan D. and Wheate, Lauren and Wheeler, Thomas P. and White, Christy L. and Whitehouse, Paul and Whiteleather, Jennifer Margaret and Whitman, William Russell and Williams, Christina C. and Willmer, Christopher N. A. and Willott, Chris J. and Willoughby, Scott P. and Wilson, Andrew and Wilson, Debra and Wilson, Donna V. and Windhorst, Rogier and Wislowski, Emily Christine and Wolfe, David J. and Wolfe, Michael A. and Wolff, Schuyler and Wondel, Amancio and Woo, Cindy and Woods, Robert T. and Worden, Elaine and Workman, William and Wright, Gillian S. and Wu, Carl and Wu, Chi-Rai and Wun, Dakin D. and Wymer, Kristen B. and Yadetie, Thomas and Yan, Isabelle C. and Yang, Keith C. and Yates, Kayla L. and Yeager, Christopher R. and Yerger, Ethan John and Young, Erick T. and Young, Gary and Yu, Gene and Yu, Susan and Zak, Dean S. and Zeidler, Peter and Zepp, Robert and Zhou, Julia and Zincke, Christian A. and Zonak, Stephanie and Zondag, Elisabeth},
  year = {2023},
  month = jun,
  journal = {PASP},
  volume = {135},
  number = {1048},
  pages = {068001},
  publisher = {The Astronomical Society of the Pacific},
  issn = {1538-3873},
  doi = {10.1088/1538-3873/acd1b5},
  langid = {english}
}

@article{gelman_2014,
  title = {Understanding Predictive Information Criteria for {{Bayesian}} Models},
  author = {Gelman, Andrew and Hwang, Jessica and Vehtari, Aki},
  year = {2014},
  month = nov,
  journal = {Stat Comput},
  volume = {24},
  number = {6},
  pages = {997--1016},
  issn = {1573-1375},
  doi = {10.1007/s11222-013-9416-2},
  langid = {english}
}

@article{gordon_2022,
  title = {The {{HITRAN2020}} Molecular Spectroscopic Database},
  author = {Gordon, I. E. and Rothman, L. S. and Hargreaves, R. J. and Hashemi, R. and Karlovets, E. V. and Skinner, F. M. and Conway, E. K. and Hill, C. and Kochanov, R. V. and Tan, Y. and Wcis{\l}o, P. and Finenko, A. A. and Nelson, K. and Bernath, P. F. and Birk, M. and Boudon, V. and Campargue, A. and Chance, K. V. and Coustenis, A. and Drouin, B. J. and Flaud, J. --M. and Gamache, R. R. and Hodges, J. T. and Jacquemart, D. and Mlawer, E. J. and Nikitin, A. V. and Perevalov, V. I. and Rotger, M. and Tennyson, J. and Toon, G. C. and Tran, H. and Tyuterev, V. G. and Adkins, E. M. and Baker, A. and Barbe, A. and Can{\`e}, E. and Cs{\'a}sz{\'a}r, A. G. and Dudaryonok, A. and Egorov, O. and Fleisher, A. J. and Fleurbaey, H. and Foltynowicz, A. and Furtenbacher, T. and Harrison, J. J. and Hartmann, J. --M. and Horneman, V. --M. and Huang, X. and Karman, T. and Karns, J. and Kassi, S. and Kleiner, I. and Kofman, V. and {Kwabia--Tchana}, F. and Lavrentieva, N. N. and Lee, T. J. and Long, D. A. and Lukashevskaya, A. A. and Lyulin, O. M. and Makhnev, V. {\relax Yu}. and Matt, W. and Massie, S. T. and Melosso, M. and Mikhailenko, S. N. and Mondelain, D. and M{\"u}ller, H. S. P. and Naumenko, O. V. and Perrin, A. and Polyansky, O. L. and Raddaoui, E. and Raston, P. L. and Reed, Z. D. and Rey, M. and Richard, C. and T{\'o}bi{\'a}s, R. and Sadiek, I. and Schwenke, D. W. and Starikova, E. and Sung, K. and Tamassia, F. and Tashkun, S. A. and Vander Auwera, J. and Vasilenko, I. A. and Vigasin, A. A. and Villanueva, G. L. and Vispoel, B. and Wagner, G. and Yachmenev, A. and Yurchenko, S. N.},
  year = {2022},
  month = jan,
  journal = {JQSRT},
  volume = {277},
  pages = {107949},
  issn = {0022-4073},
  doi = {10.1016/j.jqsrt.2021.107949},
  langid = {english}
}

@article{grant_2022,
  title = {Exo-{{TiC}}/{{ExoTiC-LD}}: {{ExoTiC-LD}} v3.0.0},
  author = {Grant, David and Wakeford, Hannah R.},
  year = {2022},
  month = dec,
  journal = {Zenodo},
  publisher = {Zenodo},
  doi = {10.5281/zenodo.7437681}
}

@article{harris_2020,
  title = {Array Programming with {{NumPy}}},
  author = {Harris, Charles R. and Millman, K. Jarrod and {van der Walt}, St{\'e}fan J. and Gommers, Ralf and Virtanen, Pauli and Cournapeau, David and Wieser, Eric and Taylor, Julian and Berg, Sebastian and Smith, Nathaniel J. and Kern, Robert and Picus, Matti and Hoyer, Stephan and {van Kerkwijk}, Marten H. and Brett, Matthew and Haldane, Allan and {del R{\'i}o}, Jaime Fern{\'a}ndez and Wiebe, Mark and Peterson, Pearu and {G{\'e}rard-Marchant}, Pierre and Sheppard, Kevin and Reddy, Tyler and Weckesser, Warren and Abbasi, Hameer and Gohlke, Christoph and Oliphant, Travis E.},
  year = {2020},
  month = sep,
  journal = {Nature},
  volume = {585},
  number = {7825},
  pages = {357--362},
  publisher = {Nature Publishing Group},
  issn = {1476-4687},
  doi = {10.1038/s41586-020-2649-2},
  copyright = {2020 The Author(s)},
  langid = {english}
}

@article{horne_1986,
  title = {An Optimal Extraction Algorithm for {{CCD}} Spectroscopy.},
  author = {Horne, K.},
  year = {1986},
  month = jun,
  journal = {PASP},
  volume = {98},
  pages = {609--617},
  issn = {0004-6280},
  doi = {10.1086/131801}
}

@article{hunter_2007,
  title = {Matplotlib: {{A 2D Graphics Environment}}},
  author = {Hunter, John D.},
  year = {2007},
  month = may,
  journal = {Computing in Science \& Engineering},
  volume = {9},
  number = {3},
  pages = {90--95},
  issn = {1558-366X},
  doi = {10.1109/MCSE.2007.55}
}

@article{jakobsen_2022,
  title = {The {{Near-Infrared Spectrograph}} ({{NIRSpec}}) on the {{James Webb Space Telescope}} - {{I}}. {{Overview}} of the Instrument and Its Capabilities},
  author = {Jakobsen, P. and Ferruit, P. and de Oliveira, C. Alves and Arribas, S. and Bagnasco, G. and Barho, R. and Beck, T. L. and Birkmann, S. and B{\"o}ker, T. and Bunker, A. J. and Charlot, S. and de Jong, P. and de Marchi, G. and Ehrenwinkler, R. and Falcolini, M. and Fels, R. and Franx, M. and Franz, D. and Funke, M. and Giardino, G. and Gnata, X. and Holota, W. and Honnen, K. and Jensen, P. L. and Jentsch, M. and Johnson, T. and Jollet, D. and Karl, H. and Kling, G. and K{\"o}hler, J. and Kolm, M.-G. and Kumari, N. and Lander, M. E. and Lemke, R. and {L{\'o}pez-Caniego}, M. and L{\"u}tzgendorf, N. and Maiolino, R. and Manjavacas, E. and Marston, A. and Maschmann, M. and Maurer, R. and Messerschmidt, B. and Moseley, S. H. and Mosner, P. and Mott, D. B. and Muzerolle, J. and Pirzkal, N. and Pittet, J.-F. and Plitzke, A. and Posselt, W. and Rapp, B. and Rauscher, B. J. and Rawle, T. and Rix, H.-W. and R{\"o}del, A. and Rumler, P. and Sabbi, E. and Salvignol, J.-C. and Schmid, T. and Sirianni, M. and Smith, C. and Strada, P. and te Plate, M. and Valenti, J. and Wettemann, T. and Wiehe, T. and Wiesmayer, M. and Willott, C. J. and Wright, R. and Zeidler, P. and Zincke, C.},
  year = {2022},
  month = may,
  journal = {A\&A},
  volume = {661},
  pages = {A80},
  publisher = {EDP Sciences},
  issn = {0004-6361, 1432-0746},
  doi = {10.1051/0004-6361/202142663},
  copyright = {{\copyright} ESO 2022},
  langid = {english}
}

@article{jtecers_2023,
  title = {Identification of Carbon Dioxide in an Exoplanet Atmosphere},
  author = {{JWST Transiting Exoplanet Community Early Release Science Team} and Ahrer, Eva-Maria and Alderson, Lili and Batalha, Natalie M. and Batalha, Natasha E. and Bean, Jacob L. and Beatty, Thomas G. and Bell, Taylor J. and Benneke, Bj{\"o}rn and {Berta-Thompson}, Zachory K. and Carter, Aarynn L. and Crossfield, Ian J. M. and Espinoza, N{\'e}stor and Feinstein, Adina D. and Fortney, Jonathan J. and Gibson, Neale P. and Goyal, Jayesh M. and Kempton, Eliza M.-R. and Kirk, James and Kreidberg, Laura and {L{\'o}pez-Morales}, Mercedes and Line, Michael R. and Lothringer, Joshua D. and Moran, Sarah E. and Mukherjee, Sagnick and Ohno, Kazumasa and Parmentier, Vivien and Piaulet, Caroline and Rustamkulov, Zafar and Schlawin, Everett and Sing, David K. and Stevenson, Kevin B. and Wakeford, Hannah R. and Allen, Natalie H. and Birkmann, Stephan M. and Brande, Jonathan and Crouzet, Nicolas and Cubillos, Patricio E. and Damiano, Mario and D{\'e}sert, Jean-Michel and Gao, Peter and Harrington, Joseph and Hu, Renyu and Kendrew, Sarah and Knutson, Heather A. and Lagage, Pierre-Olivier and Leconte, J{\'e}r{\'e}my and Lendl, Monika and MacDonald, Ryan J. and May, E. M. and Miguel, Yamila and Molaverdikhani, Karan and Moses, Julianne I. and Murray, Catriona Anne and Nehring, Molly and Nikolov, Nikolay K. and {Petit dit de la Roche}, D. J. M. and Radica, Michael and Roy, Pierre-Alexis and Stassun, Keivan G. and Taylor, Jake and Waalkes, William C. and Wachiraphan, Patcharapol and Welbanks, Luis and Wheatley, Peter J. and Aggarwal, Keshav and Alam, Munazza K. and Banerjee, Agnibha and Barstow, Joanna K. and Blecic, Jasmina and Casewell, S. L. and Changeat, Quentin and Chubb, K. L. and Col{\'o}n, Knicole D. and Coulombe, Louis-Philippe and Daylan, Tansu and {de Val-Borro}, Miguel and Decin, Leen and Dos Santos, Leonardo A. and Flagg, Laura and France, Kevin and Fu, Guangwei and Garc{\'i}a Mu{\~n}oz, A. and Gizis, John E. and Glidden, Ana and Grant, David and Heng, Kevin and Henning, Thomas and Hong, Yu-Cian and Inglis, Julie and Iro, Nicolas and Kataria, Tiffany and Komacek, Thaddeus D. and Krick, Jessica E. and Lee, Elspeth K. H. and Lewis, Nikole K. and {Lillo-Box}, Jorge and {Lustig-Yaeger}, Jacob and Mancini, Luigi and Mandell, Avi M. and Mansfield, Megan and Marley, Mark S. and {Mikal-Evans}, Thomas and Morello, Giuseppe and Nixon, Matthew C. and Ortiz Ceballos, Kevin and Piette, Anjali A. A. and Powell, Diana and Rackham, Benjamin V. and {Ramos-Rosado}, Lakeisha and Rauscher, Emily and Redfield, Seth and Rogers, Laura K. and Roman, Michael T. and Roudier, Gael M. and Scarsdale, Nicholas and Shkolnik, Evgenya L. and Southworth, John and Spake, Jessica J. and Steinrueck, Maria E. and Tan, Xianyu and Teske, Johanna K. and Tremblin, Pascal and Tsai, Shang-Min and Tucker, Gregory S. and Turner, Jake D. and Valenti, Jeff A. and Venot, Olivia and Waldmann, Ingo P. and Wallack, Nicole L. and Zhang, Xi and Zieba, Sebastian},
  year = {2023},
  month = feb,
  journal = {Nature},
  volume = {614},
  number = {7949},
  pages = {649--652},
  publisher = {Nature Publishing Group},
  issn = {1476-4687},
  doi = {10.1038/s41586-022-05269-w},
  copyright = {2022 The Author(s), under exclusive licence to Springer Nature Limited},
  langid = {english}
}

@article{kass_1995,
  title = {Bayes {{Factors}}},
  author = {Kass, Robert E. and Raftery, Adrian E.},
  year = {1995},
  month = jun,
  journal = {Journal of the American Statistical Association},
  volume = {90},
  number = {430},
  pages = {773--795},
  publisher = {Taylor \& Francis},
  issn = {0162-1459},
  doi = {10.1080/01621459.1995.10476572}
}

@article{keers_2024,
  title = {Reliable {{Transmission Spectrum Extraction}} with a {{Three-parameter Limb-darkening Law}}},
  author = {Keers, Rosa E. and Shapiro, Alexander I. and Kostogryz, Nadiia M. and Glidden, Ana and Niraula, Prajwal and Rackham, Benjamin V. and Seager, Sara and Solanki, Sami K. and Unruh, Yvonne C. and Vasilyev, Valeriy and {de Wit}, Julien},
  year = {2024},
  month = dec,
  journal = {ApJL},
  volume = {977},
  number = {1},
  pages = {L7},
  publisher = {The American Astronomical Society},
  issn = {2041-8205},
  doi = {10.3847/2041-8213/ad8b51},
  langid = {english}
}

@article{kempton_2023,
  title = {A Reflective, Metal-Rich Atmosphere for {{GJ}} 1214b from Its {{JWST}} Phase Curve},
  author = {Kempton, Eliza M.-R. and Zhang, Michael and Bean, Jacob L. and Steinrueck, Maria E. and Piette, Anjali A. A. and Parmentier, Vivien and Malsky, Isaac and Roman, Michael T. and Rauscher, Emily and Gao, Peter and Bell, Taylor J. and Xue, Qiao and Taylor, Jake and Savel, Arjun B. and Arnold, Kenneth E. and Nixon, Matthew C. and Stevenson, Kevin B. and Mansfield, Megan and Kendrew, Sarah and Zieba, Sebastian and Ducrot, Elsa and Dyrek, Achr{\`e}ne and Lagage, Pierre-Olivier and Stassun, Keivan G. and Henry, Gregory W. and Barman, Travis and Lupu, Roxana and Malik, Matej and Kataria, Tiffany and Ih, Jegug and Fu, Guangwei and Welbanks, Luis and McGill, Peter},
  year = {2023},
  month = aug,
  journal = {Nature},
  volume = {620},
  number = {7972},
  pages = {67--71},
  publisher = {Nature Publishing Group},
  issn = {1476-4687},
  doi = {10.1038/s41586-023-06159-5},
  copyright = {2023 The Author(s), under exclusive licence to Springer Nature Limited},
  langid = {english}
}

@article{kreidberg_2015,
  title = {Batman: {{BAsic Transit Model cAlculatioN}} in {{Python}}},
  author = {Kreidberg, Laura},
  year = {2015},
  month = nov,
  journal = {PASP},
  volume = {127},
  number = {957},
  pages = {1161},
  publisher = {IOP Publishing},
  issn = {1538-3873},
  doi = {10.1086/683602},
  langid = {english}
}

@article{li_2015,
  title = {{{ROVIBRATIONAL LINE LISTS FOR NINE ISOTOPOLOGUES OF THE CO MOLECULE IN THE X1$\Sigma$}}+ {{GROUND ELECTRONIC STATE}}},
  author = {Li, Gang and Gordon, Iouli E. and Rothman, Laurence S. and Tan, Yan and Hu, Shui-Ming and Kassi, Samir and Campargue, Alain and Medvedev, Emile S.},
  year = {2015},
  month = jan,
  journal = {ApJS},
  volume = {216},
  number = {1},
  pages = {15},
  publisher = {The American Astronomical Society},
  issn = {0067-0049},
  doi = {10.1088/0067-0049/216/1/15},
  langid = {english}
}

@article{lueber_2024,
  title = {Information Content of {{JWST}} Spectra of {{WASP-39b}}},
  author = {Lueber, Anna and Novais, Aline and Fisher, Chloe and Heng, Kevin},
  year = {2024},
  month = jul,
  journal = {A\&A},
  volume = {687},
  pages = {A110},
  issn = {0004-6361},
  doi = {10.1051/0004-6361/202348802}
}

@article{lustig-yaeger_2023,
  title = {A {{JWST}} Transmission Spectrum of the Nearby {{Earth-sized}} Exoplanet {{LHS}} 475 b},
  author = {{Lustig-Yaeger}, Jacob and Fu, Guangwei and May, E. M. and Ceballos, Kevin N. Ortiz and Moran, Sarah E. and Peacock, Sarah and Stevenson, Kevin B. and Kirk, James and {L{\'o}pez-Morales}, Mercedes and MacDonald, Ryan J. and Mayorga, L. C. and Sing, David K. and Sotzen, Kristin S. and Valenti, Jeff A. and Redai, J{\'e}a I. Adams and Alam, Munazza K. and Batalha, Natasha E. and Bennett, Katherine A. and {Gonzalez-Quiles}, Junellie and Kruse, Ethan and Lothringer, Joshua D. and Rustamkulov, Zafar and Wakeford, Hannah R.},
  year = {2023},
  month = aug,
  journal = {Nature Astronomy},
  pages = {1--12},
  publisher = {Nature Publishing Group},
  issn = {2397-3366},
  doi = {10.1038/s41550-023-02064-z},
  copyright = {2023 The Author(s), under exclusive licence to Springer Nature Limited},
  langid = {english}
}

@article{macdonald_2020,
  title = {Why {{Is}} It {{So Cold}} in {{Here}}? {{Explaining}} the {{Cold Temperatures Retrieved}} from {{Transmission Spectra}} of {{Exoplanet Atmospheres}}},
  author = {MacDonald, Ryan J. and Goyal, Jayesh M. and Lewis, Nikole K.},
  year = {2020},
  month = apr,
  journal = {ApJ},
  volume = {893},
  pages = {L43},
  issn = {0004-637X},
  doi = {10.3847/2041-8213/ab8238}
}

@article{madhusudhan_2019,
  title = {Exoplanetary {{Atmospheres}}: {{Key Insights}}, {{Challenges}}, and {{Prospects}}},
  author = {Madhusudhan, Nikku},
  year = {2019},
  journal = {ARA\&A},
  volume = {57},
  number = {1},
  pages = {617--663},
  doi = {10.1146/annurev-astro-081817-051846}
}

@article{madhusudhan_2023,
  title = {Carbon-Bearing {{Molecules}} in a {{Possible Hycean Atmosphere}}},
  author = {Madhusudhan, Nikku and Sarkar, Subhajit and Constantinou, Savvas and Holmberg, M{\aa}ns and Piette, Anjali A. A. and Moses, Julianne I.},
  year = {2023},
  month = oct,
  journal = {ApJL},
  volume = {956},
  pages = {L13},
  issn = {0004-637X},
  doi = {10.3847/2041-8213/acf577}
}

@article{magic_2015,
  title = {The {{Stagger-grid}}: {{A}} Grid of {{3D}} Stellar Atmosphere Models - {{IV}}. {{Limb}} Darkening Coefficients},
  author = {Magic, Z. and Chiavassa, A. and Collet, R. and Asplund, M.},
  year = {2015},
  month = jan,
  journal = {A\&A},
  volume = {573},
  pages = {A90},
  publisher = {EDP Sciences},
  issn = {0004-6361, 1432-0746},
  doi = {10.1051/0004-6361/201423804},
  copyright = {{\copyright} ESO, 2014},
  langid = {english}
}

@article{mancini_2018,
  title = {The {{GAPS}} Programme with {{HARPS-N}} at {{TNG}} - {{XVI}}. {{Measurement}} of the {{Rossiter}}--{{McLaughlin}} Effect of Transiting Planetary Systems {{HAT-P-3}}, {{HAT-P-12}}, {{HAT-P-22}}, {{WASP-39}}, and {{WASP-60}}},
  author = {Mancini, L. and Esposito, M. and Covino, E. and Southworth, J. and Biazzo, K. and Bruni, I. and Ciceri, S. and Evans, D. and Lanza, A. F. and Poretti, E. and Sarkis, P. and Smith, A. M. S. and Brogi, M. and Affer, L. and Benatti, S. and Bignamini, A. and Boccato, C. and Bonomo, A. S. and Borsa, F. and Carleo, I. and Claudi, R. and Cosentino, R. and Damasso, M. and Desidera, S. and Giacobbe, P. and {Gonz{\'a}lez-{\'A}lvarez}, E. and Gratton, R. and Harutyunyan, A. and Leto, G. and Maggio, A. and Malavolta, L. and Maldonado, J. and {Martinez-Fiorenzano}, A. and Masiero, S. and Micela, G. and Molinari, E. and Nascimbeni, V. and Pagano, I. and Pedani, M. and Piotto, G. and Rainer, M. and Scandariato, G. and Smareglia, R. and Sozzetti, A. and Andreuzzi, G. and Henning, Th},
  year = {2018},
  month = may,
  journal = {A\&A},
  volume = {613},
  pages = {A41},
  publisher = {EDP Sciences},
  issn = {0004-6361, 1432-0746},
  doi = {10.1051/0004-6361/201732234},
  copyright = {{\copyright} ESO 2018},
  langid = {english}
}

@article{mant_2018,
  title = {{{ExoMol}} Molecular Line Lists -- {{XXVII}}. {{Spectra}} of {{C2H4}}},
  author = {Mant, Barry P and Yachmenev, Andrey and Tennyson, Jonathan and Yurchenko, Sergei N},
  year = {2018},
  month = aug,
  journal = {MNRAS},
  volume = {478},
  number = {3},
  pages = {3220--3232},
  issn = {0035-8711},
  doi = {10.1093/mnras/sty1239}
}

@article{molliere_2022,
  title = {Interpreting the {{Atmospheric Composition}} of {{Exoplanets}}: {{Sensitivity}} to {{Planet Formation Assumptions}}},
  author = {Molli{\`e}re, Paul and Molyarova, Tamara and Bitsch, Bertram and Henning, Thomas and Schneider, Aaron and Kreidberg, Laura and Eistrup, Christian and Burn, Remo and Nasedkin, Evert and Semenov, Dmitry and Mordasini, Christoph and Schlecker, Martin and Schwarz, Kamber R. and Lacour, Sylvestre and Nowak, Mathias and Schulik, Matth{\"a}us},
  year = {2022},
  month = jul,
  journal = {ApJ},
  volume = {934},
  pages = {74},
  issn = {0004-637X},
  doi = {10.3847/1538-4357/ac6a56}
}

@article{moran_2023,
  title = {High {{Tide}} or {{Riptide}} on the {{Cosmic Shoreline}}? {{A Water-rich Atmosphere}} or {{Stellar Contamination}} for the {{Warm Super-Earth GJ}} 486b from {{JWST Observations}}},
  author = {Moran, Sarah E. and Stevenson, Kevin B. and Sing, David K. and MacDonald, Ryan J. and Kirk, James and {Lustig-Yaeger}, Jacob and Peacock, Sarah and Mayorga, L. C. and Bennett, Katherine A. and {L{\'o}pez-Morales}, Mercedes and May, E. M. and Rustamkulov, Zafar and Valenti, Jeff A. and Redai, J{\'e}a I. Adams and Alam, Munazza K. and Batalha, Natasha E. and Fu, Guangwei and {Gonzalez-Quiles}, Junellie and Highland, Alicia N. and Kruse, Ethan and Lothringer, Joshua D. and Ceballos, Kevin N. Ortiz and Sotzen, Kristin S. and Wakeford, Hannah R.},
  year = {2023},
  month = may,
  journal = {ApJL},
  volume = {948},
  number = {1},
  pages = {L11},
  publisher = {The American Astronomical Society},
  issn = {2041-8205},
  doi = {10.3847/2041-8213/accb9c},
  langid = {english}
}

@article{morello_2017,
  title = {High-Precision {{Stellar Limb-darkening}} in {{Exoplanetary Transits}}},
  author = {Morello, G. and Tsiaras, A. and Howarth, I. D. and Homeier, D.},
  year = {2017},
  month = aug,
  journal = {AJ},
  volume = {154},
  number = {3},
  pages = {111},
  publisher = {The American Astronomical Society},
  issn = {1538-3881},
  doi = {10.3847/1538-3881/aa8405},
  langid = {english}
}

@article{morello_2022-1,
  title = {Is Binning Always Sinning? {{The}} Impact of Time-Averaging for Exoplanet Phase Curves},
  author = {Morello, Giuseppe and Dyrek, Achr{\`e}ne and Changeat, Quentin},
  year = {2022},
  month = oct,
  journal = {MNRAS},
  volume = {517},
  number = {2},
  eprint = {2210.14194},
  primaryclass = {astro-ph},
  pages = {2151--2164},
  issn = {0035-8711, 1365-2966},
  doi = {10.1093/mnras/stac2828},
  archiveprefix = {arXiv}
}

@article{mugnai_2024,
  title = {Comparing Transit Spectroscopy Pipelines at the Catalogue Level: Evidence for Systematic Differences},
  author = {Mugnai, Lorenzo V and Swain, Mark R and Estrela, Raissa and Roudier, Gael M},
  year = {2024},
  month = jun,
  journal = {MNRAS},
  volume = {531},
  number = {1},
  pages = {35--51},
  issn = {0035-8711},
  doi = {10.1093/mnras/stae1073}
}

@article{murphy_2025,
  title = {{{HST Transmission Spectra}} of the {{Hot Neptune HD}} 219666 b: {{Detection}} of {{Water}} and the {{Challenge}} of {{Constraining Both Water}} and {{Methane}}},
  author = {Murphy, Matthew M. and Beatty, Thomas G. and Welbanks, Luis and Fu, Guangwei},
  year = {2025},
  month = may,
  journal = {AJ},
  volume = {169},
  number = {6},
  pages = {286},
  publisher = {The American Astronomical Society},
  issn = {1538-3881},
  doi = {10.3847/1538-3881/adc684},
  langid = {english}
}

@article{niraula_2022,
  title = {The Impending Opacity Challenge in Exoplanet Atmospheric Characterization},
  author = {Niraula, Prajwal and {de Wit}, Julien and Gordon, Iouli E. and Hargreaves, Robert J. and {Sousa-Silva}, Clara and Kochanov, Roman V.},
  year = {2022},
  month = nov,
  journal = {Nat Astron},
  volume = {6},
  number = {11},
  pages = {1287--1295},
  publisher = {Nature Publishing Group},
  issn = {2397-3366},
  doi = {10.1038/s41550-022-01773-1},
  copyright = {2022 The Author(s), under exclusive licence to Springer Nature Limited},
  langid = {english}
}

@article{niraula_2023,
  title = {Origin and {{Extent}} of the {{Opacity Challenge}} for {{Atmospheric Retrievals}} of {{WASP-39}} b},
  author = {Niraula, Prajwal and de Wit, Julien and Gordon, Iouli E. and Hargreaves, Robert J. and {Sousa-Silva}, Clara},
  year = {2023},
  month = jun,
  journal = {ApJL},
  volume = {950},
  number = {2},
  pages = {L17},
  publisher = {The American Astronomical Society},
  issn = {2041-8205},
  doi = {10.3847/2041-8213/acd6f8},
  langid = {english}
}

@article{pluriel_2022,
  title = {Toward a Multidimensional Analysis of Transmission Spectroscopy - {{II}}. {{Day-night-induced}} Biases in Retrievals from Hot to Ultrahot {{Jupiters}}},
  author = {Pluriel, William and Leconte, J{\'e}r{\'e}my and Parmentier, Vivien and Zingales, Tiziano and Falco, Aur{\'e}lien and Selsis, Franck and Bord{\'e}, Pascal},
  year = {2022},
  month = feb,
  journal = {A\&A},
  volume = {658},
  pages = {A42},
  publisher = {EDP Sciences},
  issn = {0004-6361, 1432-0746},
  doi = {10.1051/0004-6361/202141943},
  copyright = {{\copyright} W. Pluriel et al. 2022},
  langid = {english}
}

@article{polyansky_2018,
  title = {{{ExoMol}} Molecular Line Lists {{XXX}}: A Complete High-Accuracy Line List for Water},
  author = {Polyansky, Oleg L and Kyuberis, Aleksandra A and Zobov, Nikolai F and Tennyson, Jonathan and Yurchenko, Sergei N and Lodi, Lorenzo},
  year = {2018},
  month = oct,
  journal = {MNRAS},
  volume = {480},
  number = {2},
  pages = {2597--2608},
  issn = {0035-8711},
  doi = {10.1093/mnras/sty1877}
}

@article{powell_2024,
  title = {Sulfur Dioxide in the Mid-Infrared Transmission Spectrum of {{WASP-39b}}},
  author = {Powell, Diana and Feinstein, Adina D. and Lee, Elspeth K. H. and Zhang, Michael and Tsai, Shang-Min and Taylor, Jake and Kirk, James and Bell, Taylor and Barstow, Joanna K. and Gao, Peter and Bean, Jacob L. and Blecic, Jasmina and Chubb, Katy L. and Crossfield, Ian J. M. and Jordan, Sean and Kitzmann, Daniel and Moran, Sarah E. and Morello, Giuseppe and Moses, Julianne I. and Welbanks, Luis and Yang, Jeehyun and Zhang, Xi and Ahrer, Eva-Maria and {Bello-Arufe}, Aaron and Brande, Jonathan and Casewell, S. L. and Crouzet, Nicolas and Cubillos, Patricio E. and Demory, Brice-Olivier and Dyrek, Achr{\`e}ne and Flagg, Laura and Hu, Renyu and Inglis, Julie and Jones, Kathryn D. and Kreidberg, Laura and {L{\'o}pez-Morales}, Mercedes and Lagage, Pierre-Olivier and Meier Vald{\'e}s, Erik A. and Miguel, Yamila and Parmentier, Vivien and Piette, Anjali A. A. and Rackham, Benjamin V. and Radica, Michael and Redfield, Seth and Stevenson, Kevin B. and Wakeford, Hannah R. and Aggarwal, Keshav and Alam, Munazza K. and Batalha, Natalie M. and Batalha, Natasha E. and Benneke, Bj{\"o}rn and {Berta-Thompson}, Zach K. and Brady, Ryan P. and Caceres, Claudio and Carter, Aarynn L. and D{\'e}sert, Jean-Michel and Harrington, Joseph and Iro, Nicolas and Line, Michael R. and Lothringer, Joshua D. and MacDonald, Ryan J. and Mancini, Luigi and Molaverdikhani, Karan and Mukherjee, Sagnick and Nixon, Matthew C. and Oza, Apurva V. and Palle, Enric and Rustamkulov, Zafar and Sing, David K. and Steinrueck, Maria E. and Venot, Olivia and Wheatley, Peter J. and Yurchenko, Sergei N.},
  year = {2024},
  month = feb,
  journal = {Nature},
  volume = {626},
  number = {8001},
  pages = {979--983},
  publisher = {Nature Publishing Group},
  issn = {1476-4687},
  doi = {10.1038/s41586-024-07040-9},
  copyright = {2024 The Author(s)},
  langid = {english}
}

@article{rackham_2018,
  title = {The {{Transit Light Source Effect}}: {{False Spectral Features}} and {{Incorrect Densities}} for {{M-dwarf Transiting Planets}}},
  author = {Rackham, Benjamin V. and Apai, D{\'a}niel and Giampapa, Mark S.},
  year = {2018},
  month = jan,
  journal = {ApJ},
  volume = {853},
  number = {2},
  pages = {122},
  publisher = {The American Astronomical Society},
  issn = {0004-637X},
  doi = {10.3847/1538-4357/aaa08c},
  langid = {english}
}

@article{rothman_2010,
  title = {{{HITEMP}}, the High-Temperature Molecular Spectroscopic Database},
  author = {Rothman, L. S. and Gordon, I. E. and Barber, R. J. and Dothe, H. and Gamache, R. R. and Goldman, A. and Perevalov, V. I. and Tashkun, S. A. and Tennyson, J.},
  year = {2010},
  month = oct,
  journal = {JQSRT},
  series = {{{XVIth Symposium}} on {{High Resolution Molecular Spectroscopy}} ({{HighRus-2009}})},
  volume = {111},
  number = {15},
  pages = {2139--2150},
  issn = {0022-4073},
  doi = {10.1016/j.jqsrt.2010.05.001},
  langid = {english}
}

@article{rustamkulov_2023,
  title = {Early {{Release Science}} of the Exoplanet {{WASP-39b}} with {{JWST NIRSpec PRISM}}},
  author = {Rustamkulov, Z. and Sing, D. K. and Mukherjee, S. and May, E. M. and Kirk, J. and Schlawin, E. and Line, M. R. and Piaulet, C. and Carter, A. L. and Batalha, N. E. and Goyal, J. M. and {L{\'o}pez-Morales}, M. and Lothringer, J. D. and MacDonald, R. J. and Moran, S. E. and Stevenson, K. B. and Wakeford, H. R. and Espinoza, N. and Bean, J. L. and Batalha, N. M. and Benneke, B. and {Berta-Thompson}, Z. K. and Crossfield, I. J. M. and Gao, P. and Kreidberg, L. and Powell, D. K. and Cubillos, P. E. and Gibson, N. P. and Leconte, J. and Molaverdikhani, K. and Nikolov, N. K. and Parmentier, V. and Roy, P. and Taylor, J. and Turner, J. D. and Wheatley, P. J. and Aggarwal, K. and Ahrer, E. and Alam, M. K. and Alderson, L. and Allen, N. H. and Banerjee, A. and Barat, S. and Barrado, D. and Barstow, J. K. and Bell, T. J. and Blecic, J. and Brande, J. and Casewell, S. and Changeat, Q. and Chubb, K. L. and Crouzet, N. and Daylan, T. and Decin, L. and D{\'e}sert, J. and {Mikal-Evans}, T. and Feinstein, A. D. and Flagg, L. and Fortney, J. J. and Harrington, J. and Heng, K. and Hong, Y. and Hu, R. and Iro, N. and Kataria, T. and Kempton, E. M.-R. and Krick, J. and Lendl, M. and {Lillo-Box}, J. and Louca, A. and {Lustig-Yaeger}, J. and Mancini, L. and Mansfield, M. and Mayne, N. J. and Miguel, Y. and Morello, G. and Ohno, K. and Palle, E. and {Petit dit de la Roche}, D. J. M. and Rackham, B. V. and Radica, M. and {Ramos-Rosado}, L. and Redfield, S. and Rogers, L. K. and Shkolnik, E. L. and Southworth, J. and Teske, J. and Tremblin, P. and Tucker, G. S. and Venot, O. and Waalkes, W. C. and Welbanks, L. and Zhang, X. and Zieba, S.},
  year = {2023},
  month = feb,
  journal = {Nature},
  volume = {614},
  number = {7949},
  pages = {659--663},
  publisher = {Nature Publishing Group},
  issn = {1476-4687},
  doi = {10.1038/s41586-022-05677-y},
  copyright = {2023 The Author(s)},
  langid = {english}
}

@article{saba_2022,
  title = {The {{Transmission Spectrum}} of {{WASP-17}} b {{From}} the {{Optical}} to the {{Near-infrared Wavelengths}}: {{Combining STIS}}, {{WFC3}}, and {{IRAC Data Sets}}},
  author = {Saba, Arianna and Tsiaras, Angelos and Morvan, Mario and Thompson, Alexandra and Changeat, Quentin and Edwards, Billy and Jolly, Andrew and Waldmann, Ingo and Tinetti, Giovanna},
  year = {2022},
  month = jun,
  journal = {AJ},
  volume = {164},
  number = {1},
  pages = {2},
  publisher = {The American Astronomical Society},
  issn = {1538-3881},
  doi = {10.3847/1538-3881/ac6c01},
  langid = {english}
}

@article{sarkar_2021,
  title = {{{JexoSim}} 2.0: End-to-End {{JWST}} Simulator for Exoplanet Spectroscopy -- Implementation and Case Studies},
  author = {Sarkar, Subhajit and Madhusudhan, Nikku},
  year = {2021},
  month = nov,
  journal = {MNRAS},
  volume = {508},
  number = {1},
  pages = {433--452},
  issn = {0035-8711},
  doi = {10.1093/mnras/stab2472}
}

@article{sarkar_2024,
  title = {Exoplanet Transit Spectroscopy with {{JWST NIRSpec}}: Diagnostics and Homogeneous Case Study of {{WASP-39}} b},
  author = {Sarkar, Subhajit and Madhusudhan, Nikku and Constantinou, Savvas and Holmberg, M{\aa}ns},
  year = {2024},
  month = jun,
  journal = {MNRAS},
  volume = {531},
  number = {2},
  pages = {2731--2756},
  issn = {0035-8711},
  doi = {10.1093/mnras/stae1230}
}

@article{schleich_2024,
  title = {Knobs and Dials of Retrieving {{JWST}} Transmission Spectra - {{I}}. {{The}} Importance of p--{{T}} Profile Complexity},
  author = {Schleich, S. and Boro Saikia, S. and Changeat, Q. and G{\"u}del, M. and Voigt, A. and Waldmann, I.},
  year = {2024},
  month = oct,
  journal = {A\&A},
  volume = {690},
  pages = {A336},
  publisher = {EDP Sciences},
  issn = {0004-6361, 1432-0746},
  doi = {10.1051/0004-6361/202451845},
  copyright = {{\copyright} The Authors 2024},
  langid = {english}
}

@article{schmidt_2025,
  title = {A {{Comprehensive Reanalysis}} of {{K2-18}} b's {{JWST NIRISS}}+{{NIRSpec Transmission Spectrum}}},
  author = {Schmidt, Stephen P. and MacDonald, Ryan J. and Tsai, Shang-Min and Radica, Michael and Wang, Le-Chris and Ahrer, Eva-Maria and Bell, Taylor J. and Fisher, Chloe and Thorngren, Daniel P. and Wogan, Nicholas and May, Erin M. and Ferrari, Piero and Bennett, Katherine A. and Rustamkulov, Zafar and {L{\'o}pez-Morales}, Mercedes and Sing, David K.},
  year = {2025},
  month = jan,
  number = {arXiv:2501.18477},
  eprint = {2501.18477},
  primaryclass = {astro-ph},
  publisher = {arXiv},
  doi = {10.48550/arXiv.2501.18477},
  archiveprefix = {arXiv}
}

@article{tennyson_2020,
  title = {The 2020 Release of the {{ExoMol}} Database: {{Molecular}} Line Lists for Exoplanet and Other Hot Atmospheres},
  author = {Tennyson, Jonathan and Yurchenko, Sergei N. and {Al-Refaie}, Ahmed F. and Clark, Victoria H. J. and Chubb, Katy L. and Conway, Eamon K. and Dewan, Akhil and Gorman, Maire N. and Hill, Christian and {Lynas-Gray}, A. E. and Mellor, Thomas and McKemmish, Laura K. and Owens, Alec and Polyansky, Oleg L. and Semenov, Mikhail and Somogyi, Wilfrid and Tinetti, Giovanna and Upadhyay, Apoorva and Waldmann, Ingo and Wang, Yixin and Wright, Samuel and Yurchenko, Olga P.},
  year = {2020},
  month = nov,
  journal = {JQSRT},
  volume = {255},
  pages = {107228},
  issn = {0022-4073},
  doi = {10.1016/j.jqsrt.2020.107228}
}

@article{tsai_2023,
  title = {Photochemically Produced {{SO2}} in the Atmosphere of {{WASP-39b}}},
  author = {Tsai, Shang-Min and Lee, Elspeth K. H. and Powell, Diana and Gao, Peter and Zhang, Xi and Moses, Julianne and H{\'e}brard, Eric and Venot, Olivia and Parmentier, Vivien and Jordan, Sean and Hu, Renyu and Alam, Munazza K. and Alderson, Lili and Batalha, Natalie M. and Bean, Jacob L. and Benneke, Bj{\"o}rn and Bierson, Carver J. and Brady, Ryan P. and Carone, Ludmila and Carter, Aarynn L. and Chubb, Katy L. and Inglis, Julie and Leconte, J{\'e}r{\'e}my and Line, Michael and {L{\'o}pez-Morales}, Mercedes and Miguel, Yamila and Molaverdikhani, Karan and Rustamkulov, Zafar and Sing, David K. and Stevenson, Kevin B. and Wakeford, Hannah R. and Yang, Jeehyun and Aggarwal, Keshav and Baeyens, Robin and Barat, Saugata and {de Val-Borro}, Miguel and Daylan, Tansu and Fortney, Jonathan J. and France, Kevin and Goyal, Jayesh M. and Grant, David and Kirk, James and Kreidberg, Laura and Louca, Amy and Moran, Sarah E. and Mukherjee, Sagnick and Nasedkin, Evert and Ohno, Kazumasa and Rackham, Benjamin V. and Redfield, Seth and Taylor, Jake and Tremblin, Pascal and Visscher, Channon and Wallack, Nicole L. and Welbanks, Luis and Youngblood, Allison and Ahrer, Eva-Maria and Batalha, Natasha E. and Behr, Patrick and {Berta-Thompson}, Zachory K. and Blecic, Jasmina and Casewell, S. L. and Crossfield, Ian J. M. and Crouzet, Nicolas and Cubillos, Patricio E. and Decin, Leen and D{\'e}sert, Jean-Michel and Feinstein, Adina D. and Gibson, Neale P. and Harrington, Joseph and Heng, Kevin and Henning, Thomas and Kempton, Eliza M.-R. and Krick, Jessica and Lagage, Pierre-Olivier and Lendl, Monika and Lothringer, Joshua D. and Mansfield, Megan and Mayne, N. J. and {Mikal-Evans}, Thomas and Palle, Enric and Schlawin, Everett and Shorttle, Oliver and Wheatley, Peter J. and Yurchenko, Sergei N.},
  year = {2023},
  month = may,
  journal = {Nature},
  volume = {617},
  number = {7961},
  pages = {483--487},
  publisher = {Nature Publishing Group},
  issn = {1476-4687},
  doi = {10.1038/s41586-023-05902-2},
  copyright = {2023 The Author(s)},
  langid = {english}
}

@article{tsiaras_2018,
  title = {A {{Population Study}} of {{Gaseous Exoplanets}}},
  author = {Tsiaras, A. and Waldmann, I. P. and Zingales, T. and Rocchetto, M. and Morello, G. and Damiano, M. and Karpouzas, K. and Tinetti, G. and McKemmish, L. K. and Tennyson, J. and Yurchenko, S. N.},
  year = {2018},
  month = mar,
  journal = {AJ},
  volume = {155},
  number = {4},
  pages = {156},
  publisher = {The American Astronomical Society},
  issn = {1538-3881},
  doi = {10.3847/1538-3881/aaaf75},
  langid = {english}
}

@article{underwood_2016,
  title = {{{ExoMol}} Molecular Line Lists -- {{XIV}}. {{The}} Rotation--Vibration Spectrum of Hot {{SO2}}},
  author = {Underwood, Daniel S. and Tennyson, Jonathan and Yurchenko, Sergei N. and Huang, Xinchuan and Schwenke, David W. and Lee, Timothy J. and Clausen, S{\o}nnik and Fateev, Alexander},
  year = {2016},
  month = jul,
  journal = {MNRAS},
  volume = {459},
  number = {4},
  pages = {3890--3899},
  issn = {0035-8711},
  doi = {10.1093/mnras/stw849}
}

@article{vehtari_2017,
  title = {Practical {{Bayesian}} Model Evaluation Using Leave-One-out Cross-Validation and {{WAIC}}},
  author = {Vehtari, Aki and Gelman, Andrew and Gabry, Jonah},
  year = {2017},
  month = sep,
  journal = {Stat Comput},
  volume = {27},
  number = {5},
  pages = {1413--1432},
  issn = {1573-1375},
  doi = {10.1007/s11222-016-9696-4},
  langid = {english}
}

@article{virtanen_2020,
  title = {{{SciPy}} 1.0: Fundamental Algorithms for Scientific Computing in {{Python}}},
  author = {Virtanen, Pauli and Gommers, Ralf and Oliphant, Travis E. and Haberland, Matt and Reddy, Tyler and Cournapeau, David and Burovski, Evgeni and Peterson, Pearu and Weckesser, Warren and Bright, Jonathan and {van der Walt}, St{\'e}fan J. and Brett, Matthew and Wilson, Joshua and Millman, K. Jarrod and Mayorov, Nikolay and Nelson, Andrew R. J. and Jones, Eric and Kern, Robert and Larson, Eric and Carey, C. J. and Polat, {\.I}lhan and Feng, Yu and Moore, Eric W. and VanderPlas, Jake and Laxalde, Denis and Perktold, Josef and Cimrman, Robert and Henriksen, Ian and Quintero, E. A. and Harris, Charles R. and Archibald, Anne M. and Ribeiro, Ant{\^o}nio H. and Pedregosa, Fabian and {van Mulbregt}, Paul},
  year = {2020},
  month = mar,
  journal = {Nat Methods},
  volume = {17},
  number = {3},
  pages = {261--272},
  publisher = {Nature Publishing Group},
  issn = {1548-7105},
  doi = {10.1038/s41592-019-0686-2},
  copyright = {2020 The Author(s)},
  langid = {english}
}

@article{voyer_2025,
  title = {{{MIRI-LRS Spectrum}} of a {{Cold Exoplanet}} around a {{White Dwarf}}: {{Water}}, {{Ammonia}}, and {{Methane Measurements}}},
  author = {Voyer, Ma{\"e}l and Changeat, Quentin and Lagage, Pierre-Olivier and Tremblin, Pascal and Waters, Rens and G{\"u}del, Manuel and Henning, Thomas and Absil, Olivier and Barrado, David and Boccaletti, Anthony and Bouwman, Jeroen and Coulais, Alain and Decin, Leen and Glauser, Adrian M. and Pye, John and Glasse, Alistair and Gastaud, Ren{\'e} and Kendrew, Sarah and Patapis, Polychronis and Rouan, Daniel and {van Dishoeck}, Ewine F. and {\"O}stlin, G{\"o}ran and Ray, Tom P. and Wright, Gillian},
  year = {2025},
  month = mar,
  journal = {ApJL},
  volume = {982},
  number = {2},
  pages = {L38},
  publisher = {The American Astronomical Society},
  issn = {2041-8205},
  doi = {10.3847/2041-8213/adbd46},
  langid = {english}
}

@article{wakeford_2018,
  title = {The {{Complete Transmission Spectrum}} of {{WASP-39b}} with a {{Precise Water Constraint}}},
  author = {Wakeford, H. R. and Sing, D. K. and Deming, D. and Lewis, N. K. and Goyal, J. and Wilson, T. J. and Barstow, J. and Kataria, T. and Drummond, B. and Evans, T. M. and Carter, A. L. and Nikolov, N. and Knutson, H. A. and Ballester, G. E. and Mandell, A. M.},
  year = {2018},
  month = jan,
  journal = {AJ},
  volume = {155},
  number = {1},
  pages = {29},
  publisher = {The American Astronomical Society},
  issn = {1538-3881},
  doi = {10.3847/1538-3881/aa9e4e},
  langid = {english}
}

@article{waldmann_2015-1,
  title = {Tau-{{REx I}}: {{A}} next Generation Retrieval Code for Exoplanetary Atmospheres},
  author = {Waldmann, I. P. and Tinetti, G. and Rocchetto, M. and Barton, E. J. and Yurchenko, S. N. and Tennyson, J.},
  year = {2015},
  month = mar,
  journal = {ApJ},
  volume = {802},
  number = {2},
  pages = {107},
  publisher = {American Astronomical Society},
  issn = {0004-637X},
  doi = {10.1088/0004-637X/802/2/107},
  langid = {english}
}

@article{welbanks_2019,
  title = {Mass--{{Metallicity Trends}} in {{Transiting Exoplanets}} from {{Atmospheric Abundances}} of {{H2O}}, {{Na}}, and {{K}}},
  author = {Welbanks, Luis and Madhusudhan, Nikku and Allard, Nicole F. and Hubeny, Ivan and Spiegelman, Fernand and Leininger, Thierry},
  year = {2019},
  month = dec,
  journal = {ApJL},
  volume = {887},
  number = {1},
  pages = {L20},
  publisher = {The American Astronomical Society},
  issn = {2041-8205},
  doi = {10.3847/2041-8213/ab5a89},
  langid = {english}
}

@article{welbanks_2023,
  title = {On the {{Application}} of {{Bayesian Leave-one-out Cross-validation}} to {{Exoplanet Atmospheric Analysis}}},
  author = {Welbanks, Luis and McGill, Peter and Line, Michael and Madhusudhan, Nikku},
  year = {2023},
  month = feb,
  journal = {AJ},
  volume = {165},
  number = {3},
  pages = {112},
  publisher = {The American Astronomical Society},
  issn = {1538-3881},
  doi = {10.3847/1538-3881/acab67},
  langid = {english}
}

@article{welbanks_2025,
  title = {The {{Challenges}} of {{Detecting Gases}} in {{Exoplanet Atmospheres}}},
  author = {Welbanks, Luis and Nixon, Matthew C. and McGill, Peter and Tilke, Lana J. and Wiser, Lindsey S. and Rotman, Yoav and Mukherjee, Sagnick and Feinstein, Adina and Line, Michael R. and Seager, Sara and Beatty, Thomas G. and Seligman, Darryl Z. and Parmentier, Vivien and Sing, David},
  year = {2025},
  month = apr,
  number = {arXiv:2504.21788},
  eprint = {2504.21788},
  primaryclass = {astro-ph},
  publisher = {arXiv},
  doi = {10.48550/arXiv.2504.21788},
  archiveprefix = {arXiv}
}

@article{yurchenko_2017,
  title = {A Hybrid Line List for {{CH4}} and Hot Methane Continuum},
  author = {Yurchenko, Sergei N. and Amundsen, David S. and Tennyson, Jonathan and Waldmann, Ingo P.},
  year = {2017},
  month = sep,
  journal = {A\&A},
  volume = {605},
  pages = {A95},
  publisher = {EDP Sciences},
  issn = {0004-6361, 1432-0746},
  doi = {10.1051/0004-6361/201731026},
  copyright = {{\copyright} ESO, 2017},
  langid = {english}
}

@article{yurchenko_2020,
  title = {{{ExoMol}} Line Lists -- {{XXXIX}}. {{Ro-vibrational}} Molecular Line List for {{CO2}}},
  author = {Yurchenko, S N and Mellor, Thomas M and Freedman, Richard S and Tennyson, J},
  year = {2020},
  month = aug,
  journal = {MNRAS},
  volume = {496},
  number = {4},
  pages = {5282--5291},
  issn = {0035-8711},
  doi = {10.1093/mnras/staa1874}
}

@article{espinoza_2021,
  title = {Constraining {{Mornings}} and {{Evenings}} on {{Distant Worlds}}: {{A}} New {{Semianalytical Approach}} and {{Prospects}} with {{Transmission Spectroscopy}}},
  author = {Espinoza, N{\'e}stor and Jones, Kathryn},
  year = {2021},
  month = sep,
  journal = {AJ},
  volume = {162},
  number = {4},
  pages = {165},
  publisher = {The American Astronomical Society},
  issn = {1538-3881},
  doi = {10.3847/1538-3881/ac134d},
  langid = {english}
}

@article{blecic_2017,
  title = {The {{Implications}} of {{3D Thermal Structure}} on {{1D Atmospheric Retrieval}}},
  author = {Blecic, Jasmina and {Dobbs-Dixon}, Ian and Greene, Thomas},
  year = {2017},
  month = oct,
  journal = {ApJ},
  volume = {848},
  number = {2},
  pages = {127},
  publisher = {The American Astronomical Society},
  issn = {0004-637X},
  doi = {10.3847/1538-4357/aa8171},
  langid = {english}
}

@article{caldas_2019,
  title = {Effects of a Fully {{3D}} Atmospheric Structure on Exoplanet Transmission Spectra: Retrieval Biases Due to Day--Night Temperature Gradients},
  author = {Caldas, A. and Leconte, J. and Selsis, F. and Waldmann, I. P. and Bord{\'e}, P. and Rocchetto, M. and Charnay, B.},
  year = {2019},
  month = mar,
  journal = {A\&A},
  volume = {623},
  pages = {A161},
  publisher = {EDP Sciences},
  issn = {0004-6361, 1432-0746},
  doi = {10.1051/0004-6361/201834384},
  copyright = {{\copyright} A. Caldas et al. 2019},
  langid = {english}
}

@article{barstow_2020-1,
  title = {A Comparison of Exoplanet Spectroscopic Retrieval Tools},
  author = {Barstow, Joanna K and Changeat, Quentin and Garland, Ryan and Line, Michael R and Rocchetto, Marco and Waldmann, Ingo P},
  year = {2020},
  month = apr,
  journal = {MNRAS},
  volume = {493},
  number = {4},
  pages = {4884--4909},
  issn = {0035-8711},
  doi = {10.1093/mnras/staa548}
}

@article{nixon_2024,
  title = {Methods for {{Incorporating Model Uncertainty}} into {{Exoplanet Atmospheric Analysis}}},
  author = {Nixon, Matthew C. and Welbanks, Luis and McGill, Peter and Kempton, Eliza M.-R.},
  year = {2024},
  month = may,
  journal = {ApJ},
  volume = {966},
  number = {2},
  pages = {156},
  publisher = {The American Astronomical Society},
  issn = {0004-637X},
  doi = {10.3847/1538-4357/ad354e},
  langid = {english}
}
\bibliographystyle{aa}

\begin{appendix}

\section{Opacity sources}

\begin{table}[!ht]    
    \centering
    \caption{References for opacity data used in this work.}
    \begin{tabularx}{\hsize}{lY}
        \toprule\toprule
        Name    & Reference \\
        \midrule
        \multicolumn{2}{c}{Molecular cross-sections} \\
        \midrule
        \ce{CH3}  & \citet{adam_2019}      \\
        \ce{CH4}  & \citet{yurchenko_2017} \\
        \ce{C2H2} & \citet{chubb_2020}     \\
        \ce{C2H4} & \citet{mant_2018}      \\
        \ce{CO}   & \citet{li_2015}        \\
        \ce{CO2}  & \citet{yurchenko_2020} \\
        \ce{HCN}  & \citet{barber_2014}    \\
        \ce{H2O}  & \citet{polyansky_2018} \\
        \ce{H2S}  & \citet{azzam_2016}     \\
        \ce{NH3}  & \citet{coles_2019}     \\
        \ce{SO2}  & \citet{underwood_2016} \\
        \midrule
        \multicolumn{2}{c}{Collision-induced absorption (CIA)} \\
        \midrule
        \ce{H2-H2} & \citet{abel_2011,fletcher_2018} \\
        \ce{H2-He} & \citet{abel_2012} \\
        \bottomrule
    \end{tabularx}
    \label{tab:opacity-sources}
\end{table}

\FloatBarrier
\section{Model tuning process and metrics}\label{app:model-tuning}
The baseline for our atmospheric forward model contains opacity contributions from CIA of \ce{H2-H2} and \ce{H2-He} pairs, Rayleigh scattering of all molecules, a flat-opacity cloud deck, and molecular absorption from \ce{H2O, CO2, CO, and CH4}.
We assign the index `0' to the baseline model, $M_0$. 
This model is extended through an investigation into model preference under variations of the considered molecular species.

We evaluate the preference of individual models using multiple metrics.
Firstly, we consider the Bayes factor, $B_{\mathrm{m}0}$, between forward models including new molecules, and the baseline model,
\begin{align}
    B_{m0} = \frac{E_m}{E_0}.
\end{align}
In this equation, $E_0$ and $E_\mathrm{m}$ represent the Bayesian evidence of the baseline model, and the extended model, respectively. 
We evaluate the natural logarithm of the Bayes factor, $\ln B_{\mathrm{m}0}$, based on the formalism suggested in \citet{kass_1995}, which compares the posterior odds of two models under the assumption of equal prior odds of the models.
The threshold values associated with this are given in Table~\ref{tab:bayes-factor-thresholds}.
If an extended forward model shows $\ln B_{\mathrm{m}0} > 3$ (corresponding to a posterior odds ratio of more than 20:1), we consider it as significant in our selection process and therefore include it in our finalised model setup.

Secondly, we compare the corrected Akaike Information Criterion (cAIC, henceforth referred to as $\Psi$) values for all models,
\begin{align}
    \Psi = -2 \log(\hat{\mathcal{L}}) + 2K + \frac{2K(K+1)}{n - K - 1},
\end{align}
where $n$ is the sample size (number of transmission spectrum data points in our case), $K$ is the number of free parameters (between 9 and 14 for the models we consider), and $\hat{\mathcal{L}}$ the maximised log-likelihood\footnote{
    We note that \texttt{multinest} reports the maximised log-likelihood values as part of the sampling run summary statistics in \texttt{[root]summary.txt} \citep{feroz_2009, buchner_2014}.
}.
$\Psi$ provides a second-order correction to the standard AIC for small sample sizes \citep[$n < 40K$,][]{burnham_2004}. 
Like the Bayes factor, the $\Psi$ is a relative metric, but uses a point-estimate (the maximised likelihood) rather then the marginalised likelihood value of each model.
In our model ensemble, we evaluate $\Delta_m = \Psi_\mathrm{min} - \Psi_i$,
%
%
where $\Psi_\mathrm{min}$ is the minium cAIC value within the ensemble, and $\Psi_m$ the cAIC value for the model $M_\mathrm{m}$.
Following the prescription of \citet{burnham_2004}, we judge that models with values of $\Delta_m < 2$ have considerable support compared to the model with $\Psi_\mathrm{min}$ (translating into a likelihood ratio of approximately 1:3).

Lastly, we calculate the reduced $\chi$-square metric, $\bar{\chi}_\nu^2$, connected to each model.
As with $\Psi$, $\bar{\chi}_\nu^2$ represents a model performance metric judged on a point-estimate as a result of the inference process. In our case, we calculate $\bar{\chi}_\nu^2$ for the median model,
\begin{align}
    \bar{\chi}^2_\nu = \frac{1}{\nu} \cdot \sum_{i=1}^n\frac{(O_i - \bar{M}_i)^2}{\sigma_i^2},
\end{align}
where $(O_i, \bar{M}_i, \sigma^2_i)$ represent the measurement, median model value, and variance associated with data point $i$, $n$ represents the sample size, and $\nu = n - K$ the degrees of freedom of a model with $K$ free parameters.
Compared to the other two metrics, $\bar{\chi}_\nu^2$ is an absolute metric measuring the weighted sum of square deviations normalised to the number of degrees of freedom for each model. A summary of the individual model tuning metric values used in our work is provided in Table~\ref{tab-app:model-tuning-metric}. In total, we run 23 retrievals during the model tuning.

We point out that a bottom-up model parameter space selection stands in contrast to the top-down model tuning process described in \citet{benneke_2013}.
Adding contributions to a small baseline model can exacerbate the value of the Bayes factor and lead to the spurious identification of molecules \citep{welbanks_2025}.
Our model tuning process starts from a constrained set of molecules based on prior analyses of WASP-39~b (e.g. \citealt{tsiaras_2018}, \citealt{wakeford_2018}, \citetalias{jtecers_2023}).
We then evaluate necessary changes to the small baseline model.
This model extension process does indeed lead to large changes in the Bayes factor (Table~\ref{tab-app:model-tuning-metric}).
However, we supplement our initial selection by an analysis of the model performance containing all iterations of the molecules selected in the first step.
With this, we confirm that a step-by-step extended model is still preferred over the initial baseline to a sufficiently large degree.
Additionally, the Bayes factor is also a purely relative metric.
The values of $\ln B_{\mathrm{m}0}$ for models 18, 19, 20, and 21 can be compared with model 22 to look at this selection from a top-down view.

The tuning process we perform in our work is also anchored to the data set it is performed on.
In Fig.~\ref{fig-app:bf-comparison}, we show the values of $\ln B_{\mathrm{m0}}$ for model cases run on RU-23 and CA-24.
We find that, while the relative behaviour of all three Bayes factor cases shows similar increases for the same models, the absolute values of $\ln B_{\mathrm{m}0}$ are significantly smaller for the retrievals performed on CA-24.
The overlapping favoured model setup cases based on each of the input spectra share the inclusion of \ce{H2S} and \ce{SO2}.
Only model comparisons based on SP-TW and RU-23 additionally favour the inclusion of \ce{C2H2} and \ce{HCN}.
However, as all three spectra considered in our work are based on the same underlying observation, we assume that they contain the same information about the nature of the atmosphere of WASP-39~b.
Therefore, we use the model setup selected from the model tuning performed on SP-TW (model $M_{22}$) homogeneously in the comparison with retrievals performed on RU-23 and CA-24.
A comprehensive list of all values for the evidence of each model, $\ln E$, can be found in the associated online repository.

\begin{onecolumn}
\begin{table*}[!ht]
    \centering
    \caption{Model evaluation metrics for the model tuning process based on the SP-TW data set.}
    \begin{tabularx}{\textwidth}{XcXXXXXX}
        \toprule\toprule
        ID & Model description &  $\ln E$ & $\ln B_\mathrm{m0}$ & $\Psi$ & $\Delta_m$ & \# FP & $\bar{\chi}^2_\nu$ \\
        \midrule
        \multicolumn{8}{c}{\textbf{Initial opacity search}} \\
        \midrule
        0  & Baseline (\ce{CO, CO2, CH4, H2O}) & 642.02 & --      & -1306.02 & 55.00  & 10 & 3.26  \\
        1  & Removed \ce{CO}                   & 634.52 & -7.50   & -1304.88 & 56.14  & 9  & 3.26  \\
        2  & Removed \ce{H2O}                  & 515.48 & -126.54 & -1063.76 & 297.26 & 9  & 6.00  \\
        3  & Removed \ce{CH4}                  & 642.73 & 0.71    & -1309.45 & 51.57  & 9  & 3.21  \\
        4  & Removed \ce{CO2}                  & 193.96 & -448.06 & -423.07  & 937.95 & 9  & 13.28 \\
        5  & Added \ce{SO2}                    & 647.65 & \textbf{5.63}    & -1323.51 & 37.51  & 11 & 3.06  \\
        6  & Added \ce{H2S}                    & 661.40 & \textbf{19.38}   & -1350.08 & 10.95  & 11 & 2.75  \\
        7  & Added \ce{HCN}                    & 647.20 & \textbf{5.18}    & -1322.10 & 38.92  & 11 & 3.08  \\
        8  & Added \ce{C2H2}                   & 648.95 & \textbf{6.92}    & -1321.43 & 39.60  & 11 & 3.09  \\
        9  & Added \ce{C2H4}                   & 640.31 & -1.72   & -1303.85 & 57.17  & 11 & 3.29  \\
        10 & Added \ce{CH3}                    & 641.84 & -0.18   & -1303.80 & 57.22  & 11 & 3.29  \\
        11 & Added \ce{NH3}                    & 641.02 & -1.00   & -1303.70 & 57.32  & 11 & 3.29  \\
        \midrule
        \multicolumn{8}{c}{\textbf{Combinations of \{\ce{C2H2, H2S, HCN, SO2}\}}} \\
        \midrule
        12 & Added \{\ce{H2S, HCN}\}             & 663.67 & 21.65   & -1348.69 & 12.33  & 12 & 2.77  \\
        13 & Added \{\ce{H2S, SO2}\}             & 662.52 & 20.49   & -1353.36 & 7.66   & 12 & 2.72  \\
        14 & Added \{\ce{HCN, SO2}\}             & 651.16 & 9.14    & -1329.87 & 31.15  & 12 & 2.99  \\
        15 & Added \{\ce{C2H2, SO2}\}            & 656.27 & 14.24   & -1345.07 & 15.95  & 12 & 2.81  \\
        16 & Added \{\ce{C2H2, HCN}\}            & 649.54 & 7.52    & -1327.72 & 33.30  & 12 & 3.02  \\
        17 & Added \{\ce{C2H2, H2S}\}            & 663.65 & 21.63   & -1354.81 & 6.21   & 12 & 2.70  \\
        18 & Added \{\ce{H2S, SO2, HCN}\}        & 664.08 & 22.06   & -1351.26 & 9.76   & 13 & 2.74  \\
        19 & Added \{\ce{C2H2, HCN, SO2}\}       & 657.63 & 15.60   & -1344.19 & 16.83  & 13 & 2.83  \\
        20 & Added \{\ce{C2H2, H2S, SO2}\}       & 666.06 & 24.03   & \textbf{-1361.02} & --   & 13 & 2.63 \\
        21 & Added \{\ce{C2H2, H2S, HCN}\}       & 664.01 & 21.99   & -1352.96 & 8.06   & 13 & 2.72  \\
        22 & Added \{\ce{C2H2, H2S, HCN, SO2}\}  & 665.88 & 23.85   & \textbf{-1358.32} & 2.70 & 14 & 2.66  \\
        \bottomrule
    \end{tabularx}
    \label{tab-app:model-tuning-metric}
    \tablefoot{
        The columns show the ID and corresponding molecular inventory of the forward models used for model tuning. The Bayesian evidence ($\ln E$) is used to calculate the Bayes factor ($\ln B_\mathrm{m0}$) in reference to the baseline model ($M_0$). The cAIC ($\Psi$) is used to calculate the relative metric $\Delta_m = \Psi_\mathrm{min} - \Psi_m$ (where $\Psi_\mathrm{min}$ corresponds to $M_{20}$). The last two columns indicate the number of free parameters in the model, as well as the reduced $\chi$-square metric ($\bar{\chi}^2_\nu$) of the median model.
    }
\end{table*}

\begin{figure*}[!ht]
    \includegraphics[width=\hsize]{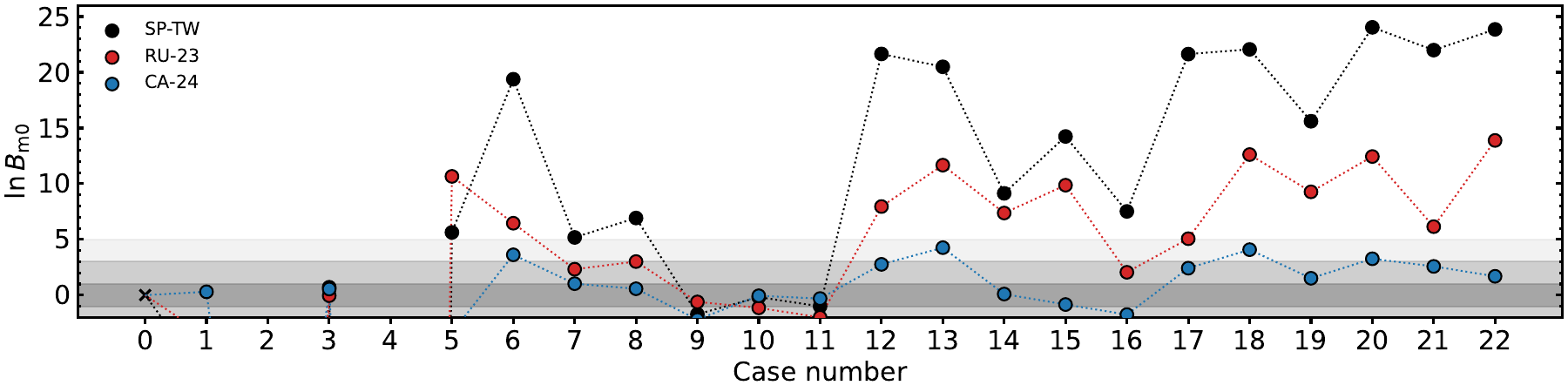}
    \caption{Values of the Bayes factor, $\ln B_{\mathrm{m}0}$, for all model setups considered in this work based on the data set SP-TW (black), RU-23 (red), and CA-24 (blue). All Bayes factors are computed in reference to the baseline model ($M_0$, see also Table~\ref{tab-app:model-tuning-metric}). Values of $\ln B_{\mathrm{m}0}$ are cut off for readability, and dotted lines are inserted as visual guides. The grey areas marked in the figure denote, in decreasing opacity, the Bayes factor threshold values given in Table~\ref{tab:bayes-factor-thresholds}.}
    \label{fig-app:bf-comparison}
\end{figure*}

\FloatBarrier
\clearpage\newpage
\begin{figure*}
    \section{Comparison of marginalised posteriors from scattered instances of SP-TW}
    \centering
    \includegraphics[width=\hsize]{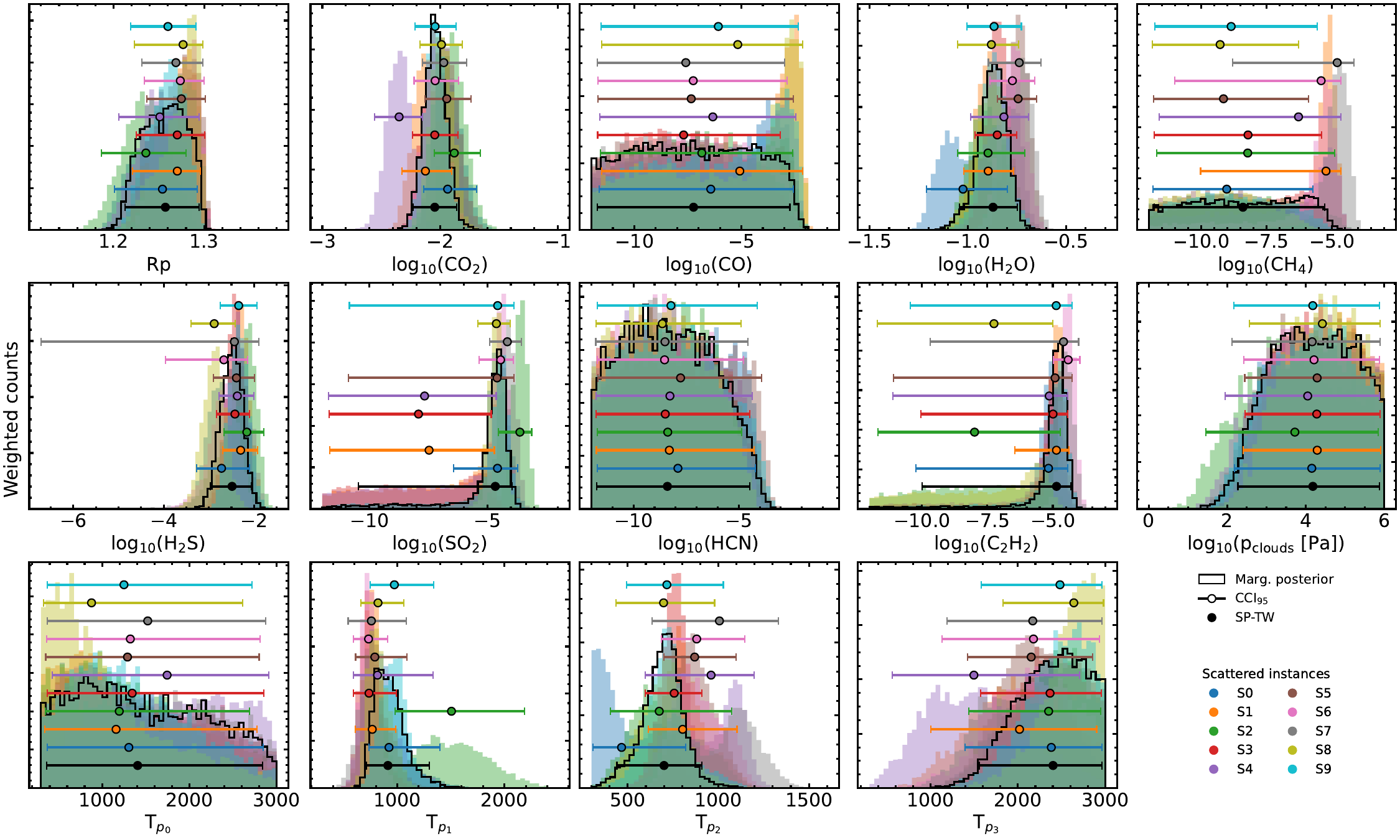}
    \caption{
        Same as Fig.~\ref{fig:posterior_scatter_comparison}, but for the marginalised posterior distributions of all parameters.
    }
    \label{fig-app:scatter_posterior_full}
\end{figure*}

\begin{table*}
    \centering
    \caption{
        $p$-value of the K-S statistic for all marginalised posterior distributions from retrievals of scattered instances of SP-TW.
    }
    \begin{tabularx}{\textwidth}{lYcYcYcYYcYYYcY}
        \toprule\toprule
        \multirow{2}{*}{} & \multicolumn{14}{c}{$\log_{10}(p_\mathrm{KS})$} \\
        \cmidrule(lr){2-15}
         & $R_\mathrm{p}$ & $\ce{CO2}$ & $\ce{CO}$ & $\ce{H2O}$ & $\ce{CH4}$ & $\ce{H2S}$ & $\ce{SO2}$ & $\ce{HCN}$ & $\ce{C2H2}$ & $p_\mathrm{cloud}$ & $T_{p_0}$ & $T_{p_1}$ & $T_{p_2}$ & $T_{p_3}$ \\
        \midrule
        0 & -45.1 & -- & -76.0 & -- & -138.9 & -- & -144.0 & -59.4 & -- & -1.5 & -16.9 & -23.5 & -- & -10.3 \\ 
        1 & -40.0 & -- & -75.5 & -- & -141.2 & -- & -145.4 & -55.0 & -- & -1.9 & -16.9 & -24.5 & -- & -9.6 \\ 
        2 & -41.3 & -- & -77.8 & -- & -141.1 & -- & -140.3 & -51.5 & -- & -1.5 & -17.1 & -25.0 & -- & -9.8 \\ 
        3 & -41.4 & -- & -76.0 & -- & -142.0 & -- & -141.0 & -54.0 & -- & -2.7 & -16.2 & -22.6 & -- & -10.0 \\ 
        4 & -44.1 & -- & -78.1 & -- & -140.2 & -- & -142.2 & -57.2 & -- & -2.2 & -16.1 & -24.1 & -- & -10.3 \\ 
        5 & -42.6 & -- & -75.1 & -- & -141.5 & -- & -144.1 & -52.9 & -- & -1.7 & -20.1 & -23.3 & -- & -10.5 \\ 
        6 & -43.0 & -- & -75.2 & -- & -141.0 & -- & -143.3 & -53.6 & -- & -1.7 & -18.9 & -24.2 & -- & -9.7 \\ 
        7 & -43.7 & -- & -76.6 & -- & -140.5 & -- & -140.7 & -54.9 & -- & -1.5 & -19.3 & -23.1 & -- & -10.4 \\ 
        8 & -39.2 & -- & -74.7 & -- & -140.7 & -- & -144.5 & -54.4 & -- & -1.2 & -20.4 & -22.1 & -- & -10.3 \\ 
        9 & -43.1 & -- & -77.3 & -- & -144.2 & -- & -139.6 & -54.3 & -- & -1.8 & -16.6 & -22.4 & -- & -10.4 \\ 
        \bottomrule
    \end{tabularx}
    \tablefoot{
        Two-sample K-S statistics are calculated between the marginalised posteriors of the original instance of SP-TW, and of the scattered instances for each parameter.
        We have used the \texttt{resample\_equal()} method as implemented in the \texttt{nestle} Python package to resample the samples to have equal weights\footnote{\url{https://github.com/kbarbary/nestle}}.
        Values reported in the table are the base-10 logarithm of the $p$-value associated with the K-S statistic, where empty entries correspond to values of $p_\mathrm{KS} = 0$.
    }
    \label{tab-app:scatter-posterior-pval}
\end{table*}
\end{onecolumn}

\FloatBarrier
\clearpage\newpage
\begin{twocolumn}
\section{Data reduction}\label{app:data-reduction}

To perform end-to-end data reduction, we use the open-source data reduction pipeline \texttt{Eureka!}\footnote{Version \texttt{0.10.dev0+g3c10926.d20230426}} \citep{bell_2022}.
\texttt{Eureka!} acts as both a wrapper for the official \texttt{jwst} pipeline \citep{bushouse_2024}, as well as a framework to perform light-curve fitting.
It has been used to successfully perform data reduction and spectra extraction on several JWST observations, including the observations conducted during the ERS programme \citep[e.g.][]{ahrer_2023, alderson_2023, feinstein_2023, rustamkulov_2023}, and multiple other exoplanets \citep[e.g.][]{august_2023, bell_2023, kempton_2023, lustig-yaeger_2023, moran_2023, dyrek_2024}.
\texttt{Eureka!} is also highly modular, supporting the fine-tuning of data reduction steps to ensure optimal precision in the reduced data products.

Stage 1 of \texttt{Eureka!} acts mostly as a wrapper for steps from the official \texttt{jwst} pipeline, which performs ramp-to-slope fitting.
Unless otherwise stated, we use the standard steps recommended for this stage, with the version \texttt{jwst=1.8.2}.
We deviate from it by using a 10\,$\sigma$ threshold for the \texttt{jump\_rejection} step, instead of the more constraining default threshold of 4\,$\sigma$.
As pointed by \citet{jtecers_2023} (referred to as \citetalias{jtecers_2023} from here onwards) and \citet{rustamkulov_2023}, a threshold this small, combined with the small number of groups for this observation, leads to excessive fractions of detector pixels being flagged as outliers.
We therefore choose the larger threshold of 10\,$\sigma$.
Additionally, we perform the group-level background subtraction (GLBS) step introduced in \texttt{Eureka!} v0.7, to account for 1/f-noise introduced during the detector read-out \citepalias[][]{jtecers_2023}.
We mask several pixel coordinates by hand that have, in this step and at time of data reduction, not yet been flagged as bad pixels\footnote{We refer to the online repository for this publication for the corresponding pixel map: \url{https://doi.org/10.5281/zenodo.15697941}}.
The background region for this step is defined as being outside of $y \in [5;22]$.
For the background, we fit a second-order polynomial, with an outlier rejection threshold of 5\,$\sigma$.
We skip the \texttt{refpix} step, as there are no reference pixels in this sub-array of the detector \citep{birkmann_2022}, as well as the \texttt{gain\_scale} step, as the relative flux-measurements of interest here do not necessitate flux calibration.

Similarly to stage 1, stage 2 of \texttt{Eureka!} is primarily a wrapper for the underlying \texttt{jwst} pipeline, which performs additional calibration and unit conversion steps.
Notably, we skip the \texttt{flat\_field} step, which at the time of our data reduction did not properly work due to incomplete reference files \citep[e.g.][]{alderson_2023,sarkar_2024}
We also skip the \texttt{photom} (count-rate to flux-density conversion) and \texttt{extract\_1d} steps (1D signal extraction), which are not necessary for our purposes.

In stage 3, we constrain the spectral data in the dispersion direction within the range $x \in [160;512]$.
We do this to exclude the saturated lower end of the spectrum.
Saturation in the wavelength region was intentionally achieved to increase the signal-to-noise ratio (S/N) for longer wavelengths \citepalias[][]{jtecers_2023}.
We define the saturation threshold conservatively as 60\% of the full-well capacity, to avoid potentially strong influences of the detector non-linearity.
This is illustrated in Fig.~\ref{fig-app:saturation}, which shows the mean count-rate after the first and fifth group the observation for detector row 16, which receives the majority of the signal.
As visible in this figure, the median signal over all integrations indicates that the columns below $y=160$ are above this threshold, which is where we set the lower limit (we note the exception of column 162, which oscillates around our limit and was included in our accepted range for consistency).
We perform a double-iteration outlier-rejection step along the time axis with a threshold of 5\,$\sigma$, and follow the optimal spectral extraction routine included in \texttt{Eureka!}.
For this, we use an aperture with a half-width of 6 pixels, centred at row 16, and a background half-width of 9 pixels.
We find to be the best combination in an effort to keep the spectral aperture half-width as large as possible \citep{horne_1986}, while maximising the overall precision of the spectrum.
We construct the spatial profile through the median frame, and perform outlier rejection with thresholds of $10\,\sigma$ during the construction of the spatial profile, and a $60\,\sigma$ during the optimal spectral extraction.

\begin{figure}
    \centering
    \includegraphics[width=\hsize]{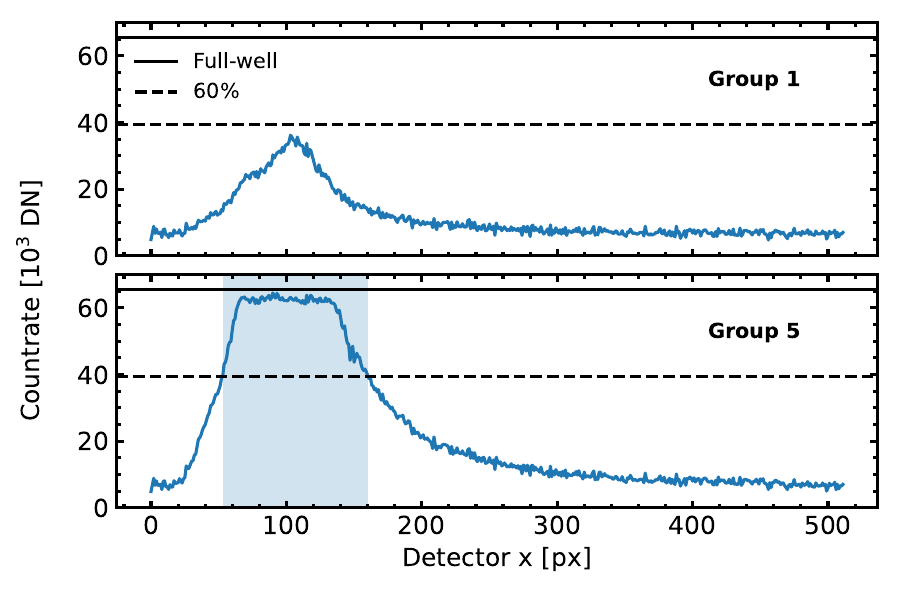}
    \caption{Median signal for row 16, showing dispersion-direction pixel position on the x-axis, and received signal in DN on the y-axis, for the first (top panel) and fifth (bottom panel) group. The solid black line denotes the full-well value, while the dashed black line marks the 60\,\% saturation threshold. The blue shaded area indicates the range in dispersion direction that falls above the 60\,\% saturation threshold.}
    \label{fig-app:saturation}
\end{figure}

\subsection{Extraction of spectroscopic LCs}\label{sec:lc-extraction}

In stage 4, we restrict the data to wavelengths below \SI{5.3}{\micro\meter}, as the throughput becomes negligible beyond this wavelength regime \citep{jakobsen_2022}.
We perform a final round of outlier-rejection in the temporal direction in this stage, using a rolling median from a box-car filter with a width of 100 data points and a rejection threshold of $5\,\sigma$, using a maximum of 10 iterations.
To identify outliers in the expected precision of the spectroscopic light-curves, we compare the median absolute deviation (MAD) value for each individual light-curve extracted by \texttt{Eureka!} to simulations performed with the JWST Exoplanet Simulator \citep[\texttt{JexoSim},][]{sarkar_2021}.
We perform a noise-budget simulation including all noise-sources considered in \texttt{JexoSim} and 10 realisations.
We use a simulation setup according the actual observational parameters, including NIRSpec PRISM instrument setup with 5 groups per integration and a fixed-bin size of one pixel.
System parameters are queried within \texttt{JexoSim}.
We then compare the calculated spectroscopic precision for our data reduction with the noise-budget from \texttt{JexoSim}, and flag individual detector-column positions exceeding a threshold of \num{1.75} times the photon-noise as deviations.
These detector-level channels are excluded from further analysis.
We show this in Fig.~\ref{fig:dr_precision}.
The integrated white-light curve is constructed by excluding these flagged detector columns, and performing an additional time-series outlier rejection, using a rolling median with an outlier rejection threshold of $3\,\sigma$.

To calculate the limb-darkening coefficients corresponding to each spectroscopic, as well as the white-light curve, we use the \texttt{ExoTiC-LD} Python package \citep{grant_2022} incorporated in \texttt{Eureka!}.
We use the stellar parameters given in Table~\ref{tab:wasp39-system-pars}, and the {\sc Stagger} grid of stellar models \citep{magic_2015}.

\begin{figure}
    \centering
    \includegraphics[width=\hsize]{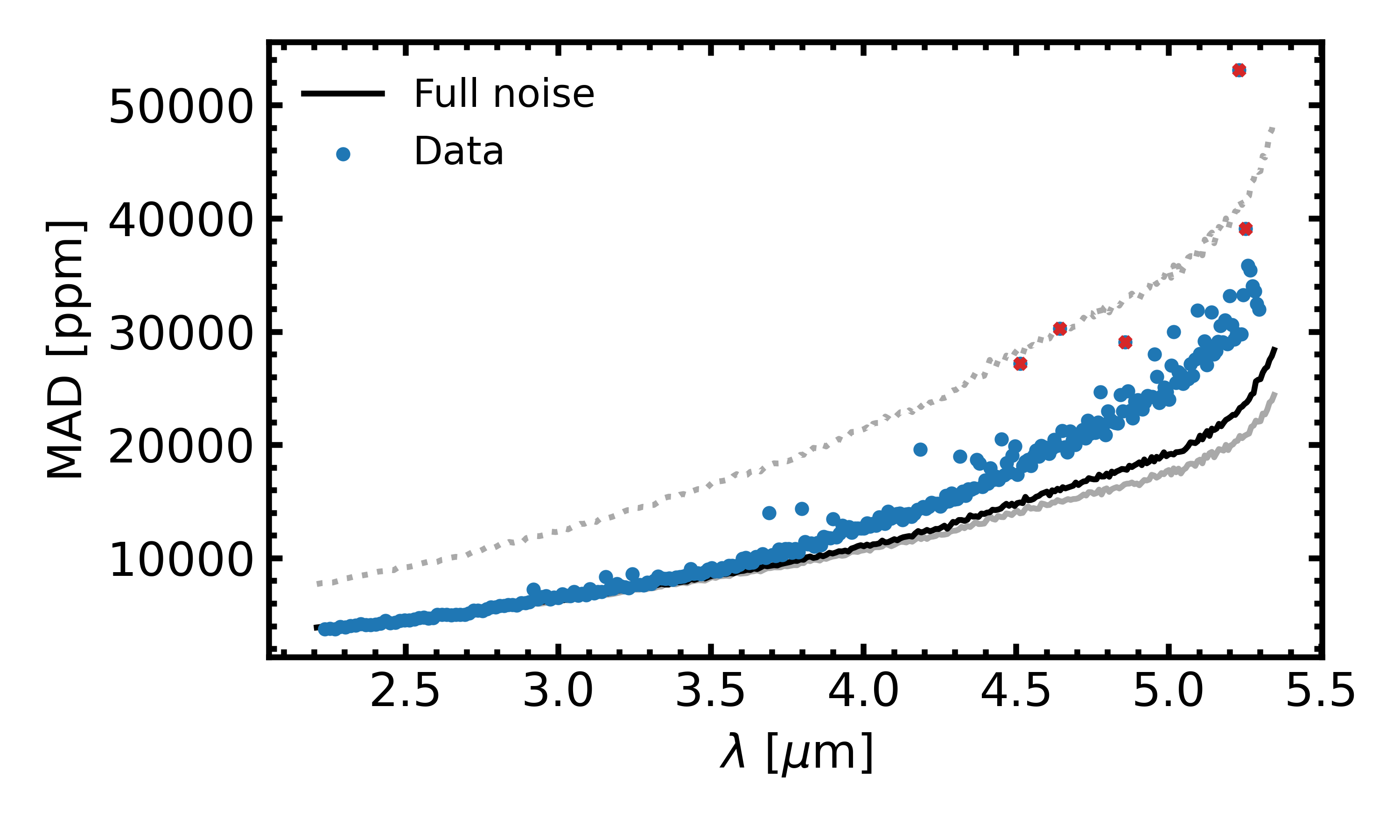}
    \caption{Spectroscopic light-curve precision, showing wavelength (in $\si{\micro\meter}$) on the x-axis, and the median-absolute deviation (MAD, in ppm) on the y-axis. The MAD-values from our data reduction (blue points) are compared to the photon-noise contribution (grey solid line) from a \texttt{JexoSim} simulation. The black solid line shows the corresponding full noise-budget, and the grey dashed line represents twice the photon-noise contribution. Individual channels deviation by a factor of 1.75 from the photon-noise contribution (red crosses) are not considered in further analysis.}
    \label{fig:dr_precision}
\end{figure}

\subsection{Light-curve fitting}\label{sec:lc-fitting}

We use a combined astrophysical and systematics model to fit both the pixel-resolution spectroscopic light-curves, and the integrated white light-curve resulting from stage 4 of \texttt{Eureka!}.
The Python-package \texttt{batman} \citep{kreidberg_2015} is used within \texttt{Eureka!} to calculate transit light-curves, based on a set of astrophysical parameters for the star-planet system through the fraction of blocked stellar flux by the transiting planet.
The area of the stellar disk obscured by the transiting planet is determined by the planet radius in units of the host star radius ($R_\mathrm{p} / R_*$). 
The amount of missing flux depends on the projected location of the planet on the stellar disk, and is determined through the time of inferior conjunction ($t_0$), the orbital period ($P$), the semi-major axis in units of the host star radius ($a/R_*$), as well as the orbital eccentricity ($e$) and longitude of periastron ($w$). 

An additional influencing factor in the astrophysical transit model is the limb-darkening description.
Stellar limb-darkening characterises the intensity-gradient of the sky-projected stellar disk.
In the case of our fitting routine, we use pre-calculated limb-darkening coefficients from the \texttt{ExoTiC-LD} framework \citep{grant_2022} corresponding to a 4-parameter non-linear limb-darkening law \citep{claret_2000}.
We choose the 4-parameter limb-darkening prescription over the commonly employed quadratic limb-darkening law, which has been shown to introduce biases in the retrieved transit depth \citep[e.g.][]{morello_2017, keers_2024}.
Limb-darkening parameters are calculated for the integrated white-light curve, as well as for the individual spectroscopic light-curves based on the stellar parameters listed in Table~\ref{tab:wasp39-system-pars}.

As a model for instrument-dependent systematics, we fit a quadratic global trend with three polynomial coefficients ($c_0$, $c_1$, $c_2$) in time to the median-normalised spectroscopic and integrated white-light curves to produce a combined model.
We fit for a total of seven free parameters in the case of the integrated white-light curve.
For the spectroscopic light-curves, we fit four free parameters, assuming that the inclination, $i$, time of inferior conjunction, $t_0$, and scaled semi-major axis, $a / R_*$ are wavelength-independent. 
These parameters, as well as their associated priors, and the fixed parameters of the astrophysical model are given in Table~\ref{tab:mcmc-pars}.

\begin{table}
    \caption{Light-curve fitting parameters}
    \label{tab:mcmc-pars}
    \centering
    \begin{tabularx}{\hsize}{lcYY}
        \toprule\toprule
        Parameter & Unit & Prior / Value & Application \\
        \midrule
        \multicolumn{4}{c}{\textbf{Astrophysical model}} \\
        \midrule
        $R_\mathrm{p}$		& $R_\mathrm{J}$		& $\mathcal{N}(0.148, 0.015^2)$		& all \\
        $t_0$				& \si{\day}				& $\mathcal{U}(59770.81, 59770.86)$	& white \\
        $i$					& \si{\deg}				& $\mathcal{N}(87.83, 0.25^2)$		& white \\
        $a$					& $R_*$					& $\mathcal{N}(11.4, 1^2)$			& white \\
        $P$					& \si{\day}				& $4.0552765$						& fixed \\
        $e$					& --					& $0$								& fixed \\
        $\omega$ 			& --					& $90$								& fixed \\
        \midrule 
        \multicolumn{4}{c}{\textbf{Systematics model}} \\
        \midrule
        $c_0$	& --	& $\mathcal{N}(1, 0.05^2)$	& all \\
        $c_1$	& --	& $\mathcal{N}(0, 0.01^2)$	& all \\
        $c_2$	& --	& $\mathcal{N}(0, 0.01^2)$	& all \\
        \bottomrule
    \end{tabularx}
    \tablefoot{
        For the application of our combined model parameters, `all' refers to the parameter being free in both the fitting routines for the spectroscopic and white light-curves; `white' refers to the parameter being free in the integrated white light-curve fit, and then fixed in the spectroscopic light-curve fit; `fixed' denotes parameters that are fixed in all cases. For Gaussian priors, we list the mean and standard deviation, $\mathcal{N}(\mu, \sigma^2)$. For uniform priors, we list the upper and lower boundaries of the prior space, $\mathcal{U}(a, b)$. Gaussian-prior parameters in the astrophysical model are taken from the values of \citet{mancini_2018}.
    }
\end{table}

To fit the combined model to each light-curve, we use the Markov chain Monte Carlo (MCMC) sampling algorithm \texttt{emcee} \citep{foreman-mackey_2013}.
In all cases, we run the MCMC chain with \num{5000} steps using a burn-in phase of \num{500} steps and \num{20} walkers. 
We show the mean auto-correlation time for each spectroscopic channel in Fig.~\ref{fig:lcfit-autocor}, and find a mean auto-correlation time of approximately \num{60} steps for the white-light curve fit, which we judge as sufficient for convergence of our MCMC chains \citep{foreman-mackey_2013}.
A summary of the light-curve fitting is shown in Fig.~\ref{fig:data_reduction_multiplot}, comparing the two-dimensional, pixel-resolution light-curve data to the model fit results and associated fit residuals. Individual columns of increased noise can be seen in the figure, which are flagged as excessively noisy and excluded from the final transmission spectrum.

\begin{figure}
    \centering
    \includegraphics[width=\hsize]{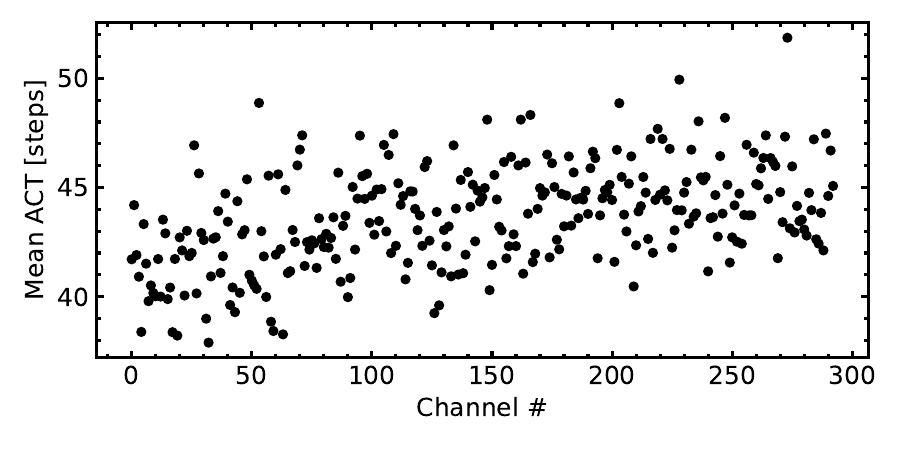}
    \caption{Convergence of the MCMC chains for each spectroscopic light-curve fit, showing the channel number on the x-axis, and mean auto-correlation time (in number of steps) on the y-axis. We point out that the associated mean auto-correlation time for the white light-curve fit is \num{60} steps.}
    \label{fig:lcfit-autocor}
\end{figure}

\begin{figure}
    \centering
    \includegraphics[width=\hsize]{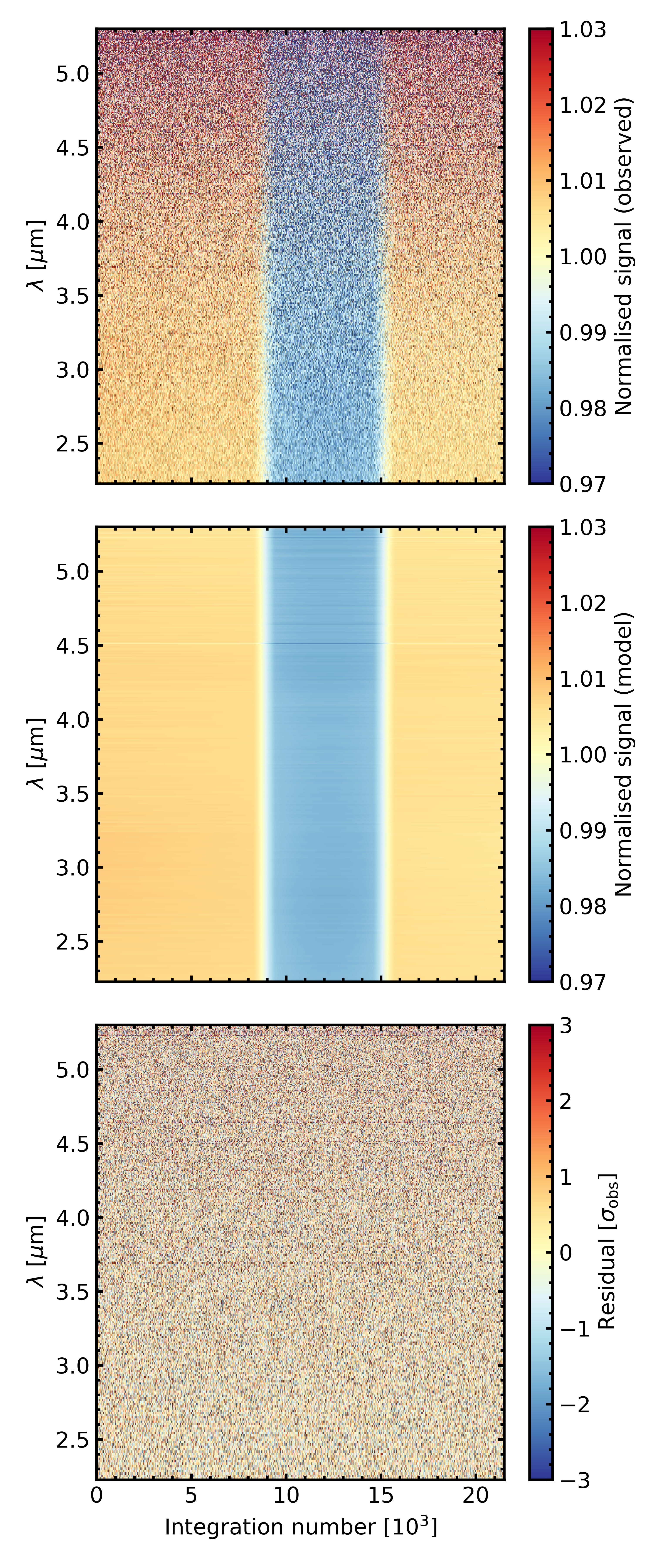}
    \caption{Data reduction results from \texttt{Eureka!}, showing integration number (equivalent to time) on the x-axis, and wavelength (in \si{\micro\meter}) on the y-axis.
        Top: Dynamic light-curve as resulting from stage 4 of \texttt{Eureka!}, with the colour-bar indicating mean-normalised signal.
        Middle: Combined systematic and astrophysical light-curve fits from stage 5 of \texttt{Eureka!}, with the colour-map representing the same parameters as in the left panel.
        Bottom: Fit residuals, normalised to the measurement-associated uncertainty.}
    \label{fig:data_reduction_multiplot}
\end{figure}

\subsection{Transmission spectrum binning}\label{sec:td-binning}
			
To produce the finalised transmission spectrum, we bin the pixel-resolution transit depth values from our spectroscopic light-curve fits using a weighted arithmetic mean (WAM) with a fixed bin size of three,
\begin{align}
    \bar{\delta} = \frac{\sum_{i=0}^{2} w_i \delta_i}{\sum_{i=0}^{2} w_i},
\end{align}
where $\bar{\delta}$ represents the binned transit depth value, with is derived from pixel-resolution transit depth values, $\delta_i$.
The corresponding weights, $w_i$, are defined through the inverse of the associated variance, $\sigma_i^2$. 
We note that the light-curve fitting determines the transit depth value (as all other parameters in the transit model) through Bayesian inference using MCMC sampling.
The reported error bars on $\delta_i$ are defined as median-centred credible intervals, which are not necessarily symmetric.
To calculate the weights, we therefore determine the arithmetic mean of the positive and negative error bars for each transit depth point.

A constant bin size of three pixels accounts for the typically assumed resolution element size of \num{2.2} pixels for NIRSpec \citep{jakobsen_2022}. We show the finalised transmission spectrum in Fig.~\ref{fig:finalised-tm-spec}, together with the underlying values on a pixel-resolution basis, where flagged pixel-columns (see also Fig.~\ref{fig:dr_precision}) are marked as red data points, and excluded from the binning process.
\end{twocolumn}

\begin{onecolumn}
\begin{figure*}
	\centering
	\includegraphics[width=\hsize]{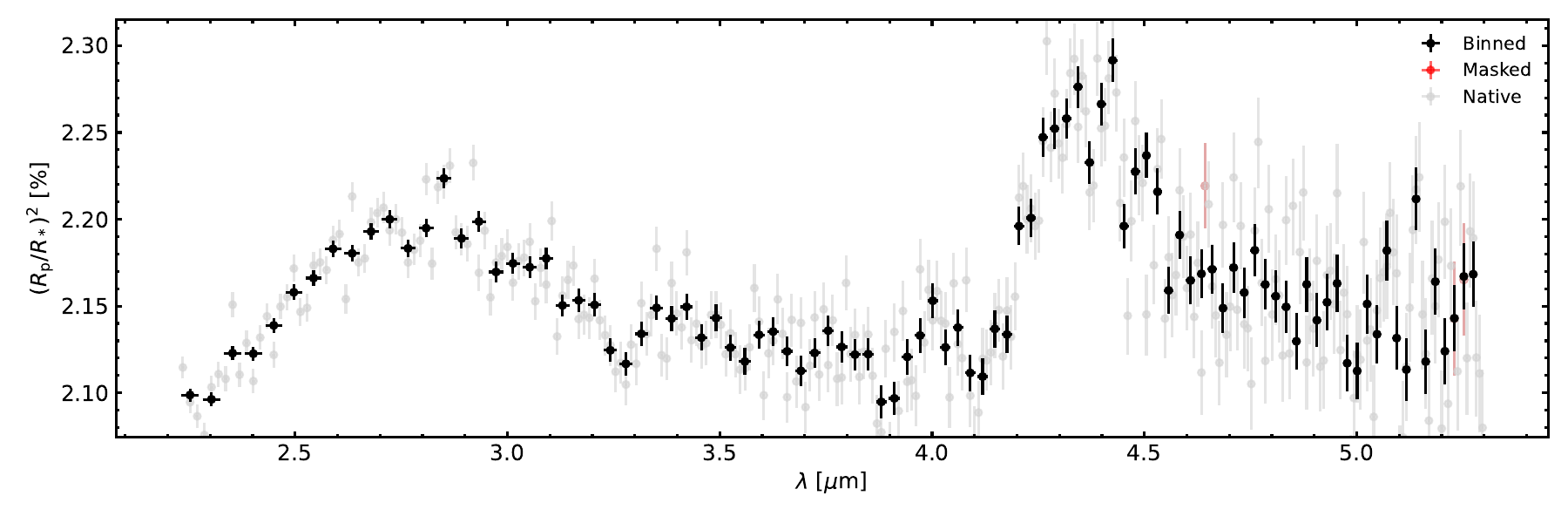}
	\caption{
        Transmission spectrum of WASP-39~b produced in this work, showing wavelength (in $\si{\micro\meter}$) on the x-axis against transit depth (in \%) on the y-axis. 
        Grey data points represent the transit depth values extracted at pixel-resolution level.
        Black data points show the finalised transmission spectrum, binned in 3-pixel intervals. Data points marked in red are excluded from this based on Fig.~\protect\ref{fig:dr_precision}.
    }
	\label{fig:finalised-tm-spec}
\end{figure*}


\subsection{Comparison with existing reductions}
Table~\ref{tab-app:dr-diff} lists the differences between the data reduction steps applied in this work to derive SP-TW, the \texttt{Eureka!}-\texttt{ExoTEP} data reduction presented in \citet{rustamkulov_2023} to derive RU-23, and the \texttt{FIREFLy}-based data reduction presented in \citetalias{carter_2024} to derive CA-24.
Steps not listed are performed with equal assumptions in all three cases.
\begin{table*}[!ht]
    \centering
    \caption{Main data reduction differences between SP-TW, RU-23, and CA-24.}
    \begin{tabularx}{\textwidth}{lYYY}
    \toprule\toprule
    Step & SP-TW & RU-23 & CA-24 \\
    \midrule
    \texttt{jwst} pipeline version & 1.8.2 & 1.6.0 & 1.6.2 \\
    \midrule
    \multicolumn{4}{c}{\textbf{Stage 1 and 2}} \\
    \midrule
    Bias subtraction & \texttt{jwst\_nirspec\_superbias\_0299} & Custom & Custom \\
    Dark current subtraction & Yes & Yes & No \\
    Reference pixel correction & No & Top / bottom 6 px & Top / bottom 6 px \\
    Jump rejection & 10$\,\sigma$ & No & No \\
    GLBS & 2nd order polynomial & Median (top / bottom 6 px) & Yes, but unspecified \\
    \midrule
    \multicolumn{4}{c}{\textbf{Stage 3}} \\
    \midrule
    Spectral half-width & 6 px & 4 px & Variable \\
    Time series outliers & 100 px box-car (5$\,\sigma$) & 20 px box-car (3$\,\sigma$) & unspecified \\
    \midrule
    \multicolumn{4}{c}{\textbf{Light-curve fitting}} \\
    \midrule
    Systematic trend & Quadratic & Linear & Linear \\
    Limb-darkening & 4-parameter (fixed) & Quadratic (fitted) & Quadratic (fitted) \\
    Error bar inflation & No & Yes ($\bar{\chi}^2_\nu$ to 1) & unspecified \\
    Pre-fit binning & No (pixel-level) & No (pixel-level) & Instrument resolution \\
    \bottomrule
    \end{tabularx}
    \label{tab-app:dr-diff}
    \tablefoot{
        We note that the CRDS context \texttt{jwst\_1202.pmap} was used to derive SP-TW.
    }
\end{table*}
\end{onecolumn}

\end{appendix}
\end{document}